%% file: header-free-format.tex
\documentclass{article}
\usepackage[utf8]{inputenc}
\usepackage{fullpage}
\usepackage[numbers]{natbib}
\usepackage{amsmath}
\usepackage{amsfonts}
\usepackage{hyperref}
\usepackage{mathtools}
\usepackage{graphicx}
\usepackage{verbatim}
\usepackage{subcaption}
\usepackage{adjustbox}
\usepackage{booktabs}
\usepackage{multirow}
\RequirePackage{etex}
\usepackage{pifont}

\usepackage{xcolor}
\definecolor{darkblue}{RGB}{0,0,128}
\definecolor{darkgreen}{RGB}{143, 39, 39}
\definecolor{correctcolor}{RGB}{0, 128, 0}
\definecolor{darkred}{RGB}{128, 0, 0}
\definecolor{black}{RGB}{0, 0, 0}
\definecolor{errorcolor}{RGB}{217, 63, 63}
\definecolor{viridisgreen}{HTML}{55C667}
\definecolor{amortizer}{RGB}{169,209,142}

\usepackage{tikz}
\usepackage{tkz-euclide}
\usetikzlibrary{positioning, shapes, arrows, fit, shapes.multipart}
\usepackage{adjustbox}

\makeatletter
\newcommand{\hathat}[1]{%
\begingroup%
  \let\macc@kerna\z@%
  \let\macc@kernb\z@%
  \let\macc@nucleus\@empty%
  \widehat{\raisebox{.35ex}{\vphantom{\ensuremath{#1}}}\smash{\widehat{#1}}}%
\endgroup%
}
\makeatother

\usepackage{todonotes}

\makeatletter
\renewcommand\@seccntformat[1]{\csname the#1\endcsname.\quad}
\makeatother

\newcommand{\newcontent}[1]{{\color{black}#1}}

\newcommand{\post}{p(\theta \mid y)}
\newcommand{\postpsi}{p\left(\psi(\theta) \mid y\right)}
\newcommand{\postpsir}{p\left(\psi \mid y\right)}
\newcommand{\posta}{p_{\text{A}}(\theta \mid y)}
\newcommand{\postd}{p(\theta \mid \tilde{y})}
\newcommand{\postad}{p_\text{A}(\theta \mid \tilde{y})}
\newcommand{\prior}{p(\theta)}
\newcommand{\priorpsi}{p\left(\psi(\theta)\right)}
\newcommand{\priorpsir}{p\left(\psi\right)}
\newcommand{\lik}{p(y \mid \theta)}
\newcommand{\likd}{p(\tilde{y} \mid \theta)}
\newcommand{\joint}{p(y, \theta)}

\newcommand{\data}{\tilde{y}}
\newcommand{\E}{\mathbb{E}}
\newcommand{\explicit}{\textrm{P}_E}
\newcommand{\implicit}{\textrm{P}_I}
\newcommand{\enp}{\textrm{ENP}}
\newcommand{\enc}{\textrm{ENC}}
\newcommand\monte{\stackrel{\mathclap{\normalfont\tiny\mbox{\textrm{MC}}}}{\approx}}

\newcommand\numberthis{\addtocounter{equation}{1}\tag{\theequation}}

\newcommand{\cmark}{\ding{51}}%
\newcommand{\xmark}{\ding{55}}%

\makeatletter
\def\@maketitle{%
\begin{center}   \let \footnote \thanks     {\large \@title \par}     {\normalsize       \begin{tabular}[t]{c}         \@author       \end{tabular}\par}     {\small \@date}   \end{center}
}
\makeatother

\begin{document}

\title{Some models are useful, but how do we know which ones? \\Towards a unified Bayesian model taxonomy
\vspace{.2in}}
\author{Paul-Christian Bürkner$^{1,2,*}$ \and Maximilian Scholz$^2$ \and Stefan T. Radev$^{3,4}$
\vspace{.2in}}
\date{$^1$ Department of Statistics, TU Dortmund University, Germany \break
$^2$ Cluster of Excellence SimTech, University of Stuttgart, Germany \break
$^3$ Cluster of Excellence STRUCTURES, University of Heidelberg, Germany \break
$^4$ Cognitive Science Department, Rensselaer Polytechnic Institute, NY, USA \break
$^*$ Corresponding author, Email: paul.buerkner@gmail.com}

\maketitle

\begin{abstract}
  \noindent
  Probabilistic (Bayesian) modeling has experienced a surge of applications in almost all quantitative sciences and industrial areas. This development is driven by a combination of several factors, including better probabilistic estimation algorithms, flexible software, increased computing power, and a growing awareness of the benefits of probabilistic learning. However, a principled Bayesian model building workflow is far from complete and many challenges remain. To aid future research and applications of a principled Bayesian workflow, we ask and provide answers for what we perceive as two fundamental questions of Bayesian modeling, namely (a) ``What actually \emph{is} a Bayesian model?'' and (b) ``What makes a \emph{good} Bayesian model?''. As an answer to the first question, we propose the PAD model taxonomy that defines four basic kinds of Bayesian models, each representing some combination of the assumed joint distribution of all (known or unknown) variables (P), a posterior approximator (A), and training data (D). As an answer to the second question, we propose ten utility dimensions according to which we can evaluate Bayesian models holistically, namely, (1) causal consistency, (2) parameter recoverability, (3) predictive performance, (4) fairness, (5) structural faithfulness, (6) parsimony, (7) interpretability, (8) convergence, (9) estimation speed, and (10) robustness. Further, we propose two example utility decision trees that describe hierarchies and trade-offs between utilities depending on the inferential goals that drive model building and testing. \hspace{2cm} \linebreak\linebreak 
\end{abstract}



\tableofcontents 
\newpage

\sloppy
\input{content}

\section*{Acknowledgments}

We thank Rudolf Debelak for helpful feedback on earlier versions of this paper.
Our work was partially funded by the Deutsche Forschungsgemeinschaft (DFG, German Research Foundation) under Germany's Excellence Strategy -- EXC-2075 - 390740016 (the Stuttgart Cluster of Excellence SimTech) and EXC-2181 - 390900948 (the Heidelberg Cluster of Excellence STRUCTURES). The authors gratefully acknowledge the support and funding.

\bibliography{bibfile}

\end{document}

%% file: content.tex
\section{Introduction}

Probabilistic (Bayesian) modeling has seen a surge of applications in almost all quantitative sciences and industrial areas \citep{BDA3, mcelreath2020statistical, gelman2020workflow, cranmer_sbi_2020, lavin_simulation_2021, izmailov_BNN_2021}. 
This development is driven by a combination of several factors, including powerful probabilistic estimation algorithms \citep{hoffman2014no, betancourt_hmc_2017, greenberg2019automatic, papamakarios2021normalizing, radev_amortized_2020}, efficient post-processing \citep{vehtari_2017_loo, gronau_bridgesampling_2020}, flexible open-source software \citep{stan_2022, duerr_deeplearning_2020, brms1}, and increased information processing capacity.
Furthermore, these factors are coupled with a growing awareness of the benefits of probabilistic modeling, such as inclusion of prior knowledge \citep{ohagan_expert_2019, mikkola_prior_2021}, regularization \citep{gelman_data_2006, brms2, bhadra_default_2016, piironen_sparsity_2017},  or uncertainty quantification and propagation \citep{hullermeier2021aleatoric, mcelreath2020statistical, gelman2020workflow}.

Despite these advances, creating and improving Bayesian models in the context of a principled Bayesian workflow \citep{schad_workflow_2021, gelman2020workflow} remains a complicated endeavor that requires expertise in various domains; these include subject matter knowledge about the system and the data it generates, statistical learning expertise, programming and understanding of software development, as well as knowledge of numerical approximation and simulation methods \citep{lavin_simulation_2021, gelman2020workflow}.
Thus, to aid future research on and applications of a principled Bayesian workflow, we ask and provide answers to what we hold to be two fundamental questions:

\begin{enumerate}
\itemsep0.5em 
    \item What actually \emph{is} a Bayesian model?
    \item What makes a \emph{good} Bayesian model?
\end{enumerate}

In current practice, the term \emph{Bayesian model} is highly overloaded and used to describe a wide range of objects with potentially very different properties.
Moreover, modern Bayesian models are more than just a likelihood and a prior -- rather, they resemble complex simulation programs coupled with black-box approximators, interacting with various data structures and context variables, embedded within iterative workflows with multiple feedback loops \cite{mayo2021computational, duran2020simulation, lavin_simulation_2021, gelman2020workflow, schad_workflow_2021}.
Thus, we aim to disambiguate and structure the different meanings of a Bayesian model by proposing the PAD model taxonomy (see Section \ref{bayesian_model}).
Our taxonomy aims to accommodate modern uses of Bayesian models and provides an answer to Question 1.
With a clear definition of Bayesian models in hand, we describe a collection of ten \emph{utility dimensions} that can be used to quantify the goodness of Bayesian models holistically (see Section~\ref{good_bayesian_model}), thus providing an answer for Question 2. 
We then continue with a discussion of importance hierarchies and common trade-offs between utilities in Section~\ref{hierarchies_tradeoffs} and end with a conclusion in Section~\ref{conclusion}.

\begin{table*}
\centering
\caption{Table of important symbols and their corresponding description.}
\begin{tabular}{ll}
\toprule
Notation (Symbol) & Meaning (Description)\\
\midrule
P, A, D & Joint distribution, approximator, training data \\
$\theta$, $y$, $\data$ & Latent parameters, unrealized observables, realized observables \\
$z, \xi$ & Random state, random noise (nuisance or exogenous variables) \\
$\varphi$, $\psi$ & Quantity of interest, its model-based estimator (function of $\theta$) \\
$p(\theta)$ & Prior distribution of parameters \\
$p(y \mid \theta)$ & Likelihood function (explicit or implicit/simulation-based)\\
$p(\theta, y)$ & Joint distribution of parameters and observables \\
$p(\theta \mid y)$ & Posterior distribution of parameters given observables \\
$p_{\text{A}}(\theta \mid y)$ & Approximate representation of posterior by approximator A \\
$\mathbb{G}(\cdot)$, $p^*(y)$ & True data generator, true data-generating distribution\\
$\mathbb{E}_p[\cdot]$ & Expected value of a quantity with respect to density $p$ \\
$T, H$ & Summary statistics of posterior, summary statistics of data\\
\bottomrule
\end{tabular}
\label{table:sym}
\end{table*}

This paper started as an attempt to organize our thoughts and provide a unifying and consistent language of Bayesian model building. 
To a certain extent, it is inevitably opinionated. 
Nevertheless, we aim to be comprehensive in the utility dimensions we discuss, such that all the goals we can sensibly ask from a Bayesian model to achieve have their place in this paper. 
In contrast, due to the large number of different topics we touch on in the process, the amount of details and cited literature per topic are necessarily non-exhaustive. 
The cited literature is only meant as a starting point for the interested reader to dive in deeper if they wish. 
In terms of the target audience, we hope that this paper will be helpful to both methodological researchers developing Bayesian models as well as users applying Bayesian models in practice.

\begin{table*}
\centering
\caption{List of important abbreviations and their corresponding definitions.}
\newcontent{
\begin{tabular}{lll}
\toprule
Abbreviation & Definition & Section\\
\midrule
BNN & Bayesian neural network & \ref{non-parametrics} \\ 
MCMC & Markov chain Monte Carlo & \ref{approx-explicit-models} \\ 
HMC & Hamiltonian Monte Carlo & \ref{approx-explicit-models} \\ 
VI & variational inference & \ref{approx-explicit-models} \\ 
KL (divergence) & Kullback-Leibler (divergence) & \ref{approx-explicit-models} \\ 
ABC & approximate Bayesian computation & \ref{approx-implicit-models} \\ 
SMC & sequential Monte Carlo & \ref{approx-implicit-models} \\ 
KDE & kernel density estimation & \ref{approx-implicit-models} \\ 
NDE & neural density estimation &  \ref{approx-implicit-models} \\ 
NPE & neural posterior estimation & \ref{approx-implicit-models} \\ 
SNPE & sequential neural posterior estimation & \ref{abi-vs-non-abi} \\ 
SCM & structural causal model & \ref{structural-causal-models} \\ 
DAG & directed acyclic graph & \ref{structural-causal-models} \\ 
HDI & highest density interval & \ref{ground-truth-comparisons} \\ 
SBC & simulation-based calibration & \ref{calibration} \\ 
ECDF & empirical cumulative distribution function & \ref{calibration} \\ 
ELPD & expected log predictive density & \ref{pp-prior-post} \\ 
ENP & effective number of parameters & \ref{P-parsimony} \\ 
LOO-CV & leave-one-out cross-validation & \ref{P-parsimony} \\ 
GLS & global-local shrinkage & \ref{P-parsimony} \\ 
ENC & effective number of coefficients & \ref{P-parsimony} \\ 
ESS & effective sample size & \ref{convergence-MCMC} \\ 
MCSE & Monte Carlo standard error & \ref{convergence-MCMC} \\ 
MAP (estimate) & maximum a posteriori (estimate) & \ref{convergence-optimization} \\ 
\bottomrule
\end{tabular}
}
\label{table:abb}
\end{table*}

\section{What is a Bayesian Model?}
\label{bayesian_model}

\begin{figure}[t]
\centering
\includegraphics[width=0.99\textwidth]{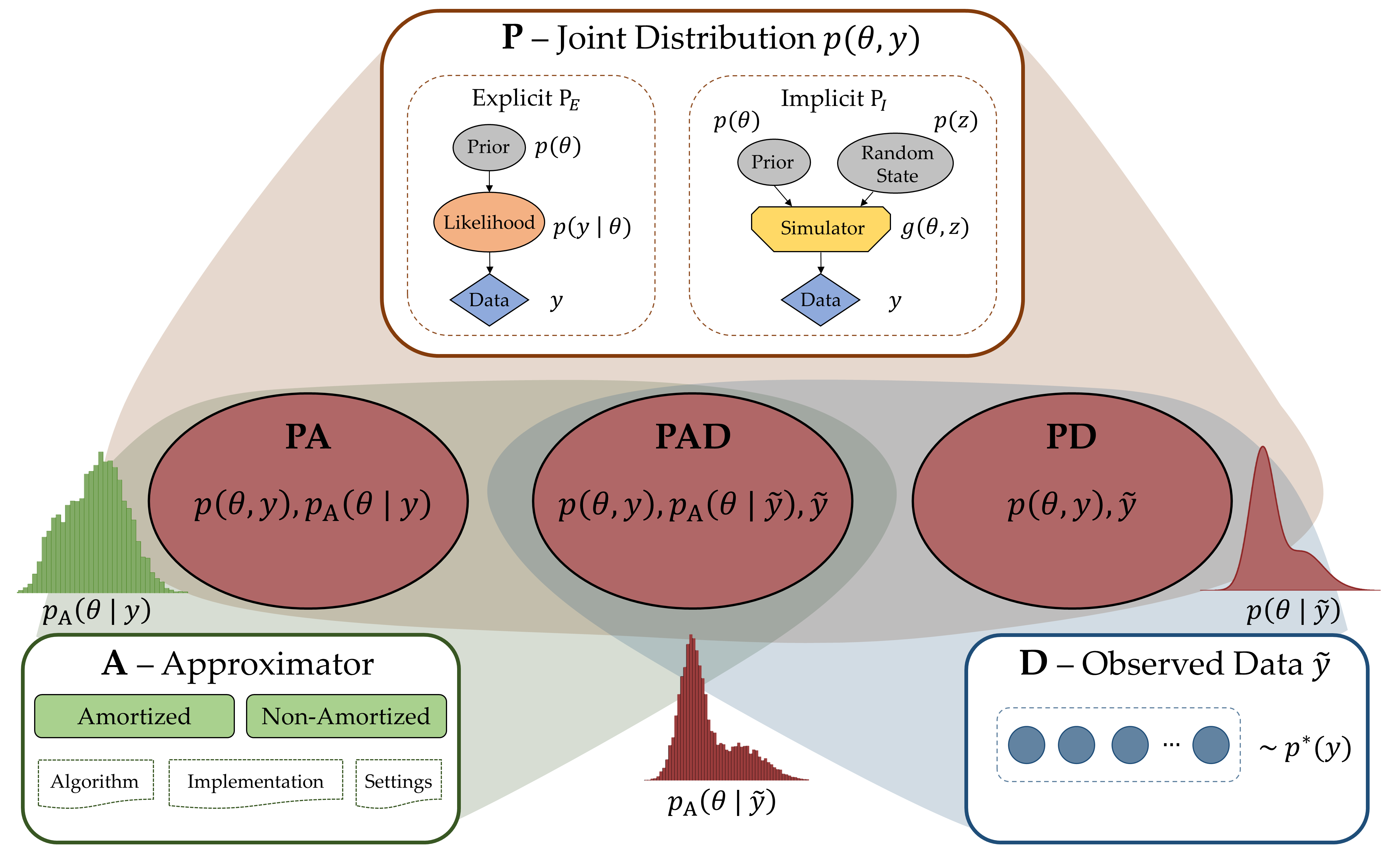}
\caption{The PAD Bayesian model taxonomy defines four basic kinds of Bayesian models. Each model kind represents a combination of the joint distribution of all random quantities (P), a posterior approximator (A), and observed data (D).} \label{fig:pad_model_taxonomie}
\end{figure}

As the term \textit{Bayesian model} (or just \textit{model} for that matter) can sustain multiple meanings depending on context, it can prove incredibly difficult to talk about models with sufficient clarity. 
As we will see later, different kinds of models may have different kinds of properties which need to be considered and prioritized by an analyst. 
Without clearly communicating the essential kind of model one has in mind, a discussion about its properties only contributes to the conceptual entropy in quantitative research.
In this section, we attempt to resolve this issue by proposing the PAD taxonomy for Bayesian models (see Figure \ref{fig:pad_model_taxonomie} for an overview; see also Table~\ref{table:sym} for a quick reference of key concepts and corresponding notation).
We will define four basic model classes and explain how they relate to each other. 
While the PAD taxonomy might be applicable and useful in other contexts, we will specifically expand on it from a Bayesian perspective.

\subsection{P Models} 
\label{sec:p_models}

We define P models by a joint probability distribution $\joint$ over all quantities of interest whose potential variation or uncertainty we express in terms of probability theory.
We assume that $y$ represents all observable quantities (i.e., data, observations, or measurements) and $\theta$ represents all unobservable quantities (i.e., parameters, latent states, or system variables) within a particular modeling context.
In most cases, the joint distribution factorizes into a likelihood $\lik$ and a prior $\prior$ via the chain rule of probability:
\begin{equation}
    \joint = \lik \; \prior
    \label{eq:joint_fact}
\end{equation}
\newcontent{This conceptually simple factorization serves as the basis for the common generative (forward) notation used to denote a ``probabilistic recipe'' for creating synthetic data by sequentially sampling from the prior and the likelihood:
\begin{align}
    \theta &\sim \prior\\
    y &\sim \lik
\end{align}
The generative notation overloads the semantics of the ``$\sim$'' operator, which attains a dual meaning of ``distributed as'' and ``sampled from''.}

Not all P models are created equal, but most are built to mimic a real-world process or a system, $\mathbb{G}$, whose behavior we can observe or measure.
Having some properties that are of interest to the analyst, the opaque generator $\mathbb{G}$ induces an unknown (true) data distribution $p^*(y)$, typically available only through finite observations $\data \sim p^*(y)$ (i.e., real-world data).
Accordingly, P models strive to encode probabilistic information about the true distribution $p^*(y)$ and/or structural information about the true generator $\mathbb{G}$.
The former means that our model matches the statistical properties of $p^*$ either \textit{a priori}, $p(y) \approx p^*(y)$, or \textit{a posteriori} $p(y \mid \data) \approx p^*(y)$, where $p(y)$ and $p(y \mid \data)$ are the prior and posterior predictive distributions of P, respectively.
The latter means that our parameters $\theta$ correspond to some relevant (hidden) properties $\varphi$ of $\mathbb{G}$, for which we endeavor to learn something by analyzing $\data$. 
We will expand on these goals in more detail in Section \ref{hierarchies_tradeoffs}.

P models are typically \emph{generative}, that is, we can obtain pseudo-random parameter and data draws via Monte Carlo simulations from Equation~\eqref{eq:joint_fact}.
The generative property presupposes that the prior is proper (i.e., its density function has a finite integral) and that efficient algorithms for sampling random draws from both $\prior$ and $\lik$ exist.

\newcontent{P models are the basic building blocks of all further model classes described in the upcoming sections. Moreover, due to their generative properties, standalone P models can be useful on their own for various \textit{forward inference} tasks. These include, for instance, exploring the stability of complex mechanistic equations \cite{koren2017exploring}, testing different prior assumptions before a model sees any real-world data \cite{best2020prior}, or venturing into computational philosophy using simulation \cite{mayo2021computational}.
As part of a Bayesian workflow, the plausibility of P models can already be evaluated through prior predictive or prior pushfoward checks \cite{schad2021toward}, which ultimately aim to determine whether the generative behavior of a P model is consistent with the available domain expertise.

\subsubsection{Non-Parametric P Models}
\label{non-parametrics}

In contrast to the above introduced parametric formulation, non-parametric P models replace the finite joint model $\joint$ with an infinite dimensional (functional) expression \cite{maceachern2016nonparametric}. 
Practically speaking, the number of parameters in such models simply grows with the number of observed data points \cite{mackay1998introduction, rasmussen_gps_2003, maceachern2016nonparametric}.
We may still assume that the observed data $\data$ is drawn from some unknown distribution $\data \sim p^*$, but then place a prior $p(f)$ over the set of all possible generating functions $f$, instead of over a finite-dimensional parameter space.
The forward (generative) model is thus given by:
\begin{align}
    f &\sim p(f)\\
    y &\sim p(y \mid f),
\end{align}
where the ``likelihood'' describes the probability of the data given a realization of the function $f$. 
For non-parametric regression models (e.g., Gaussian processes, \cite{rasmussen_gps_2003}), the function $f$ would also depend on additional inputs (i.e., predictors or covariates) and is thus restricted by the problem design.
The corresponding prior typically prescribes some properties of $f$, for instance, smoothness or certain frequency characteristics \cite{mackay1998introduction}, but may itself be non-analytic; still, it is often possible to obtain random draws from the generative model and compute marginal and conditional distributions. 

In between the parametric and non-parametric worlds, we can encounter high-dimensional P models, such as Bayesian neural networks (BNNs) \cite{mackay1995bayesian, izmailov_BNN_2021}.
The parameters $\theta$ of BNNs represent the set of trainable network weights and biases or a subset thereof, such as the weights and biases of the last hidden layer (for a practical overview of recent techniques, see \citep{jospin2022hands}).
The prior over network weights is typically chosen out of computational convenience \cite{foong2020expressiveness}, since there is hardly any domain expertise which can yield informative priors.
The likelihood of BNNs can also be an ostensibly simple distribution (e.g., a Gaussian) whose parameters are obtained through a highly nonlinear transformation defined by the computational graph of the network. Thus, even though BNNs are formally parametric P models, their high-dimensionality and non-linearity makes them behave more like non-parametric P models \cite{lee_deep_2018}.

This paper was conceptualized and written mainly with parametric P models in mind. That said, almost all of its aspects apply to non-parametric and high-dimensional P models as well, except, perhaps, for those that presuppose direct interest in the P model parameters (e.g., Section~\ref{parameter-recoverability}). Furthermore, despite theoretical differences \cite{maceachern2016nonparametric}, the practical treatment of parametric and non-parametric P models, when trained on finite data, is not radically different in the end. Finally, non-parametric P models commonly appear as local building blocks in otherwise parametric P models (e.g., a latent Gaussian process as part of an additive model; \cite{kolczynska_modeling_2021}), blurring the line even further.
\subsubsection{Explicit vs. Implicit Likelihood Models}
\label{explicit-implicit-models}

Thus far, we have emphasized that both parametric and non-parametric P models can be analyzed through the lens of their generative properties.
A common denominator in such forward inference tasks is that the P model's behavior (i.e., dynamic properties) may not be immediately obvious from the P model's specification alone (i.e., static properties).
Thus, \textit{simulation methods} bridge the gap between the specification and the realization of a P model \cite{simon1996sciences, guala2002models}. 
Indeed, from a simulation perspective, we can further draw a distinction between \textit{explicit likelihood} ($\explicit$) models and \textit{implicit likelihood} ($\implicit$) models.}

$\explicit$ models are characterized by a likelihood function that has a tractable mathematical form.
This means that the likelihood $\lik$ is known analytically (e.g., Gaussian) and its value can be evaluated directly or approximated numerically for any pair $(y, \theta)$. The same logic applies to non-parametric P models using the pair $(y, f)$.
$\explicit$ models include popular statistical models, such as (generalized) linear and additive models \citep{hastie_elements_2009}, but also (stochastic) differential equation systems with simple statistical properties \citep{jiang2011asymptotic}, finite mixture models \citep{chen2004testing}, or feedforward neural networks \citep{goodfellow2016deep}.

$\implicit$ models are defined through a Monte Carlo simulation program $y = g(\theta, z)$ and a prior $\prior$, rather than directly through an analytic likelihood function $\lik$.
The simulator $g$ transforms its inputs $\theta$ into outputs $y$ through a series of latent program states $z$.
A Monte Carlo simulator only implicitly defines the likelihood density via the relation
\begin{equation}\label{eq:implicit_lik}
    \lik = \int p(y, z \mid \theta)\,dz,
\end{equation}
where $p(y, z \mid \theta)$ is the joint distribution of observables $y$ and random latent program states $z$, if such a distribution exists.
The above integral runs over all possible execution paths of the simulation program for a given input $\theta$ and is typically intractable, that is, we cannot explicitly write down the mathematical form of the implied likelihood $\lik$.
%
$\implicit$ models are usually built upon firm theoretical assumptions and computational considerations aimed at providing a faithful representation of the modeled real-world system or process.
Common $\implicit$ models include mechanistic neural models \cite{izhikevich2003simple}, particle physics simulators \cite{de2017learning}, population genetics algorithmic models \cite{hoban2012computer}, or agent-based models \cite{grazzini2017bayesian}, to name just a few.

The distinction between $\explicit$ and $\implicit$ models is not a conceptual necessity, but rather an emerging practical convenience.
While most standard statistical models can easily be specified in terms of known density or distribution functions, the behavior of complex computational models might be easier to emulate directly using a simulation program.
Importantly, $\explicit$ and $\implicit$ models necessitate the use of different estimation methods and thus disparate modes of approximation and inference, as we will see in later sections.

\subsection{PD Models}
\label{sec:pd_models}

PD models are defined as the combination of a P model and observed data $\data$, that is, they represent a tuple $\left( \, \joint, \, \data \, \right)$.
The data can comprise any number of measurements $\tilde{y}$ with an arbitrary structure (e.g., sets, time series, graphs, etc.).
Furthermore, the number of observed data sets (conditioning quantities for the posterior) will be determined by the structure of the P model: Data on a hundred countries represents a single data set from the lens of a multilevel (hierarchical) model, but it comprises a hundred data sets for a single-level (non-hierarchical) P model.

The goal of PD models is to integrate the joint distribution and the observed data to arrive at the corresponding \emph{analytic posterior}:
\begin{equation}
    \postd = \frac{\likd \, \prior}{p(\data)} \propto \likd \, \prior,
\end{equation}
\noindent where the denominator $p(\data) = \int \likd \, \prior \, d \theta$ represents the model-implied marginal likelihood (aka evidence) evaluated at $\data$ and typically treated as a normalizing constant due to its independence of the model parameters.

If the P model is generative, the analytic posterior exists for every $\data$ that satisfies the expected data structure of the P model, regardless of whether or not it represents the true real-world generator $\mathbb{G}$.
In the non-representative case, the P model is said to be \textit{misspecified}.
In most quantitative sciences, except perhaps in some areas of the natural sciences, we can expect all P models to be misspecified to some (non-negligible) degree.
This does not prevent the corresponding PD models from being useful, though, if they can at least express some relevant aspects of reality captured by $\data$.

PD models represent the ideal endpoint of Bayesian inference. 
However, because we can rarely compute the marginal likelihood $p(\data)$ analytically, we do not have access to the actual PD model outside of textbook examples with limited generality and applicability (i.e., for conjugate P models, \cite{george1993conjugate}).
In other words, for most practically relevant and non-trivial $\explicit$ and $\implicit$ models, we cannot retrieve the analytic posterior $\postd$ and can only work with an approximate representation through the lens of an intermediary A which we call a \textit{posterior approximator}.

\subsection{PA Models}
\label{sec:pa_models}

PA models are defined as the combination of a P model with a posterior approximator A, that is, they constitute a tuple $\left( \, \joint, \, \posta \, \right)$, where the latter denotes any algorithm capable of \textit{somehow} approximating the analytic posteriors of model-implied observations $y$ for a given P model.
Approximators themselves exist at both an algorithmic and an implementation level, and details on both levels can influence their behavior and performance.
In the absence of actually observed data $\data$, PA models can be useful for confirming the computational faithfulness of a workflow, for instance, via simulation-based-calibration (SBC, \citep{talts2020, modrak_simulation-based_2023}) or assessing the adequacy of a model for answering a particular research goal \citep{schad2021toward}.
Importantly, the type of P model (i.e., $\explicit$ or $\implicit$) will typically determine or necessitate the choice of a particular approximator A, as we will see shortly.

\subsubsection{What is an approximator?} 
\label{approximator?}

More precisely, we can define an approximator as a triple $\text{A} = \{\mathcal{A}, \mathcal{I}, \mathcal{H}\}$, where $\mathcal{A}$ denotes the algorithmic representation (formal computer program), $\mathcal{I}$ denotes the actual implementation in a concrete programming language, and $\mathcal{H}$ denotes the set of admissible hyperparameters (i.e., adjustable settings or inputs) of the approximator.
The first two components of A are often entangled when talking about approximators in general, but they require different levels of analysis.
For instance, we can determine the computational complexity of $\mathcal{A}$ via standard algorithmic analysis and classify approximators according to their asymptotic run time or memory requirements \cite{cormen2022introduction}.
However, the latter two will also be constrained by the particular implementation $\mathcal{I}$: Parallel computing can easily turn a scary-looking quadratic $\mathcal{O}(n^2)$ time complexity into a negligible constant run time in practice \cite{cormen2022introduction}.
Thus, we deem it important to keep the distinction between $\mathcal{A}$ and $\mathcal{I}$ explicit\footnote{Naturally, hardware specifications will further influence the actual run time and space requirements of any approximator, so these specifications should be taken into account when comparing different approximators. The utility of an approximator will also be constrained by the available hardware budget: parallelism is of little use without access to a computing cluster.}.
In addition, the performance of an approximator will heavily depend on the choice of particular hyperparameters $h \in \mathcal{H}$ and these should be explicitly specified in any PA model.

\subsubsection{Approximators for $\explicit$ models}
\label{approx-explicit-models}

Currently, the two most commonly used approximators for $\explicit$ models are Markov chain Monte Carlo (MCMC) samplers and variational inference (VI) methods, but there exist many more approximator classes, for example, integrated nested Laplace approximation (INLA, \citep{rue_inla_2009, lindgren_inla_2015}) or optimal transport applied to Bayesian inference \citep{transport1, transport2, transport3}.

MCMC sampling algorithms, such as the Metropolis-Hastings algorithm \citep{hastings1970monte}, Gibbs sampling \citep{gelfand2000gibbs}, Hamiltonian Monte Carlo \citep[HMC,][]{neal2011mcmc}, or its extension to the No-U-Turn (NUTS) sampler \citep{hoffman2014no}, belong to a family of stateful algorithms which generate a sequence of correlated draws that converge in distribution to a stationary target distribution \citep{BDA3}.
Generally, our goal in MCMC is to construct a (geometrically) ergodic Markov chain on $\theta$ whose stationary distribution is the posterior $\post$ \citep{BDA3}.
In practice, we then sample from the chain to obtain a finite number of random draws from the (hopefully accurate) stationary distribution $\posta$ and use these draws to approximate $\post$. 
More precisely, using the posterior draws, we can efficiently approximate expectations (e.g., mean or variance) and quantiles of the posterior marginals, but not the posterior density itself.

The idea of approximating a complicated distribution via dependent random draws, albeit rather straightforward in hindsight, has gradually transformed and shaped the field of Bayesian inference.
Moreover, it constitutes the main logic behind major probabilistic programming languages such as Stan \citep{carpenter2017stan} or JAGS \citep{plummer2003jags}.
A \textit{sampler} is thus a computer program which uses computer-generated randomness to generate draws from a (complicated) distribution, instead of deriving or estimating its algebraic form.

Differently, variational inference (VI) methods cast the problem of posterior inference as an optimization task.
In contrast to MCMC, the resulting posterior approximation $\posta$ is in the form of a tractable density instead of random samples from the posterior.
Our goal in VI is to specify a family of approximate densities $\mathcal{Q}$ over the parameters $\theta$ of P.
Then, we try to retrieve the density $q^* \in \mathcal{Q}$ which minimizes the Kullback-Leibler (KL) divergence to the analytic posterior.
Finally, we use $q^*(\theta)$ as our approximation $\posta$ to the analytic posterior.

MCMC and VI methods represent the two endpoints of a trade-off between theoretical guarantees and computational efficiency.
MCMC methods enjoy the guarantee that under certain regularity conditions \citep{BDA3}, the obtained draws represent the true parameter posterior $\post$. More precisely, the posterior expectations can be perfectly recovered if the MCMC chain is run infinitely long and, more practically important, expectations can be efficiently approximated already with a finite number of draws.
Despite their favorable theoretical properties and major advances in recent years, MCMC algorithms are notoriously slow, which renders estimation of some complex models or applications to really big data practically infeasible \citep{blei2017variational}.
On the other hand,  VI methods can be very fast and offer a viable alternative to MCMC in applications to large data sets or real-time inference.
However, VI approximators can suffer severe loss of posterior accuracy and, as of today, offer less guarantees for correct inference than MCMC methods (\citep{blei2017variational}, but see \citep{zhang2020convergence, zhang2020theoretical}).
Thus, the choice between an MCMC or a VI approximator for a particular PA model will largely depend on the modeling context.
In addition, highly complex $\explicit$ models might not be estimable with either MCMC or VI, in which case they might be treated as $\implicit$ models in practice and tackled via simulation-based approximators, as we discuss next.

\subsubsection{Approximators for $\implicit$ models}
\label{approx-implicit-models}

Standard MCMC and VI solutions are not applicable to statistical inference with $\implicit$ models, since the latter lack an analytic likelihood function $\lik$.
Accordingly, approximators for $\implicit$ models leverage Monte Carlo (i.e., randomized) simulations for estimating the posterior based on the implicit likelihood defined by the simulator and Equation~\eqref{eq:implicit_lik}.

Approximate Bayesian computation (ABC) comprises a broad family of asymptotically correct methods for performing inference with $\implicit$ models.
The core idea of ABC methods is to approximate the posterior by repeatedly drawing parameters from the prior and then running the simulator with the sampled parameters to obtain a synthetic data set.
Whenever a synthetic data set is sufficiently similar to an actually observed data set (as defined by a fixed similarity criterion or a distance metric), the corresponding parameters are retained as a draw from the target posterior, otherwise rejected (i.e., rejection sampling).

In practice, ABC methods are notoriously inefficient and hindered by various methodological ``curses'', such as the curse of dimensionality \citep{raynal2019abc} or the curse of insufficiency \citep{marin2018likelihood}. Several more efficient methods employ various techniques, such as sequential Monte Carlo \citep[SMC,][]{sisson2007sequential, klinger2018pyabc}) or ABC-MCMC \citep{marjoram2003markov} with kernel density estimation (KDE) \citep{turner2014generalized} to optimize sampling or correct potential deficiencies, but the core idea of using simulations to aid real-world inference remains invariant across methods.

Recently, machine learning and deep learning innovations have permeated the field of simulation-based inference with the goal of scaling up or replacing standard ABC methods altogether \cite{cranmer_sbi_2020}.
Most of these innovations require simulation-based training of an expressive machine learning algorithm (e.g., random forests or neural networks) which is then used as a standalone approximator \citep{chan2018likelihood, greenberg2019automatic, gonccalves2020training, radev2020bayesflow}, in combination with an ABC routine \cite{jiang2017learning} or an MCMC sampler \citep{hermans2020likelihood, fengler2021likelihood, lueckmann2019likelihood, boelts2022flexible}.

\newcontent{
For instance, neural density estimation (NDE) methods employ specialized neural architectures for analyzing complex high-dimensional distributions \cite[e.g., natural images,][]{dinh2016density, kingma2018glow, ardizzone2021conditional}. 
In the context of Bayesian inference, NDE methods can approximate different components of intractable $\implicit$ models and currently represent a field of active and promising development \cite{cranmer_sbi_2020, lavin_simulation_2021}.
Specifically, neural posterior estimation (NPE) methods \citep{ardizzone_inn_2018, greenberg2019automatic, radev2020bayesflow, gonccalves2020training, pacchiardi2022likelihood, avecilla2022neural} involve simulation-based training of a conditional generative neural network \citep[e.g., normalizing flows,][]{kobyzev2020normalizing, papamakarios2021normalizing}.
The trained network then acts as a \textit{functional} that can approximate the posterior across the entire prior predictive distribution of a P model without any re-training, enabling \textit{amortized inference} (to be explained shortly).
A shared feature between NPE methods is that they avoid MCMC sampling altogether and can perform exact inference under certain optimal conditions.
}

\begin{figure}
    \centering
    \includegraphics[width=0.99\textwidth]{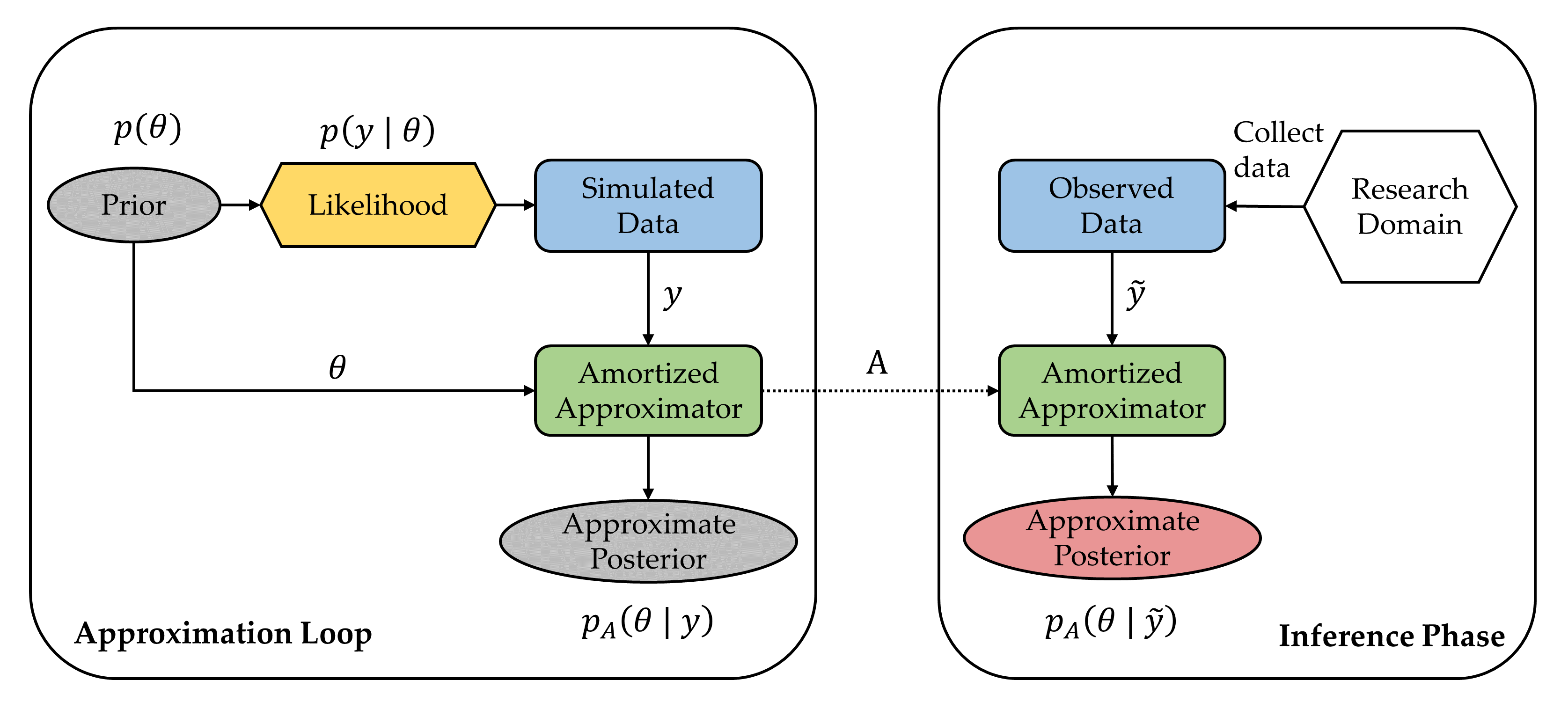}
    \caption{Amortized approximators incorporate a simulation-based approximation loop (training phase) before any real data are collected. The subsequent inference phase involves no simulations or further optimization and could be carried out almost instantly. The upfront training effort therefore amortizes over arbitrarily many observed data sets from a research domain working on the same P model family.}
    \label{fig:amortized}
\end{figure}

Ultimately, the utility of any simulation-based method will depend on a combination of various factors, such as generality, domain expertise, theoretical guarantees, efficiency, scalability, and software availability.
The amount of available data will once again play a crucial role in the choice of approximator.
In this context, the distinction between \textit{amortized} and \textit{non-amortized} posterior approximators becomes crucial.

\subsubsection{Amortized vs. Non-Amortized Approximators}
\label{abi-vs-non-abi}

\begin{figure}
    \centering
    \includegraphics[width=0.99\textwidth]{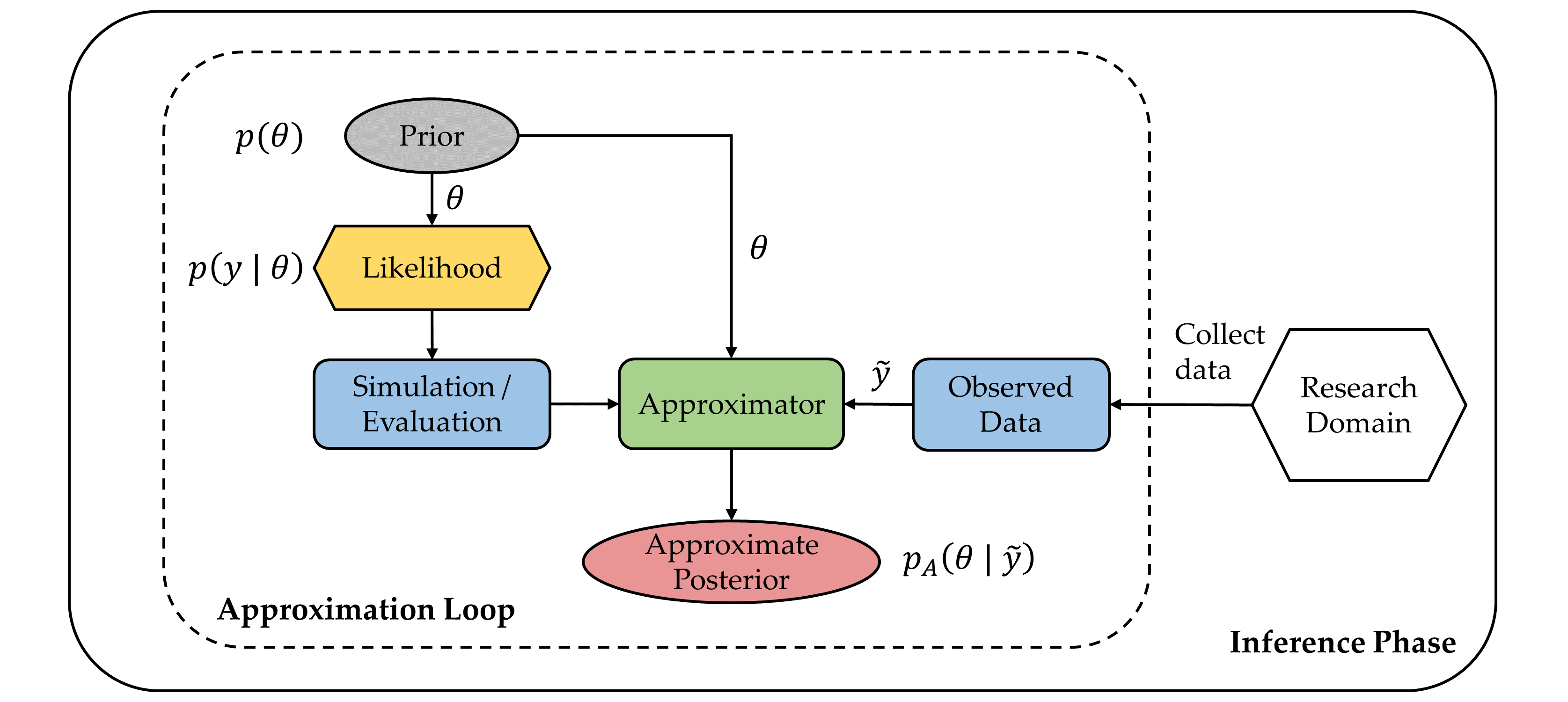}
    \caption{Non-amortized approximators perform a separate approximation loop (dashed plate) for each observed data set from a given research domain. Likelihood-based approximations, such as MCMC will evaluate the likelihood, whereas simulation-based approximators, such as ABC rejection samplers, will only use random draws from the implicit likelihood (available through stochastic simulations). Approximation and inference are tightly intertwined and the observed data enters the approximation loop.}
    \label{fig:non-amortized}
\end{figure}

Arguably, there are numerous ways to devise a taxonomy for the ever-growing zoo of posterior approximators.
A particularly useful and clear-cut classification views approximators as either \textit{amortized} or \textit{non-amortized}, with different degrees of amortization possible.
Amortized approximators involve a costly simulation-based optimization (training) phase which renders subsequent inference on simulated $y$ or real data $\data$ extremely efficient (see Figure~\ref{fig:amortized}).
In other words, the optimization/training effort \textit{amortizes} over repeated inference queries (e.g., over multiple data sets or data set sizes).
Differently, non-amortized approximators repeat all necessary computations for each data set or prior choice from scratch and utilize hardly any pooling of computational resources (see Figure~\ref{fig:non-amortized}).

\newcontent{Examples of amortized approximators include the BayesFlow method \citep{radev2020bayesflow, radev_amortized_2021, radev_amortized_2021, radev_jana_2023}, sequential neural posterior estimation (SNPE) methods operating in a single-round regime \citep{greenberg2019automatic, gonccalves2020training, durkan2020contrastive}, machine learning-enhanced ABC \citep{raynal2019abc}, or the pre-paid estimation method \cite{mestdagh2019prepaid}.
Examples of non-amortized approximators include standard explicit inference algorithms, such as MCMC or VI, but also several common ABC methods, such as ABC-SMC \citep{sisson2007sequential, klinger2018pyabc} or ABC-MCMC \citep{marjoram2003markov, turner2014generalized}.
In addition, some neural PA models might include both amortized and non-amortized components, such as multi-round SNPE methods (involving a separate training phase for each data set, \cite{papamakarios2016fast, greenberg2019automatic, durkan2020contrastive, deistler2022truncated}), likelihood approximators or surrogates (involving MCMC sampling, \cite{papamakarios2019sequential, lueckmann2019likelihood, fengler2021likelihood, boelts2022flexible}), or inference compilation methods (involving SMC, \cite{paige2016inference, le2017inference}).}

Amortized approximators are typically employed to estimate implicit PA(D)\footnote{\newcontent{Henceforth, parentheses in the PAD taxonomy denote an ``OR relationship''. For instance, P(D) would mean ``a P or a PD model'' and P(A)D would mean ``a PD or a PAD model''.}} models, but are equally applicable to explicit PA(D) models.
In the former case, their involvement often arises out of necessity, since $\implicit$ models are analytically intractable and state-of-the-art approximators, such as HMC-MCMC, are not applicable out of the box.
In the latter case, amortized approximators might be the only resort to estimate multiple PAD models in the presence of multiple data sets, where non-amortized approximators, despite being feasible, would demand an inordinate amount of a researcher's lifetime \cite{von2022mental}.

\subsection{PAD Models}
\label{sec:pad_models}

PAD models are defined as the combination of a P model, a posterior approximator A, and observed data D, that is, they constitute a triple $\left( \, \joint, \, \postad, \, \data \, \right)$.
Ultimately, PAD models aim to approximate the corresponding PD model through a suitable approximator A, whereas the amount of data D, together with the type of P model, will largely determine the choice of approximator.
As a consequence, the properties of a particular PAD model may be very different than what is expected from studying the corresponding PA model, since the observed data $\data$ may not have been generated from P itself. 
This misspecified P model case can arise for both $\explicit$ and $\implicit$ models and can have different consequences for the validity of inference depending on the particular approximator A \citep{masegosa2020learning, bissiri_general_2016, frazier2020model, frazier2021robust}.

For instance, amortized approximators face the challenge of dealing with \textit{simulation gaps} \citep{schmitt_bayesflow_2023, pacchiardi2021generalized}.
Simulation gaps occur when P model simulations do not accurately represent the real behavior of the modeled system or when they cannot adequately account for unexpected contamination of the observed data.
Simulation gaps are especially critical for amortized approximators since the latter assume that simulations are faithful proxies of reality.
Thus, simulations from misspecified P models may lead to subsequent problems for amortized inference on real data \citep{schmitt_bayesflow_2023}.
In these cases, the resulting $\postad$ will not be representative of the analytic $\postd$ and any substantive conclusions based on the former will have little validity.

In contrast, principle limitations due to model misspecification do not exist for standard, non-amortized Bayesian approximators, such as MCMC.
Under certain regularity conditions, MCMC samplers guarantee that the obtained samples represent the analytic posterior $\postd$ even when the underlying P model is misspecified \citep{BDA3}.
However, misspecified models might still cause considerable difficulties and convergence problems for MCMC methods in practice.
Thus, any trustworthy approximator should be equipped with diagnostics signaling improper convergence or invalid inference queries (see Section~\ref{convergence}).


\subsection{Intermediate Summary I}

Thus far, with our PAD taxonomy, we have defined four different classes of Bayesian models comprising different, yet interdependent, conceptual elements.
Common to all has been the joint probability model (P), which represents the core probabilistic and structural assumptions of a Bayesian model.
In addition, we proposed to treat the posterior approximator (A) and the data (D) as further constituents of Bayesian models.
We consider this warranted, since all three elements not only determine the scope and validity of the substantial conclusions derived from model-based inference but also influence which assumptions we decide to (and could!) test and which we choose to keep untouched by reality.

\section{What makes a good Bayesian model?}
\label{good_bayesian_model}

Below, we present a total of ten utility dimensions that, from our perspective, capture most relevant aspects of Bayesian models as defined by our taxonomy.
For each of these dimensions, we explain (a) its definition and meaning, (b) the reason why we deem it relevant for Bayesian model building, and (c) how to practically measure it. 
The order in which we present each utility dimension does not indicate their importance but aims to ease their presentation. 
We discuss the relative importance of utility dimensions in Section~\ref{hierarchies_tradeoffs}.

\subsection{Causal Consistency}
\label{sec:causality}

A common goal of scientific models is the investigation of a causal hypothesis, such as the improvement a certain treatment might bring to some medical condition or the effect an intervention has on an outcome of interest.
Most people are aware of the widely recited folk wisdom that \textit{correlation does not imply causation}.
Yet, this adage bears the seeds of a far-reaching and nowadays generally acknowledged opinion that statistics alone simply cannot solve questions of causality \cite{pearl2009causaloverview}.

While statistical inference can handle the static nature of associations in observational data, causality is a matter of changing conditions and handling these changing conditions requires causal assumptions to build upon \cite{pearl2009causality}.
Moreover, different P models may claim different degrees of causal sophistication.
For example, some $\explicit$ models built only to make accurate predictions may pass without a single mention of causality, while some mechanistic $\implicit$ models may directly embody causal functional relationships, such that an input variable $x$ is assumed to cause an observable $y$ by construction or by derivation from scientific theory.
Some complex P models may even hold standard unidirectional notions of causality inadequate, as the dynamics of certain natural systems appear to necessitate bidirectional or hierarchical forms of causal interplay \cite{thompson2001radical, noble2012theory}.

The scientific methods developed around the notion of causality help us determine whether a P model is a valid recipe for answering a particular causal query in principle. 
Put differently, we ask whether the probabilistic structure of a P model is consistent with a set of external causal assumptions.  
Thus, we refer to this implied model utility as \textit{Causal Consistency}.

In this section, we will briefly present the foundation of causal theory based on the work of Pearl \citep{pearl2009causality}, as it is currently the most common causal framework.
There are adoptions and adaptations for individual fields, such as the social sciences \citep{morgan2015counterfactuals, freedman2010statistical} and public health research \citep{vanderweele2015explanation}.
Moreover, recent Bayesian statistics textbooks have started discussing causality as a central aspect of statistical analysis \cite{mcelreath2020statistical}.
In addition, the fields of causal discovery \cite{hyttinen2015calculus, spirtes_causal_2016, glymour2019review} and optimal experimental design \cite{emery1998optimal, fedorov2010optimal, ivanova2021implicit} deserve a mention as well, since they tackle problems related to causality.
Finally, other promising causal frameworks have been proposed \cite{imbens2015causal} but are not discussed in detail here for reasons of brevity.




\subsubsection{Structural Causal Models}
\label{structural-causal-models}

\begin{figure}
    \centering
    \includegraphics[width=0.99\textwidth]{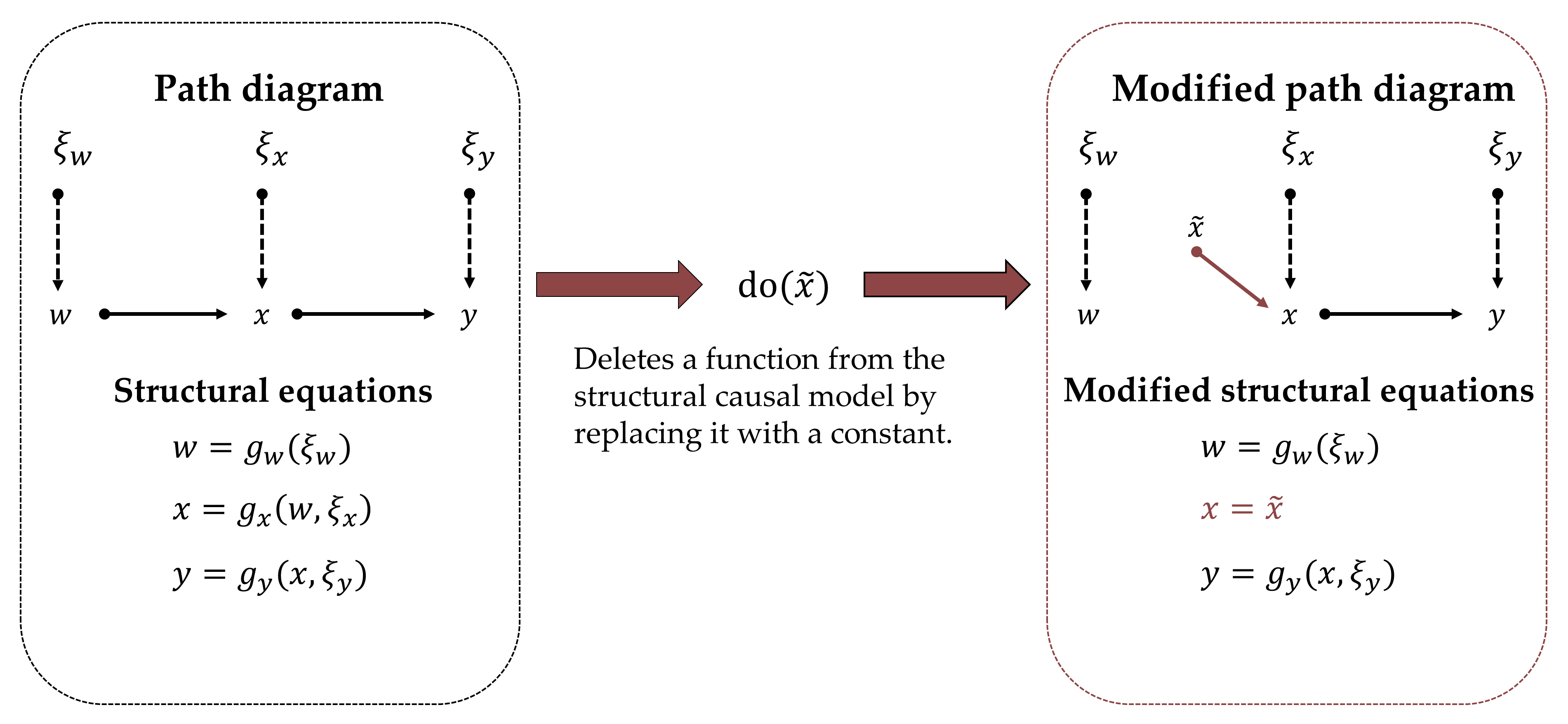}
    \caption{An example structural causal model (SCM) with three variables. The left panel depicts the pre-intervention path diagram, whereas the right panel depicts the post-intervention path diagram (see text for further clarification).}
    \label{fig:causal}
\end{figure}

Pearl \citep{pearl2009causality} proposes a framework to express causal assumptions and construct requirements on probabilistic models that make them consistent with those assumptions.
The mathematical objects that allow for causal analysis are called \textit{structural causal models} \citep[SCMs,][]{pearl2009causaloverview} and they comprise structural equations (what we express via P models), causal graphs, as well as interventional and counterfactual logic \cite{pearl2019seven}.
For instance, linear regression P models, if combined with proper causal calculus, comprise a widely used and simple form of linear SCMs.
However, vastly more complex SCM architectures are possible, such as causal generative neural networks \cite{kocaoglu2017causalgan}, where a causal graph is connected to a generative adversarial network responsible for learning interventional distributions (see Section~\ref{interventions} for details on interventions).

For the purpose of this paper, it is sufficient to discuss SCMs comprising a set of three endogenous variables whose causal relationships are to be studied. 
We refer to these variables as $w$, $x$, and $y$. 
For every endogenous variable, we assume there exists a corresponding exogenous (noise) variable, $\xi_w$, $\xi_x$, and $\xi_y$, respectively.
Under the assumption of \textit{causal sufficiency} (i.e., every exogenous variable affects no more than a single endogenous variable), a hypothesis of the form  ``$x$ \textit{causes} $y$'' means that $y$ is generated by a structural equation $y = g_y(x, \xi_y)$ for some function $g_y$. 
The corresponding causal graph is simply $x \rightarrow y$.


To extend this example, the left panel of Figure~\ref{fig:causal} illustrates a path diagram of the structural equations relating the endogenous variables $w, x$, and $y$, along with the corresponding causal graph $w \rightarrow x \rightarrow y$.
Importantly, any set of structural equations also encodes assumptions about the \textit{lack of causal influence}.
For instance, the absence of $w$ from the right-hand side of $g_y$ conveys the assumption that $y$ will remain invariant to changes in $w$, as long as variables $x$ and $\xi_y$ remain constant. 

In general, a causal graph implied by a set of structural equations will be a directed acyclic graph (DAG).
It can be constructed as follows: The variables that appear on the right-hand side of a structural equation become the parents of the variable that appears on the left-hand side of the structural equation. 
We can understand the structural equations as encoding explicit \textit{structural} assumptions about the opaque (true) data generator $\mathbb{G}$, which in turn implies a (true) joint distribution of the endogenous variables, here $p^*(x, y, z)$. 
This distribution is realized by first assuming a joint distribution of the noise variables, $p^*(\xi_w, \xi_x, \xi_y)$, and then propagating this uncertainty to $w$, $x$, and $y$ through the respective structural equations.

In P model terms, a DAG can be understood as defining a Bayesian network for the implied joint probability distribution of the endogenous variables \cite{pearl2009causaloverview}. The conditional distribution of an arbitrary endogenous variable $v$ is given by $p(v \mid \textrm{N}_{v})$, where $\textrm{N}_{v}$ denotes a set of parent variables of $v$ as implied by the DAG. For the current example, this would imply a generative likelihood that factorizes as $p(w, x, y \mid \theta) = p(y \mid x, \theta) \, p(x \mid w, \theta) \, p(w \mid \theta)$, where $\theta$ are our P model parameters (left unspecified in the DAG).

In our model taxonomy, a P model may or may not be consistent with the set of causal assumptions embodied in a DAG, which constitutes a binary metric of causal consistency.
For example, consider again the simple DAG given by $x \rightarrow y$, with structural equation $y = g_y(x, \xi_y)$. 
The concrete approximation of $g_y$ is part of the P model assumptions (see below), whereas adherence to the (external) DAG implies satisfying causal consistency. 
For example, consider the following linear P model 
\begin{align*}
\label{linear-causal-model}
    x &= \xi_x\\
    y &= \beta x + \xi_y  \numberthis
\end{align*}
with unspecified distributional forms of $\xi_x, \xi_y$, and $\beta$ for simplicity. 
The approximation $\hat{g}_y$ chosen for $g_y$ is $\hat{g}_y(x, \xi_y) = \beta x + \xi_y$ while $\hat{g}_x(\xi_x) = \xi_x$ is just the identity function. 
The above P model is clearly causally consistent with the DAG $x \rightarrow y$. 
In contrast, another linear P model in which we had swapped $x$ and $y$ (i.e., assuming $x = \beta y + \xi_x$), would be causally inconsistent with the graph $x \rightarrow y$.

In linear P models, the regression coefficients represent path coefficients of structural equations and thus quantify the linear ``causal effects'' of certain variables on others. 
However, even when a linear P model is causally consistent with a given DAG, its linear functional form $y = \beta x + \xi_y$ may still be a poor approximation of the true (potentially highly non-linear) structural equation $y = g_y(x, \xi_y)$. 
Thus, an equally causally consistent, but more flexible, non-linear P model may be a better choice in the end, depending on other utility dimensions. 
This illustrates that causal consistency, as defined here by the formal agreement with a causal DAG, is only a necessary, but not a sufficient condition for a P model to provide trustworthy causal inference. 
Further requirements will be discussed in the context of parameter recoverability (see Section~\ref{parameter-recoverability}).

Causal graphs allow for an unambiguous communication of assumptions about causal relations, but on their own, they still represent static entities.
In contrast, \textit{interventions} and \textit{counterfactuals} describe actions which enable us to answer \textit{causal queries} based on (a subset of) these assumptions. 
Below, for the sake of brevity, we will elaborate solely on interventions (see \citep{pearl2009causality} for more details of counterfactuals).



\subsubsection{Interventions}
\label{interventions}

An intervention is an operation that changes the underlying structural equations, hence the corresponding causal graph. 
Intervening on $x$ means setting it to a fixed value $\tilde{x}$, say, administering the treatment $\tilde{x}$ to a patient. 
We denote an intervention as $\textrm{do}(x = \tilde{x})$ or simply $\textrm{do}(\tilde{x})$ for short. 
The effect of an intervention on the path diagram of our example three-variable SCM is shown in the right panel of Figure~\ref{fig:causal}.
An intervention $\textrm{do}(\tilde{x})$ differs from conditioning on $\tilde{x}$ in the following way: The former removes the connections of
node $x$ to its parents, whereas the latter does not change the causal graph from which data is generated \cite{pearl2009causality, kocaoglu2017causalgan}. 
If we set the value of $x$ to some $\tilde{x}$, then it is no longer determined through the structural equation $g_x(w, \xi_x)$, that is, we have intervened in the generative mechanism. 
Importantly, the interventional distribution of interest, say, $p(y \mid \text{do}(\tilde{x}))$ may differ from the corresponding conditional distribution $p(y \mid \tilde{x})$.

However, when we only have access to observational data because we cannot intervene in the causal graph (e.g., an experiment is too expensive to perform), our resort is to estimate conditional distributions. 
Thus, an important question arises: ``Which causal queries can we answer (i.e., which interventions' effects can we estimate) based on observational data alone?'' 
In the language of $\text{do}$-calculus, this translates to the question of whether we can circumvent the $\text{do}$ operator and express the interventional distribution of interest $p(y \mid \text{do}(\tilde{x}))$ via a conditional distribution \cite{pearl2012docalculus}. 
For this purpose, we can use three basic rules of $\text{do}$-calculus that specify the conditions under which we can 1) ignore observations, 2) treat interventions as equivalent to observations, and 3) ignore interventions \cite{pearl2012docalculus}. 

Against this background, we say that a P model is \textit{causally consistent for a given causal query}, if that query can be answered by applying the rules of $\text{do}$-calculus to the underlying DAG and all necessary conditional distributions are part of the P model. 
A P model which is causally consistent with a DAG is also causally consistent for all valid causal queries of that DAG. 
In practice, however, we can rarely attain (or care about) the former but are only concerned with causal consistency for a few queries of interest.
To illustrate this point, let us again consider the DAG $w \rightarrow x \rightarrow y$ from Figure~\ref{fig:causal}. 
The linear P model~\eqref{linear-causal-model} is not causally consistent with this DAG, since it does not include the structural equation $x = g_x(w, \xi_x)$ but only $y = g_y(x, \xi_y)$. 
However, it is causally consistent for the specific query $p(y \mid \text{do}(\tilde{x}))$ because, after applying the second rule of do-calculus, we find that $p(y \mid \text{do}(\tilde{x})) = p(y \mid \tilde{x})$ for this DAG. 
Correspondingly, the latter conditional distribution is part of the P model in the form of $y = \beta x + \xi_y$.
Naturally, the conditions under which we can answer causal queries using conditional distributions become harder to test for causal graphs containing more than just three variables, but the underlying principles remain the same \cite{cinelliCrashCourseGood2020}.

\subsection{Parameter Recoverability}
\label{parameter-recoverability}

A central goal of Bayesian modeling is to perform parameter inference, that is, to draw conclusions directly from the posterior of the latent parameters or other pushforward quantities of interest.
But how can we assess whether our inferences are informative and capture all relevant layers of uncertainty?
The \emph{Parameter Recoverability} dimension captures the ability of P models (and of PA models; see Section~\ref{calibration}) to gain information from data and perform faithful uncertainty quantification.
Moreover, recoverability is a concept where frequentist statistics inevitably play a role, even in the context of purely Bayesian models.

For the purpose of generality, consider the task of estimating a quantity of interest $\varphi$ based on a P model and (yet to be realized) data $y$ using an estimator $\psi = \psi(\theta)$ of $\varphi$ where $\theta \sim p(\theta \mid y)$.
The epistemic uncertainty implied by the posterior $p(\theta \mid y)$ is naturally propagated to the posterior of $\psi$.
Based on the implied posterior $p(\psi(\theta) \mid y)$, we can derive both point and uncertainty estimates, among other things, as detailed further below. 

\begin{figure}
    \centering
    \includegraphics[width=0.99\textwidth]{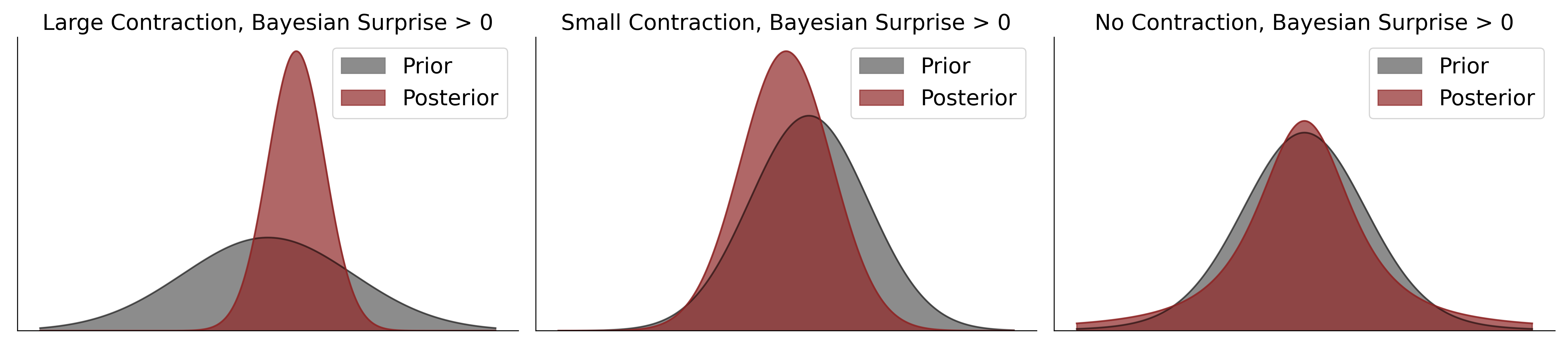}
     \caption{Three hypothetical (univariate) PD model scenarios illustrating posterior contraction and Bayesian surprise. The leftmost panel depicts a PD model which yields both large posterior contraction and large Bayesian surprise. The middle panel depicts a PD model exhibiting both small posterior contraction and small Bayesian surprise. The rightmost panel depicts a PD model which has zero posterior contraction (i.e., equal prior and posterior variances), yet non-zero Bayesian surprise (i.e., owing to a different tail exponent). Posterior contraction is easier to compute and interpret, but Bayesian surprise is more general, as it captures differences beyond second moments (i.e., variances).}
    \label{fig:bs}
\end{figure}

To make this notion more concrete, let us consider a simple example. 
Suppose we are interested in the (true) mean difference of $y$ between two groups, $\varphi = \mathbb{E}[y_1] - \mathbb{E}[y_2]$, where $y_1$ and $y_2$ represent the responses of the two groups, respectively.
One way to estimate $\varphi$ here is via a linear regression P  model with response vector $y = (y_1, y_2)$ and pointwise likelihood
\begin{align*}
    y_n &\sim \text{Normal}(\mu_n, \sigma) \\
    \mu_n &= \beta_1 \times \mathbb{I}(n \in C_1) + \beta_2 \times \mathbb{I}(n \in C_2),
\end{align*}
where $\mathbb{I}$ is the indicator function, $C_1$ is the index set of observations $n$ belonging to Group 1, and $C_2$ is the corresponding index set of Group 2. 
Then, based on this P model, we define an estimator $\psi$ of $\varphi$ as $\psi = \beta_1 - \beta_2$. Accordingly, $\psi$ does not need to be a model parameter itself but can be any pushforward quantity computable from the parameters.
The Gauss-Markov theorem tells us that the chosen estimator $\psi$ has the lowest sampling variance among the class of linear unbiased estimators in case of flat priors on $\beta_1$ and $\beta_2$.
However, the properties of any estimator in general are not always that clear: Consider another example where the true data generator is given by $y_n = f(\varphi \, x_n) + \xi_n$, with $x$ being a known continuous variable, $f$ a monotonically increasing function, and $\xi$ an additive error term.
In the absence of knowledge about the exact form of $f$, we could set up a P model with a normal likelihood that is linear in $\psi$,
\begin{align*}
    y_n &\sim \text{Normal}(\psi \, x_n, \sigma).
\end{align*}
Moreover, the properties of $\psi$ as an estimator of $\varphi$ will certainly depend on the unknown function $f$ and is likely not as favourable as in the first example.
However, through $\psi$, we can at least hope to get the sign of $\varphi$ right, which may as well turn out to be sufficient for meeting the goals of some applications.

\subsubsection{Identifiability and Information Gain}
\label{information_gain}

Oftentimes, we are interested in learning something about the (true) real-world generator $\mathbb{G}$ through the P model-dependent quantity $\psi$, justified by its resemblance to the model-independent quantity $\varphi$ that we assume to play a role in $\mathbb{G}$.
As a first step, we need to study whether the data generated by the unknown process enables the P model to extract any information about $\psi$ at all.
If the data is not informative, there is no point in further studying the recoverability of $\varphi$ through $\psi$.
In a frequentist sense, we say that a quantity $\psi$ is \emph{identified} in the given P model, if all the possible values of $\psi$ lead to unique conditional distributions, that is, for any $\psi_1 \neq \psi_2$ we have $p(y \mid \psi_1) \neq p(y \mid \psi_2)$ \citep{casella2002}.
Thus, frequentist identification implies that, in the limit of infinite data, no ambiguity remains about possible values of $\psi$ \cite{lehmann2006theory}.

In a Bayesian context, the posterior captures all information about $\psi$ gained from the data.
Thus, the posterior should be a key object for defining identifiability.
Since the posterior always exists (as long as the prior is proper), regardless of how informative the data are, the mere existence of the posterior is not a helpful measure of identifiability \citep[see also][for discussions of Bayesian identification]{lindley1956, goel1981, sanmartin2010, sanmartin2018, betancourt2018}.
Instead, we have to define identifiability by a juxtaposition of prior and posterior.
The transition from prior to posterior (i.e., Bayesian updating) essentially conveys a reduction in uncertainty brought about by observing some data.
Equivalently, it can be seen as communicating the \textit{information gain} achieved by accounting for the data.
Thus, we expect the posterior to be narrower (sharper) than the prior, as the opposite would imply a loss of information through observation -- a rather paradoxical scenario.
In other words, the data should be sufficiently informative of $\psi$, otherwise, the posterior will just resemble the prior.

\textit{Bayesian surprise} offers a way to quantify arbitrary differences between prior and posterior.
The Bayesian surprise is typically defined as the Kullback-Leibler (KL) divergence between the two distributions
\begin{align}
    \text{BS}(\psi \mid y) &:= \mathbb{KL}\left[\postpsir\,||\,\priorpsir \right]\\ &= \int \postpsi \log \left(\frac{\postpsi}{\priorpsi} \right)d\theta\label{eq:bs},
\end{align}
but other divergence or integral metrics are also possible \cite{muller1997integral}.
The Bayesian surprise, as defined above, is non-negative and equals zero if and only if $\postpsir = \priorpsir$.
Henceforth, to avoid commitment to the KL divergence, we will use the symbol $\mathbb{D}$ to denote any divergence with the above two properties.
In information theory, this particular form of the Bayesian surprise is called a \textit{relative entropy}, and, in Bayesian terms, represents the information gained by updating the prior to the posterior in units determined by the base of the logarithm.\footnote{Whenever an approximation of the Bayesian surprise is intractable because it requires access to the analytic prior and posterior \emph{densities}, we can define Bayesian surprise through an integral metric, such as the Maximum Mean Discrepancy \cite[MMD,][]{mmd}, that we can approximate efficiently from prior and posterior draws.}
Accordingly, in a Bayesian context, $\psi$ is identified if the relative entropy is non-zero.

Further, the concept of \textit{posterior contraction} provides a simpler and tractable empirical diagnostic to assess identification and degrees of informativeness  \citep{betancourt2018}.
Posterior contraction formalizes the idea that the posterior should get narrower as the amount of data increases and is computed as the ratio between posterior and prior variance:
\begin{equation}
  \text{PC}(\psi \mid y) := 1 - \frac{\text{Var}_{p(\theta \mid y)} (\psi(\theta))}{\text{Var}_{p(\theta)} (\psi(\theta))}.
\end{equation}
If $y$ contains no information about $\psi$, then $\text{PC}(\psi \mid y) = 0$.
Conversely, the more information (i.e., uncertainty reduction) we gain from $y$, the larger $\text{PC}(\psi \mid y)$ becomes, up to a maximum of $\text{PC}(\psi \mid y) = 1$.
The posterior contraction can be combined with the posterior $z$-score (i.e., the difference between the true parameter and its posterior mean) as an intuitive two-dimensional estimate of the information gain that can be achieved by a P model when combined with data D \citep{schad2021toward}.

Posterior contraction compares only the second moments (i.e., the variance) of the prior and the posterior, which means that it can be efficiently computed from random draws of those distributions. However, relevant differences between prior and posterior may manifest themselves only in higher moments: It is still possible that we learn something about a distribution, for instance, about its tail exponent or symmetry, while its variance remains largely unchanged (see Figure \ref{fig:bs} for an illustration).

So far, we have only considered posterior contraction and Bayesian surprise brought about by a single data set $y$.
Thus, Equation~\eqref{eq:bs} provides only a measure for \textit{local} (i.e., per-data) information gain.
Whenever we are interested in \textit{global} (i.e., in expectation over all possible observations) information gain, then the expected Bayesian surprise (EBS) should be considered:
\begin{align}
    \text{EBS}(\psi) &:= \mathbb{E}_{p^*(y)}\big[ \mathbb{D}\left[\postpsir\,||\,\priorpsir \right]\big]\\ &= \int \mathbb{D}\left[\postpsir\,||\,\priorpsir \right] \, p^*(y) \, dy\label{eq:ebs},
\end{align}
or, similarly, the expected posterior contraction (EPC).
Global information gain assumes access to the distribution $p^*$ of real-world generator outputs and so we can rarely compute this quantity in practice.
Instead, we can obtain a Monte Carlo estimate of Equation~\eqref{eq:ebs} over multiple observed data sets as an approximation of the true EBS.

In many scenarios (e.g., during model development), we are interested in the recoverability of $\psi$ over the full generative scope of a model P, in combination with a posterior approximator A, \textit{before collecting any data}.
In this case, we will be considering the approximate posterior $p_{\text{A}}(\psi\,\mid\,y)$ and estimating the difference between prior and approximate posterior with respect to the joint distribution $p(\theta, y)$ implied by the P model:
\begin{equation}\label{eq:ebs_post}
    \text{EBS}_{\text{P}, \text{A}}(\psi) := \mathbb{E}_{p(\theta, y)}\big[ \mathbb{D}\left[p_{\text{A}}(\psi \mid y)\,||\,\priorpsir \right]\big],
\end{equation}
In other words, we assume the P model to be a good representation of $p^*(y)$ and evaluate the identification of $\psi$ under this assumption for a given approximator A.
Note that approximating the expectation over $\joint$ will be computationally expensive for many PA models relying on non-amortized approximators (i.e., ABC or MCMC), since estimating the posterior $p_{\text{A}}(\psi \mid y)$ repeatedly will dominate almost any approach (see also Section \ref{calibration}).
Thus, well-calibrated amortized approximators \citep{greenberg2019automatic, radev2020bayesflow} can serve as remarkable catalysts for efficiently quantifying global information gain for a given PA model before committing to the (costly) process of data collection.

\subsubsection{Ground-Truth Comparisons}
\label{ground-truth-comparisons}

We have hitherto assumed that we are dealing with a black-box (true) generator $\mathbb{G}$ whose actions give rise to the data-generating distribution $p^*(y)$.
Thus, we did not require $\varphi$ to play any actual role in the process of data generation.
In this section, we will restrict our focus to scenarios where $\varphi$ does in fact represent some intrinsic properties of $\mathbb{G}$.
Thus, we assume an unknown conditional data-generating distribution $p^*(y \mid \varphi)$ and are interested in the similarity between $\varphi$ and its P-model-based estimator $\psi$.

Obtaining the posterior of $\psi$ for a single data set and verifying sufficient information gain will tell us nothing about the recoverability of $\varphi$ given a P model, (i), because the resemblance between $\varphi$ and its estimator $\psi$ remains unclear and, (ii), because we need to consider the variation in $y$, that is, variation across possible data sets as well.
This means that we ought to estimate the performance of an estimator in expectation over possible data:
\begin{equation}
  \label{par-rec-y}
  \E_{p^*(y \mid \varphi)}[f(\varphi, \psi)] = \int f(\varphi, \psi \mid y) \, p^*(y \mid \varphi) \, d y,
\end{equation}
where $f(\varphi, \psi \mid y)$ is some function comparing $\varphi$ with the posterior of $\psi$, conditional on data $y$ (see below for examples). 
If the applied P model were the actual data generator itself, then $p^*(y \mid \varphi)$ would be equal to $p(y \mid \varphi) = \int p(y, \theta \mid \varphi) \, d \theta$ and we could set $\psi = \varphi$. 
In this case, $\varphi$ could be directly estimated through its own posterior distribution induced by $\varphi(\theta)$ with $\theta \sim p(\theta \mid y)$. However, in reality, we do not know how well P represents the actual generator, and so we continue to distinguish $\varphi$ from its P model-based estimator $\psi$.

Notably, the evaluation of Equation~\eqref{par-rec-y} does not actually require any observed data and so can be done ahead of time, before commencing any data collection. Unfortunately, as for many things in Bayesian statistics, it is a lot easier to write down the target in mathematical notation than to actually compute it: The integral in \eqref{par-rec-y} is almost always intractable, even if the posterior of $\psi$ itself were analytic. 
Thus, in statistical practice, we approximate the integral with a finite sum over $M$ independently simulated data sets $y_1, \ldots, y_M \sim p^*(y \mid \psi)$:
\begin{equation}
\label{par-rec-y-approx}
   \E_{p^*(y \mid \varphi)}[f(\varphi, \psi)] \monte \frac{1}{M} \sum_{m=1}^M f(\varphi, \psi \mid y_m)
\end{equation}
This Monte Carlo estimate is now conceptually easy to compute, but potentially very time-consuming, since the P model needs to be fit $M$ times, whereby each single fit may itself demand a considerable amount of time.

In Equations \eqref{par-rec-y} and \eqref{par-rec-y-approx}, $\varphi$ is held constant, which constitutes the typical setup in simulation studies where we fix the ground-truth to a single value per simulation instance. 
However, the conclusions we can draw from such studies are naturally limited to the few investigated simulation instances (chosen ground-truths). 
If the investigated instances were non-representative in reality, then we would learn little to nothing of value from our simulations, even if the data-generating distribution $p^*(y \mid \varphi)$ itself were faithful. 
To consider this implied uncertainty, we can make the criterion \eqref{par-rec-y} fully Bayesian by adding a prior $p^*(\varphi)$ over $\varphi$.
Thereby, we can now measure recovery in expectation over data $y$ and \textit{a priori} plausible values of the quantity of interest $\varphi$:
\begin{equation}
  \label{par-rec-y-psi}
  \E_{p^*(y, \varphi)}[f(\varphi, \psi)] = \int \int f(\varphi, \psi \mid y) \, p^*(y \mid \varphi) \, p^*(\varphi) \, d y \, d \varphi,
\end{equation}
with Monte Carlo (simulation-based) approximation
\begin{equation}
\label{par-rec-y-psi-approx}
   \E_{p^*(y, \varphi)}[f(\varphi, \psi)] \monte \frac{1}{M}  \sum_{m=1}^M f(\varphi_m, \psi \mid y_m)
\end{equation}
for $M$ ground-truth simulations, each generated according to $\varphi_m \sim  p^*(\varphi)$ and $y_m \sim p^*(y \mid \varphi_m)$.

\paragraph{Point Estimation.}
One central aspect of parameter recoverability
that can be assessed in terms of expectations over the data-generating distributions is \emph{point estimation}. We write $T(\psi \mid y)$ for a point estimator derived from the posterior of $\psi$. 
Most commonly, we compute the posterior mean $\int \psi(\theta) \, p(\theta \mid y) \, d \theta$, or alternatively the posterior median or mode.
Due to aleatoric uncertainty in the data $y$, we cannot expect $T(\psi \mid y) = \varphi$ for all $y$, even if the former would be the best possible point estimator of $\varphi$. 
Instead, we can measure how far away our estimator is from the truth via a strict distance function $d$ on $T(\psi \mid y)$ and $\varphi$, such that $d(T(\psi \mid y), \psi) = 0$ holds if and only if $T(\psi \mid y) = \varphi$. 
Common distance functions are the \textit{bias} $T(\psi \mid y) - \varphi$, the \textit{squared error} $(T(\psi \mid y) - \varphi)^2$, and the \textit{absolute error} $|T(\psi \mid y) - \varphi|$. 
To estimate the performance of a point estimator in expectation over the data-generating process, we would set $f(\varphi, \psi \mid y) = d(T(\psi \mid y), \psi)$ and then apply Equations \eqref{par-rec-y} to \eqref{par-rec-y-psi-approx}.
Whenever we compare P models based on their point estimation capabilities, we would prefer the P model with the smallest expected distance of its point estimator to the assumed true $\varphi$.

\paragraph{Uncertainty Estimation.}
An uncertainty estimator is defined as a parameter region that is supposed to contain the true quantity of interest $\varphi$ with a certain (user-defined) probability $q$.
We write $U_q(\psi \mid y)$ to denote a $q$ uncertainty region derived from the posterior of $\psi$. 
Common Bayesian uncertainty regions are quantile-based credible intervals and highest density intervals (HDIs) \citep{BDA3}. 
We say that an uncertainty region is \emph{well calibrated} for a given $\varphi$ (in a frequentist sense) if the following equality holds:
\begin{equation}
\label{int-uncertainty-estimate}
     q = \E_{p^*(y \mid \varphi)}[\mathbb{I}(\varphi \in U_q(\psi)] = \int \mathbb{I}(\varphi \in U_q(\psi \mid y)) \, p^*(y \mid \varphi) \, d y,
\end{equation}
where $\mathbb{I}(\varphi \in U_q(\psi \mid y))$ is the indicator function evaluating to $1$ if $\varphi \in U_q(\psi \mid y)$ and to $0$ otherwise. 
In other words, an uncertainty region for probability $q$ is well calibrated if it contains the assumed true parameter in a fraction of $q$ data sets. 
If the above property holds for every uncertainty region $U_q(\psi \mid y)$, we say that the whole posterior of $\psi$ is well calibrated for estimation of $\varphi$.

Bayesian uncertainty regions are not generally designed to satisfy this frequentist calibration and there is no guarantee that they will \citep{BDA3, nalborczyk_pragmatism_2019}. 
Yet, it can be a perfectly valid approach to use them even to satisfy purely frequentist goals \citep{gao_priors_2021}. 
Interestingly, when considering expectations over $(y, \varphi) \sim  p^*(y \mid \varphi) \, p^*(\varphi)$ as in Equation~\eqref{par-rec-y-psi}, a P model will exhibit perfect calibration as long as its generative behavior matches the unknown data generator and posterior computation is exact \citep{talts2020}. 
This property is extensively used in diagnosing the correctness of posterior approximations, a topic we will discuss in Section \ref{calibration}.

When comparing P models based on their uncertainty estimation of $\varphi$, we would prefer the model which yields uncertainty estimates closest to the equality in Equation \eqref{int-uncertainty-estimate} for some pre-selected, application-specific uncertainty regions. 
For example, if we were primarily interested in well-calibrated 95\% credible intervals
(perhaps more precisely stated, \textit{compatible intervals}, \cite{mcelreath2020statistical}), then we would prefer the model for which these intervals had closest to $q = .95$ coverage of the assumed true $\varphi$. 
That said, for some specific analysis goals, for example in null-hypothesis significance testing \citep{krueger_null_2001}, over-coverage (higher than $q$ coverage) may be more acceptable than under-coverage, or vice versa, depending on the assigned utility values of the corresponding Type-I and Type-II errors \citep{schad_workflow_2021}.

\paragraph{Sharpness.}
\label{sharpness}
Multiple P models, say $\text{P}_1$ and $\text{P}_2$, may provide estimators $\psi_{\text{P}_1}$ and $\psi_{\text{P}_2}$ that are equally well calibrated for a quantity of interest $\varphi$, yet their uncertainty regions may differ in coverage \citep{gneiting2007}. 
This implies that calibration alone is insufficient to describe the appropriateness of uncertainty regions: Additionally, we need to introduce the concept of \emph{sharpness}. 
We say that model $\text{P}_1$
is sharper than model $\text{P}_2$ for an uncertainty region $U_q(\psi \mid y)$ with finite bounds, if that region is better or equally well calibrated in $\text{P}_1$ than for $\text{P}_2$ and
if the volume of $U_q(\psi_{\text{P}_1} \mid y)$ is smaller than the
volume of $U_q(\psi_{\text{P}_2} \mid y)$ in expectation over the data-generating distribution:

\begin{equation}
    \E_{p^*(y \mid \varphi)} \left[ \text{Vol}(U_q(\psi_{\text{P}_1} \mid y)) \right] <
    \E_{p^*(y \mid \varphi)} \left[ \text{Vol}(U_q(\psi_{\text{P}_2} \mid y)) \right],
\end{equation}

\noindent where $\text{Vol}$ indicates the volume in Euclidean space. 
For unidimensional $\varphi$ and corresponding uncertainty region, say, a 95\% credible interval, the volume is simply equal to the width of the interval. 
Of course, depending on whether we hold $\varphi$ constant or assign a generating prior $p^*(\varphi)$ to it, we can also investigate sharpness in expectation over the joint distribution $p^*(y \mid \varphi) \, p^*(\varphi)$, instead of only focusing on $p^*(y \mid \varphi)$. 
If sharpness holds for all finite-volume uncertainty regions $U_q(\psi \mid y)$, then the posterior of $\psi_{\text{P}_1}$ is sharper than the posterior of $\psi_{\text{P}_2}$. 
Well-calibrated uncertainty regions cannot be infinitely sharp and there exists a sharpest model and corresponding estimator if the set of well-calibrated models is non-empty \citep{gneiting2007}.
However, in practice, we have no access to this sharpest model.
Thus, in contrast to calibration, sharpness cannot be practically computed in an absolute sense, but can only be probed as a relative quantity in the context of two or more P models.

\subsubsection{Calibration of Posterior Approximations}
\label{calibration}

So far we have primarily focused on P models in the context of parameter recoverability and all of the estimators assumed access to the analytic posterior $p(\theta \mid y)$ to obtain the analytic posterior $p(\psi(\theta) \mid y)$ of the estimator $\psi$ of $\varphi$. 
Since we do not have access to the analytic posterior in practice, our typical estimators are based on PA models. 

\begin{figure}[h]
\centering
\includegraphics[width=0.99\textwidth]{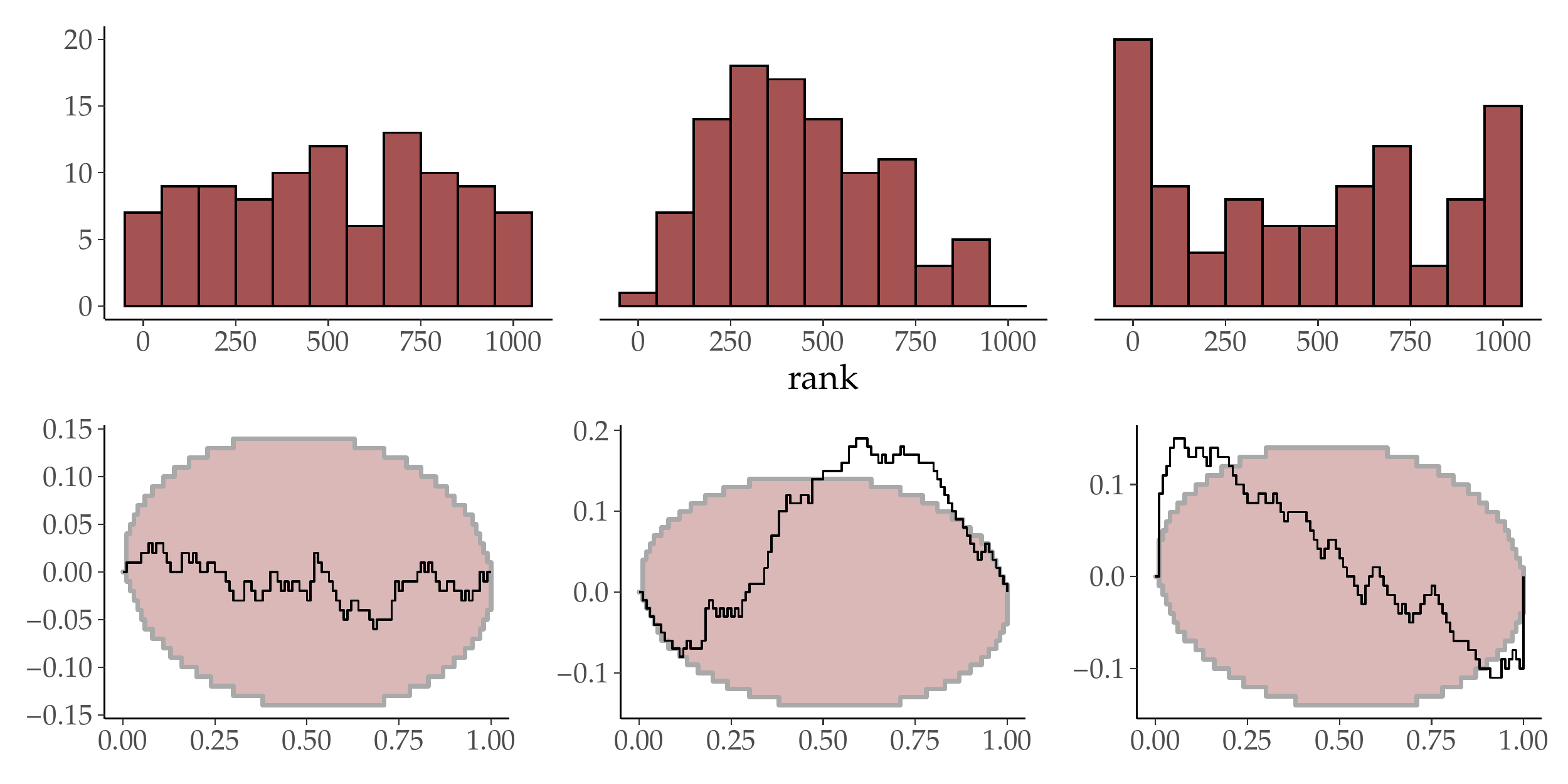}
\caption{Simulation-based rank histograms (top) and corresponding empirical cumulative distribution function (ECDF) difference plots \citep{sailynoja_graphical_2022} (bottom) for three hypothetical quantities of interest. The pink areas in the ECDF difference plots indicate 95\%-confidence intervals under the assumptions of uniformity and thus allow for a null-hypothesis significance test of self-consistent calibration. Left: A well-calibrated quantity. Center: A miscalibrated quantity with too many lower ranks indicating a positive bias in the PA model-based posteriors. Right: A miscalibrated quantity with too many extreme ranks indicating overconfident PA model-based posteriors (i.e., variance underestimated).} 
\label{fig:SBC-plots}
\end{figure}

Correspondingly, we define the estimator $\psi_{\text{A}}$ of $\varphi$ via the approximate posterior $p_{\text{A}}(\psi(\theta) \mid y)$ of $\psi$ obtained by the approximator A. 
If A were approximating the posterior via random draws $\theta^{(s)}$ from $p_\text{A}(\theta \mid y)$, the approximate posterior $p_\text{A}(\psi(\theta) \mid y)$ would be represented by the pushforward draws $\psi(\theta^{(s)})$. 
Thus, we can evaluate identifiability, point and uncertainty estimation, as well as the sharpness, of a PA model by replacing $\psi$ with $\psi_{\text{A}}$ in the corresponding equations. 
Ideally, we would like to separate the estimation of $\varphi$ via $\psi$ from the estimation of $\psi$ via $\psi_{\text{A}}$ and we can do so if we assume that the considered P model \textit{is} the true generator itself.  This is due to two related \emph{self-consistency} properties. 
The first one is
\begin{equation}
\label{prior-dap}
    p(\theta) = \int \int p(\theta \mid y) \, p(y \mid \theta^*) \, p(\theta^*) \, d y \, d \theta^*,
\end{equation}
which states that a P model's prior (left-hand side) is equal to the P model's \emph{data-averaged posterior} (right-hand side), that is, the posterior in expectation over its own generating distribution \citep{talts2020}. 
The second one states that all uncertainty regions $U_q(\psi \mid y)$ of all pushforward quantities $\psi$ are well calibrated, as long the generating distribution of the assumed P model is equal to true data-generating distribution and posterior computation is exact \citep{talts2020}. 
Writing $\psi^*$ instead of $\varphi$ to explicate the direct correspondence between the quantity of interest and its estimator $\psi$, this property can be written as
\begin{equation}
\label{prior-cond-sbc}
  q = \int \int \mathbb{I}(\psi^* \in U_q(\psi \mid y)) \, p(y \mid \psi^*) \, p(\psi^*) \, d y \, d \psi^*.
\end{equation}
Both self-consistency properties are useful, but Equality~\eqref{prior-cond-sbc} provides a particularly convenient means to diagnose the calibration of the approximated posterior $p_A(\psi(\theta) \mid y)$: Under perfect (self-consistent) calibration, the posterior probability $\Pr(\psi^* \leq \psi)$ is uniformly distributed in the unit interval \citep{talts2020,sailynoja_graphical_2022}.
If the approximate posterior can be
expressed in terms of random draws, then uniformity can be tested empirically by comparing the empirical distribution of ranks
\begin{equation}
   r(\psi^*, \psi(\theta_{1:S}) \mid y_m) := \sum_{s=1}^S \mathbb{I}(\psi^* \leq \psi(\theta^{(s)})) \quad \text{for} \quad \theta^{(s)} \sim p_{\text{A}}(\psi(\theta) \mid y_m)
\end{equation}
over $M$ simulated data sets to a uniform distribution, a procedure known an \emph{simulation-based calibration} \citep[SBC,][]{talts2020}. 
If the distribution of ranks is close enough to uniformity (e.g., according to a frequentist null-hypothesis significance test), we can conclude that the PA model is well calibrated for approximating the P model, assuming self-consistency of P. 
The required uniformity can be checked graphically, for example via histograms (top row of Figure~\ref{fig:SBC-plots}) or by plotting the empirical cumulative distribution function (ECDF) of the ranks normalized against their expected values under uniformity (bottom row of Figure~\ref{fig:SBC-plots}), a method known as ECDF difference plots \citep{sailynoja_graphical_2022}.

Even though self-consistency tested via SBC is a powerful tool to ascertain the trustworthiness of a PA model if the underlying P model is well specified, it tells us nothing about the trustworthiness of PA if P is misspecified, that is, if its joint distribution cannot accurately represent the true data generating process $p^*(y)$.
In the latter case, we currently have no general procedure to verify the trustworthiness of a posterior approximation, that is, how close a PAD model is to the PD model it attempts to approximate (but see \citep{masegosa2020learning, zhang2020convergence} for recent theoretical work).
\newcontent{
This is a subtly different problem than dealing with misspecified PD models, whose convergence characteristics have been established under certain regularity conditions \citep{kleijn2006misspecification, kleijn2012bernstein}.
In the case of PAD models, we can only hope that self-consistent calibration of PA implies \textit{good enough} calibration in a sufficiently large model neighborhood of P that also contains $p^*$. 
For posterior approximators coming with guarantees of asymptotic correctness, such as MCMC, this hope is probably better justified than for neural approximators that have been shown to perform poorly under P model misspecification \citep{schmitt_bayesflow_2023, ward2022robust}.
}

\subsection{Predictive Performance}
\label{predictive-performance}

Undoubtedly, predictive performance is the central utility in most machine learning research \citep{hastie_elements_2009} and an essential goal of computational \cite{palminteri2017importance} and scientific models in general \cite{gabaix2008seven}.
Moreover, predictive performance has recently been elevated to an indispensable condition for reproducible quantitative research in the social sciences \cite{yarkoni2017choosing}.
In deep learning, enormous amounts of computing resources are spent even for just a second decimal improvement in predictive accuracy on domain benchmark data sets \cite{gesmundo2022evolutionary}, notably at the expense of other utilities (e.g., parsimony, see Section~\ref{parsimony}, or estimation speed, see Section~\ref{speed}).
In our Bayesian model taxonomy, we treat predictive performance as just one of the ten model utilities, but we still recognize it as an important one.

In a way, predictive performance would be nothing but a special case of parameter recoverability (see Section \ref{parameter-recoverability}), if not for the fact that it targets observable variables that are comparable against observed data. 
This opens up the possibility to directly evaluate predictive performance in real-world scenarios instead of having to use simulations, as is often necessary for estimating parameter recoverability. 
Along similar lines, predictive P(D) model comparison or averaging can be seen as a form of parameter recoverability from the perspective of mixture modeling (with the individual P models as components) or in terms of continuous model expansion \citep{gelman_parameterization_2004}. However, in practice, we approach these challenges mainly based on predictions from separate P(A)D models to reduce conceptual and computational costs \citep{yao_stacking_2018}.

In the following, we denote the set of ``test'' data to be predicted as $y^*$, whereas P(A)D model ``training'' data continues to be denoted by $\data$. 
In principle, these two data sets are allowed to fully coincide, partially overlap, or be completely disjoint (see Section \ref{pp-is-oos}), and $\data$ may even be empty (see Section \ref{pp-prior-post}). 
Further, we will allow the test data to be clustered into $C$ mutually independent and exhaustive clusters $y^* = \{y^*_c\}_{c=1}^C$. In most applications, both $y^*$ and $\data$ are associated with observed input variables (aka features, predictors, or covariates), but we will keep these implicit to make the notation more readable.

The ocean of predictive performance metrics for Bayesian models is vast and we refer to \cite{vehtari_survey_2012} for a comprehensive overview. 
To illustrate some overarching points in this article, we will focus on a few important metrics that follow the general form
\begin{equation}
\label{pp-metric-general}
 \mathcal{L}(y^*, \tilde{y}) := \sum_{c = 1}^{C} l \left( \mathbb{E}_{p(\theta \mid \tilde{y})}[f(y^*_c, \theta)] \right) =
 \sum_{c = 1}^{C} l \left( \int f(y^*_c, \theta) \, p(\theta \mid \tilde{y}) \, d\theta \right),
\end{equation}
where $f(y^*_c, \theta)$ is a predictive score comparing a test data cluster $y^*_c$ with corresponding model-based predictions.
We compute the \textit{expected predictive score} by integrating over the PD model posterior $p(\theta \mid \tilde{y})$, where $l$ is some function applied to each expectation before summation over clusters.
Whenever we use a PAD model, we need to approximate the above expectation over $\postad$, either by using random draws from $\postad$ or by relying on an approximate closed-form density.
Below, we examine predictive performance along multiple dimensions: absolute versus relative, prior versus posterior, and in-sample versus out-of-sample predictive performance.

\subsubsection{Absolute and Relative Predictive Performance}
\label{pp-absolute-relative}

Evaluating absolute predictive performance requires knowing an optimally achievable value of the predictive metric, whereas relative predictive performance only involves comparing multiple P(D) models' predictions evaluated on the same test data $y^*$. 
As an example for the former, consider the per-observation squared difference $f(y^*_i, \theta) = (y^*_i - \hat{y}_i(\theta))^2$ as a predictive score, where $\hat{y}_i(\theta)$ is a P(D) model-implied prediction given parameter value $\theta$ (e.g., a single random draw or realization from the posterior predictive distribution, see \citep{vehtari_survey_2012}). 
In this case, we know that the optimal value of Equation~\eqref{pp-metric-general} is zero. 
For the sake of increased interpretability, such squared differences can be further transformed to the canonical ``percentage of explained variance'' $R^2$ measures which are bounded between $0$ and $1$, the latter indicating optimal predictions \citep{gelman_R2_2019}.

However, optimal predictions are not achievable in practice, since even a Bayes-optimal decision maker may elicit suboptimal predictions in the presence of aleatoric uncertainty \cite{hastie_elements_2009}, at least when it comes to out-of-sample predictions (see Section \ref{pp-is-oos}). 
Moreover, since we nearly never know the Bayes-optimal decisions in practice (hence the need for predictive modeling in the first place), the expected optimal achievable predictive performance is also unknown to us. 
As a result, relative predictive performance is usually our only resort in practical applications \citep{vehtari_survey_2012}. 

That said, some models produce such strikingly poor predictions that they can be ruled out without the need to find a better model first, often via visual predictive checks \cite{gabry_visualization_2019}. 
For instance, if we consider the case illustrated in Figure \ref{fig:pp-checks-epi}, it is immediately obvious that the normal likelihood P model (left-hand side) is inappropriate for the given count data. 
As another example, consider a P(A)D model for binary classification that achieves just 50\% accuracy, equal to random chance. 
Assuming a balanced data set (i.e., both classes occur with the same frequency), we would not need a competing model to conclude that the classifier is bad -- unless our goal was to demonstrate that the two categories cannot be possibly differentiated given the available information. 

\subsubsection{Prior and Posterior Predictive Performance}
\label{pp-prior-post}

The distinction between prior and posterior predictive performance has often led to confusion in the past and still remains a rather precarious one to discuss. 
Prior and posterior predictive performance are distinguished based on whether we evaluate predictions before or after conditioning on the training data $\data$, respectively \citep{lotfi2022bayesian}. 
In other words, we either compute (or approximate) expectations over the prior, $\prior$, or over the posterior $\postd$.
Since prior predictive performance does not require the training data (see Equation~\ref{pp-metric-general}), we consider it a utility of P(A) models, while we view posterior predictive performance as a utility of P(A)D models. 
We still require the test data $y^*$, but it is not a part of any model class in our taxonomy.

Statistically, the line between prior and posterior predictive performance is thin and more quantitative than qualitative \citep{ohagan_fractional_1995}. 
As an illustration, suppose we observe $N$ data points in total -- then we could choose to use none, $\tilde{y} = \emptyset$, or any number between $1$ and $N$ for model training. 
For complex P models, the predictive result implied by using one or two observations for training, rather than none at all, will be almost identical, despite everything but zero training data technically counting as ``posterior'' predictive performance \citep{ohagan_fractional_1995}. 
Yet, the metrics commonly applied to quantify prior and posterior predictive performance differ not only in the amount of available training data but also in some other non-trivial ways (to be explained below).

In general, any predictive metric should match the intended real-world prediction goals. 
Below, we will focus on certain (log-)probability metrics, which can be considered good general-purpose choices in the absence of any known task-specific option \citep{vehtari_survey_2012}.


\paragraph{Prior Predictive Performance.}

The canonical metric for evaluating prior predictive performance is the joint P model likelihood evaluated at the test data, $f(y^*, \theta) = p(y^* \mid \theta)$, with $C=1$ and $l = \text{identity}$, in which case the prior expectation above becomes the marginal likelihood:
\begin{equation}
\label{marginal-likelihood}
     p(y^*) = \mathbb{E}_{p(\theta)}[p(y^* \mid \theta)] = \int p(y^* \mid \theta) \, p(\theta) \, d \, \theta.
\end{equation}
When used for model comparison, the marginal likelihood then gives rise to well-known comparative metrics known as Bayes factors evaluated by comparing two P models $\text{P}_j$ and $\text{P}_k$ as
\begin{equation}
    \text{BF}_{jk} := \frac{p(y^* \mid \text{P}_j)}{p(y^* \mid \text{P}_k)}
\end{equation}
and posterior model probabilities over a set of $J$ models $\{\text{P}_j\}_{j=1}^J$ as
\begin{equation}
    p(\text{P}_j \mid y^*) = \frac{p(y^* \mid \text{P}_j) \, p(\text{P}_j)}{\sum_{k=1}^ J(y^* \mid \text{P}_k) \, p(\text{P}_k)},
\end{equation}
where $p(y^* \mid \text{P}_j)$ denotes the marginal likelihood of P model $\text{P}_j$ and $p(\text{P}_j)$ denotes the corresponding prior probability with $\sum_{j=1}^J p(\text{P}_j) = 1$, following a closed-world assumption \citep{bernardo_bayesian_1994, yao_stacking_2018}.

Although the marginal likelihood is formally an expectation and thus, in theory, we can approximate it arbitrarily well using sufficiently many random draws from the prior, it is practically impossible to evaluate due to its unfavorable pre-asymptotic behavior for any non-trivial model \citep{meng_simulating_1996, vehtari_survey_2012, gronau_bridgesampling_2020}. 
The main reason for this is that the parameter subset for which $p(y^* \mid \theta)$ contributes to the integral in Equation~\eqref{marginal-likelihood} (i.e., the typical parameter set implied by the test data; \citep{betancourt_hmc_2017}) is very narrow and thus we need a very high number of prior draws to ensure sufficiently many of them occupy that narrow space. 
In addition, numerical issues caused by $p(y^* \mid \theta)$, such as floating-point underflow, can also be hindering.

For these reasons, the practical computation of marginal likelihoods currently rests on bridge sampling \citep{bennett1976efficient} relying on posterior draws from a corresponding PAD model \citep{meng_simulating_1996,  gronau_bridgesampling_2020}. 
In contrast to estimating posterior expectations or quantiles, bridge sampling requires about an order of magnitude more posterior draws to yield reliable results \citep{gronau_bridgesampling_2020} and is still largely missing principled convergence diagnostics or uncertainty quantification \citep{gronau_tutorial_2017}, leaving room for future research.

An alternative prior predictive metric arises if one uses the log-likelihood $f(y^*, \theta) = \log p(y^* \mid \theta)$ as a (predictive) score instead of the likelihood itself, which leads to the Gibbs loss \citep{watanabe_algebraic_2009} that, for factorizable likelihoods \citep{burkner2021nfloo}, evaluates to
\begin{equation}
     \text{Gibbs}_{p(\theta)}(y^*) := \mathbb{E}_{p(\theta)}[\log p(y^* \mid \theta)] =
     \sum_{i=1}^{N^*} \mathbb{E}_{p(\theta)}[\log p(y^*_i \mid \theta)].
\end{equation}
The Gibbs loss is not only simpler to evaluate for exponential family models \citep{vehtari_survey_2012} and numerically more stable than the marginal likelihood but also exhibits better pre-asymptotic behavior for factorizable likelihoods when estimated via prior draws since the integrands become much simpler. 
However, the Gibbs loss cannot be used to obtain actual predictions because it does not evaluate to a predictive distribution over $y^*$ \citep{vehtari_survey_2012}.

The latter problem can be avoided by taking expectations with respect to individual test observations $y^*_i$ first and only taking the log afterwards ($C = N^*$ and $l = \log$), which leads to the expected log predictive density (ELPD) metric \citep{vehtari_survey_2012, vehtari_2017_loo}, evaluated over the prior:
\begin{equation}
\label{elpd}
    \text{ELPD}_{p(\theta)}(y^*) := \sum_{i=1}^{N^*} \log p(y^*_i) = \sum_{i=1}^{N^*} \log \mathbb{E}_{p(\theta)}[ p(y^*_i \mid \theta)].
\end{equation}
Comparing equations \eqref{marginal-likelihood} and \eqref{elpd}, we see that the marginal likelihood considers the joint predictive density of all test data $y^*$, while the ELPD considers marginal predictive densities of $y^*_i$, marginalized over all other test data.
Even though the ELPD has found wide application in the context of posterior predictive performance \citep{vehtari_2017_loo}, it does not yet seem to play a noteworthy role in the context of prior predictive performance. 
However, together with the Gibbs loss, it may become a computationally favourable competitor to metrics based on the marginal likelihood.

\paragraph{Posterior Predictive Performance.}

When assessing posterior predictive performance, we apply the same metrics we encountered in the context of prior predictive performance but evaluate expectations over the posterior induced by the training data $\tilde{y}$. 
However, the practical popularity of the metrics seems to be reversed when it comes to posterior predictions.
For example, the posterior ELPD
\begin{equation}
    \text{ELPD}_{p(\theta \mid \tilde{y})}(y^*) := \sum_{i=1}^{N^*} \log p(y^*_i \mid \tilde{y}) = \sum_{i=1}^{N^*} \log \mathbb{E}_{p(\theta \mid \tilde{y})}[ p(y^*_i \mid \theta)]
\end{equation}
finds widespread application \citep{vehtari_2017_loo}, while the ``conditional marginal likelihood''
\begin{equation}
     p(y^* \mid \tilde{y}) = \mathbb{E}_{p(\theta \mid \tilde{y})}[p(y^* \mid \theta)] = \int p(y^* \mid \theta) \, p(\theta \mid \tilde{y}) \, d\theta
\end{equation}
has not yet attained wide popularity, despite having several useful properties \citep{ohagan_fractional_1995, berger_intrinsic_1996, gu_approximated_2018, lotfi2022bayesian}. 

The choice between prior or posterior predictive performance seems to depend on the modeling goals for which a P model is specified. 
While prior predictive performance seems to be favored for the purpose of testing scientific theories \citep{wrinch_aspects_1919, haldane_note_1932, etz_bayes-factor_2017, gu_approximated_2018}, posterior predictive performance is the perspective of choice in almost all machine learning scenarios (but see \citep{yarkoni2017choosing}), where we first obtain a PAD model based on training data (and perhaps only minimal prior information) and then utilize the model in downstream predictive tasks \citep{hastie_elements_2009}.

\subsubsection{In-Sample and Out-of-Sample Predictive Performance}
\label{pp-is-oos}

We measure in-sample predictive performance if the test data is a subset of the training data, $y^* \subseteq \tilde{y}$,  but measure out-of-sample predictive performance if test and training data do not overlap, that is, $y^* \cap \tilde{y} = \emptyset$. 
Whenever we evaluate prior predictive performance, we have no training data per definition and thus always measure out-of-sample predictions. 
Accordingly, the difference between in-sample and out-of-sample predictive performance only matters in the context of posterior predictions.

From a posterior predictive perspective, the decision between using in-sample and out-of-sample predictive performance is based on whether or not we want to generalize our inferences from a data set to a wider population. 
If a given data set included the entire problem space, then in-sample predictive performance would be sufficient. 
However, as most introductory statistical courses teach, a data set is typically only a small sample from a much larger population, to which we would like to extend our inferences. 
Thus, out-of-sample predictive performance (aka generalization ability) is almost always what we are after \citep{hastie_elements_2009, vehtari_survey_2012, vehtari_2017_loo, yarkoni2017choosing}. 
That said, we can still learn from in-sample predictive performance, as it provides an upper bound for out-of-sample predictive performance in expectation, such that when in-sample predictions are poor, out-of-sample predictions are likely to be even worse \citep{hastie_elements_2009, vehtari_2017_loo, gabry_visualization_2019}.

In the presence of only a single overall data set $y_{\rm total}$, estimating out-of-sample predictions is practically realized via data splitting, such that~$y_{\rm total} = \{ \tilde{y}, y^* \}$. 
To reduce the dependency of the predictive results on a single realized data split, we typically perform cross-validation by repeating the data splitting several times (folds), evaluating out-of-sample predictions for every fold, and then aggregating the results across folds \citep{stone_cross-validation_1978, vehtari_survey_2012, vehtari_2017_loo}.
 
The type of cross-validation scheme employed should resemble the envisioned prediction goals for which the PD model has been created \citep{vehtari_survey_2012}. 
For example, the predictive goal of time series models is usually to predict future values based on past values, making leave-future-out cross-validation a sensible choice \citep{burkner_lfo_2020}. 
Regardless of the type of cross-validation employed, it involves the repeated fitting of the same P(A) model to different data sets. 
Depending on the number of such refits, the individual data sizes, and the applied approximator, the required estimation time can quickly become prohibitive for any practical use. 
As such, approximate cross-validation procedures that require no or only a few refits have proven to be highly popular in practice \citep{vehtari_2017_loo, vehtari_pareto_2021, burkner_lfo_2020}. 
However, key cross-validation schemes, such as leave-group-out cross-validation, cannot yet be robustly approximated, so there is more research needed in that direction \citep{paananen_implicitly_2021}.

Although evaluating out-of-sample predictive performance is often our best shot at preventing overfitting to the training data, it is not always sufficient to fully achieve good generalization within commonly applied model-building workflows \citep{gelman2020workflow}. 
In these workflows, we typically fit different P models to the same data in an iterative fashion. 
For example, we might first compare two models, decide which one to retain, and only then fit a third model to compare it with the winner of the first round.
Even if each model choice was based on local out-of-sample predictive performance, subsequent results can be informed by out-of-sample results from previous iterations, making it not strictly out-of-sample for any future iteration steps from the perspective of the analyst's knowledge. As such, in an iterative workflow, local out-of-sample predictive metrics may still lead to overfitting, but the degree to which this biases the end results remains a topic for future research.

\subsubsection{Predictions in a Dynamic World}

Time is one of the most precipitous sources of uncertainty and any attempt to forecast the future with a static, time-independent P(A)D model will only be meaningful if the opaque generator $\mathbb{G}$ is strictly stationary (i.e., its regularities are invariant to time).
Otherwise, a P model needs to have an appropriate temporal resolution to deliver reasonable out-of-sample predictions beyond the empirical snapshot of the collected data.
Moreover, since the precise details of temporal shifts are extremely hard to anticipate, a P(A)D model which claims universal predictive performance should regularly be subjected to the falsification of time.

This brings us to an important distinction when it comes to assessing out-of-sample predictive performance.
Whenever we make our P(A)D model ``blind'' to certain observations in the original data set D and use these observations to assess our-of-sample predictive performance (as we do in any form of cross-validation, even those built for time series data \citep{burkner_lfo_2020}), we are essentially testing the model's ability to perform \textit{induction} about the statistical regularities of $p^*(y)$ in a temporal snapshot determined by data collection.
In such a scenario, however, we are not probing the model's ability to faithfully forecast the future, since the ``left-out'' observations are new only from the perspective of the model, but not from that of the modeler.
Thus, cross-validation can sometimes be overly optimistic in estimating out-of-sample predictive performance, since a sample collected at a future date might exhibit surprisingly different properties (i.e., the P model would no longer be structurally faithful) than the sample currently at hand.

Why would the empirical distribution $p^*(y)$ change over time? 
One reason can be that the hidden properties of the generator $\mathbb{G}$ itself may change, bringing about alterations in the statistical properties of $p^*(y)$.
For instance, strong auto-correlations in financial time series are notoriously short-lived due to feedback processes and market adaptation \cite{sornette2009stock}.
Yet another reason can be that new sources of noise contaminate future data D in unexpected ways.
For instance, a sensor in a measurement device may break and yield incorrect data or case reporting policies during an ongoing pandemic may switch between waves. However, the P(A)D model may have no mechanism to adapt to any of these changes and its out-of-sample predictive performance would likely suffer.

Within our model taxonomy, prediction failures due to changes in $p^*(y)$ concern misaligned assumptions about temporal invariances embodied in the P model's structure.
One way to revise these assumptions is to include time-varying parameters $\theta_t$ in the P model, with the corresponding time-invariant parameterization being a special (and more parsimonious) case. For instance, this can be achieved within the \textit{superstatistics} framework \cite{beck2003}, which aims to represent heterogeneous dynamics through a superposition of multiple stochastic processes at different temporal scales \cite{mark2018}.  
In any case, researchers should bear in mind that static P(A)D models are not designed to deal with \textit{things that move}, so, as simple as it sounds, time remains a key arbiter of the quest for universal substantial conclusions or robust predictive systems.

\subsection{Fairness}
\label{fairness}


\emph{Fairness} in the context of model building aims to ensure that model-guided decisions are equitable, with a specific focus on groups that differ in protected attributes, such as sex, gender, or ethnic background \citep{corbett-davies_fairness_2018, barocas_fairness_2021}. 
In a relatively narrow sense, fairness is a primary concern for P(A)D models, as it applies to real-world outcomes and their real-world reverberations owing to the connection between a P model's structure and data D.
However, purely simulation-based P(A) models are not exempt from fairness considerations, especially when used to guide important public policies and decision support systems \citep{burkner_statistical_2018, bak2022computing, nussbaumer2021framework}.
In the following, due to its predominant share in the literature, we will examine the fairness of P(A)D models from two different perspectives, namely, from the perspectives of psychometric measurement and predictive modeling.

\subsubsection{Measurement Fairness}

In psychometric measurement theory, the aim is to estimate people's scores on latent psychological traits, for example, general intelligence, creativity, or aptitude for university programs \citep{drasgow_fitting_1995}.
In the model-based literature of psychometric measurement, namely Item Response Theory (IRT; \citep{van_der_linden_irt_1997, embretson_irt_2000, brms3}), two major aspects of fairness have received considerable attention. 

First, we need to ensure that the observable features (i.e., items) have been selected and administered in a fair way \citep{mccallum_nonverbal_2003, borsboom_concept_2004, area_standards_2011}. 
This aspect does not appear to be immediately model-based, since it concerns the data collection process as well as causal assumptions about the latent traits' influence on the item responses \citep{borsboom_concept_2004}. 
However, some of its requirements can be checked via P(A)D models in the form of differential item functioning (DIF) analysis \citep{holland_differential_1993, osterlind_differential_2009}. 
When investigating DIF, the item parameters $\zeta_i$ of item $i$ are allowed to vary across groups $g$ and their P(A)D model's posteriors are compared to verify their statistical equivalence. 
That is, we aim to examine whether $p(\zeta_{i} \mid \tilde{y}, g) \approx p(\zeta_{i} \mid \tilde{y}, g')$ holds for all pairs of considered groups $g$ and $g'$ and all items $i$.

Second, we need to estimate the latent traits of all individuals with a similar degree of uncertainty \citep{burkner_information_2022}. 
In the context of P(A)D models, this means that the posterior of trait $\eta_j$ for person $j$ has approximately the same entropy across all individuals being compared, that is, $\mathbb{H}(\eta_j \mid \tilde{y}) \approx \mathbb{H}(\eta_{j'} \mid \tilde{y})$ for all pairs of individuals $j$ and $j'$. 
This turns out to be a difficult, sometimes even unachievable goal: Due to floor and ceiling effects arising in almost all psychometric tests, the resulting information is non-uniform across the latent trait space in non-linear IRT models \citep{van_der_linden_irt_1997, burkner_statistical_2018, burkner_information_2022}. 
As a result, more extreme latent trait scores will be estimated less precisely than more average scores. 
As a partial remedy, one may try to ensure that the information gain about all individuals' trait scores at least exceeds a minimal, application-specific threshold \citep{burkner_information_2022}.

\subsubsection{Predictive Fairness}

What we term \textit{predictive fairness} has its origins in the field of machine learning \cite{rothwell_how_2014, barocas_fairness_2021}.
We will define predictive fairness directly on PD models because there is no hope that a P model can yield fair decisions for all possible training data; after all, training data may themselves be biased against protected groups \cite{rothwell_how_2014, barocas_fairness_2021}. And while we define it as a utility of PD models, it also automatically pertains to a corresponding PAD model, unless the posterior has a simple analytic form.

Mathematically, for individual-level decisions, we consider a PD model-specific decision rule  $d(x \mid \tilde{x}, \tilde{y})$ that outputs a decision for each admissible vector of attribute values $x$ given training data $D = (\tilde{x}, \tilde{y})$ consisting of observed attribute values $\tilde{x}$ and corresponding decision-relevant outcomes $\tilde{y}$ in a supervised learning context.
If we consider only binary decisions to simplify notation, we can write the decision rule as
\begin{equation}
  d(x \mid \tilde{x}, \tilde{y}) := \begin{cases}
1 \quad \text{if} \quad \bar{r}(x \mid \tilde{x}, \tilde{y}) > \tau \\
0 \quad \text{otherwise}
\end{cases}
\end{equation}
with
\begin{equation}
\bar{r}(x \mid \tilde{x}, \tilde{y}) := \int r(x, \theta) \, p(\theta \mid \tilde{x}, \tilde{y}) \, d \theta
\end{equation}
being a real-valued (expected) \emph{risk score} of $x$ that is obtained as an expectation over the PD model's posterior $p(\theta \mid \tilde{x}, \tilde{y})$. The decision (e.g., whether to give someone a loan or release a defendant while they await trial) is then made by comparing the risk score against a pre-defined threshold $\tau$. 
The conditional risk score $r(x, \theta)$ determines how the P model and its parameters $\theta$ are used for assessing risk.
For example, the risk score could be the mean of the PD model's predictive distribution given feature value $x$ and parameter value $\theta$:
\begin{equation}
  r(x, \theta) := \int y \, p(y \mid x, \theta) \, d y.
\end{equation}
Conditional risk scores do not necessarily have to rely on the predictive distribution. 
Rather, they may also be based on latent model quantities, such as psychometric trait scores obtained from IRT P(A)D models \citep{van_der_linden_irt_1997, embretson_irt_2000, brms3}, which bridges the gap between measurement and predictive fairness.

There are different classes of predictive fairness criteria considered in the literature, among others \emph{anti-classification} \citep{bonchi_exposing_2017, corbett-davies_fairness_2018} and \emph{classification parity} \citep{corbett-davies_fairness_2018, berk_fairness_2021} (also known as \emph{statistical parity}; \citep{chouldechova_fairness_2018}). Even within these classes, criteria are partially incompatible and neither of them can actually ensure universal fairness, but we can still learn from their limitations \citep{corbett-davies_fairness_2018, chouldechova_fairness_2018, barocas_fairness_2021, berk_fairness_2021}.
In the context of such criteria, we differentiate between protected attributes $x_p$ (e.g., sex, gender, or ethnic background) and other, unprotected attributes $x_u$ such that $x = (x_p, x_u)$. Anti-classification requires that protected attributes $x_p$ (or their proxies; \citep{bonchi_exposing_2017}) are not used in model-based decisions at all, which mathematically translates to
\begin{equation}
    d(x \mid \tilde{x}, \tilde{y}) = d(x' \mid \tilde{x}, \tilde{y})  \quad \text{for all} \quad x, x'  \quad \text{with} \quad x_u = x_u'.
\end{equation}
In our PAD model taxonomy, this can simply be realized by using a PD model with $p(\theta \mid \tilde{x}, \tilde{y}) = p(\theta \mid \tilde{x}_u, \tilde{y})$ and conditional risk score $r(x, \theta)$ that is independent of $x_p$ as well. 
Anti-classification approaches have two main drawbacks. 
First, protected attributes can often be predicted fairly well from unprotected attributes, which makes it impossible to be completely agnostic about them \citep{feldman_certifying_2015}. 
Second, empirical risk distributions (after removing all unfair risk influences) may differ across values of $x_p$, such that ignoring the latter may actually lead to unfair decisions against the groups one originally attempted to protect \citep{corbett-davies_fairness_2018}.


Differently, classification parity comprises a class of fairness criteria that requires the population distribution of certain decision metrics to be the same across all values of the protected attributes \citep{corbett-davies_fairness_2018, berk_fairness_2021}. 
Using
\emph{demographic parity} \citep{feldman_certifying_2015} as an example, we would require that the decision's distribution itself, as implied by the distribution of attributes $x$ in the considered population, to be independent of the protected attributes:
\begin{equation}
    p(d(x \mid \tilde{x}, \tilde{y}) \mid x_p) = p(d(x \mid \tilde{x}, \tilde{y})).
\end{equation}
Contrary to anti-classification, we usually have to incorporate the protected attributes into the P model in the first place to ensure any kind of classification parity \citep{berk_fairness_2021}. In the context of psychological tests, for example, this could be achieved by imposing group-specific norms of comparison \citep{rust_psychometrics_2014}. Yet, classification parity does not guarantee universal fairness either, whenever the true risk score distribution (after removing all unfair risk influences) varies between groups defined by the protected attributes \citep{corbett-davies_fairness_2018}.


The shortcomings of these predictive fairness definitions highlight that requiring a certain outcome -- the decision itself (anti-classification) or aspects of its population distribution (classification parity) -- to be independent of the protected attributes may be insufficient. 
Towards the goal of achieving fairness through a PD model, the underlying P model needs to be causally consistent (see also Section~\ref{sec:causality}) in a way that considers how the protected attributes $x_p$ relate to the causal graph that includes all the valid, unprotected attributes $x_u$ and the outcome $y$ \citep{bonchi_exposing_2017}. 
In addition, the training data D needs to be representative of the true (unbiased) outcome distribution $p^*(y)$. 
It goes without saying that these are complicated, application-specific tasks that require contributions from various scientific fields and considerable domain expertise.

What is more, fair decisions, regardless of their modeling context, need to take into account that the same decision may affect different people (and their surroundings) differently and that these differences may be related to both protected and unprotected attributes. 
More formally, we need to consider the decision $d(x \mid \tilde{x}, \tilde{y})$ in a context $C(x)$ that only together determine the output of a utility function $U(d(x \mid \tilde{x}, \tilde{y}), C(x))$, which offsets all possible gains and losses caused by the decision. 
Obtaining such a function could steer a decision towards fairness as quantified by equal utility outcomes across protected groups.

At an even higher level, we should consider taking sufficient precaution that (anticipated) political decisions or societal processes triggered by anonymous modeling results do not lead to unfair treatment of protected groups. 
However, such considerations may come into conflict with the principle of scientific freedom, in which case a careful ethical analysis of the specific situation becomes mandatory.

\subsection{Structural Faithfulness}
\label{structural_faithfulness}

In most data analysis scenarios, we have a reasonable amount of qualitative prior knowledge about the data structure and the data generating process, even if we don't know the precise analytic relation between the two.
In particular, this knowledge concerns the scales of variables to be modeled, the dependencies between observations, as well as physical constraints, such as symmetries or invariances.
The \emph{Structural Faithfulness} utility captures how well a P model incorporates such knowledge.
Structural faithfulness is at the core of statistical modeling, be it Bayesian or otherwise, as it determines the probability distributions we assign to our observed and unobserved variables, the parameters we add to our P models, and the assumptions we can justifiably make to simplify reality.

Moreover, we can roughly distinguish between probabilistic structure and functional structure, which are related to the modeler's degree of ignorance regarding the problem at hand.
Purely statistical models aim to capture the \textit{probabilistic structure} of $p^*(y)$, without making reference to \textit{functional structure} of the hidden generator $\mathbb{G}$.
Non-deterministic mechanistic models, on the other hand, aim to capture the functional structure of $\mathbb{G}$ (usually represented by physical constraints), such that the probabilistic structure of $p^*(y)$ can be reproduced or explained.
For instance, when we study the dynamics of a phenomenon via stochastic differential equations, functional faithfulness refers to the mathematical form of the differential equation and probabilistic faithfulness refers to the fidelity of the stochastic assumptions.

To us, it remains unclear how to measure structural faithfulness in an absolute sense and we see it primarily as a relative metric.
What is more, structural faithfulness consists of multiple components that may each favor a different P model.
For example, model $\text{P}_1$ might take a known symmetry into account that model $\text{P}_2$ ignores, while $\text{P}_2$ might assign a more appropriate distribution to a response variable than $\text{P}_1$ does.
In this case, none of the two P models would actually be more structurally faithful than the other, at least not uniformly so.

\subsubsection{Variable Scales}

\begin{figure}[t]
    \centering
    \includegraphics[width=0.99\textwidth]{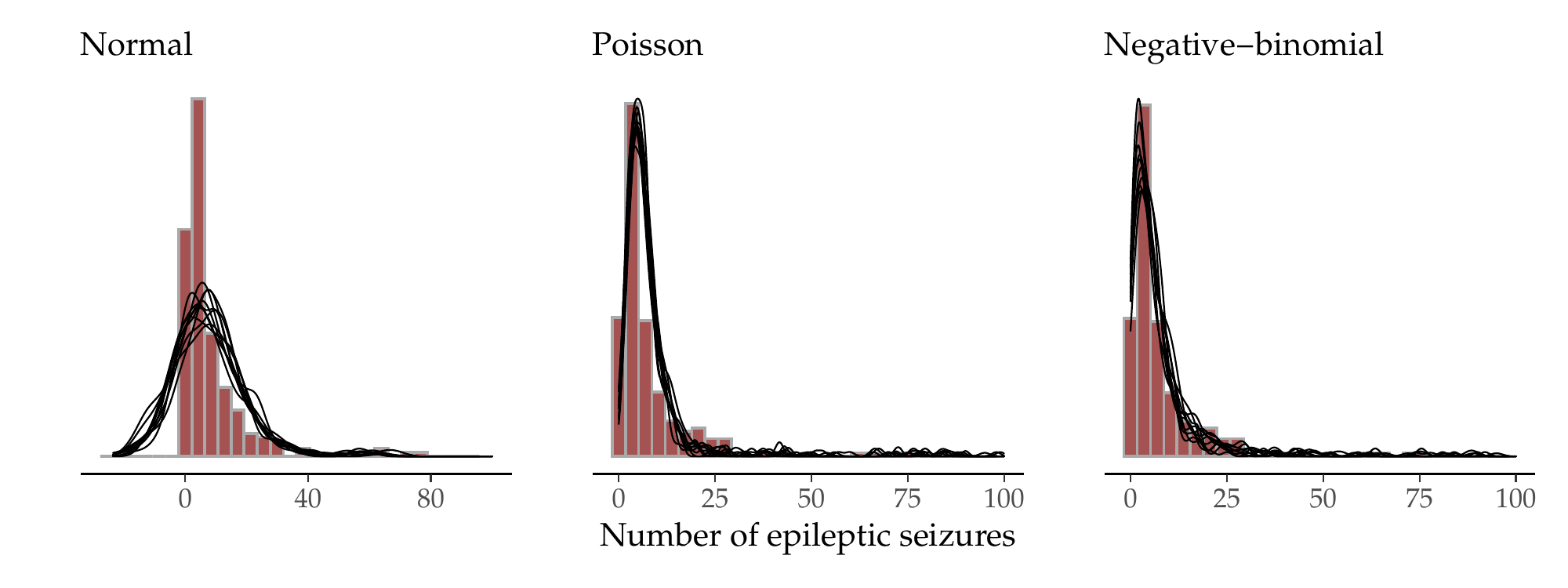}
    \caption{Posterior predictive checks \citep{gabry_visualization_2019} of epilepsy treatment data \citep{thall_count_1990}. The response variable is the number of epileptic seizures of patients in a given time interval, that is, a count variable without a known upper bound. Results are shown for three PAD models with different likelihoods (shown as facets) and posteriors approximated via MCMC in Stan \citep{stan_2022}. Histograms indicate observed data and each black line indicates one draw from the posterior predictive distribution of the corresponding PAD model, smoothed via continuous density estimation. For Poisson and negative-binomial likelihoods, posterior predictions are in fact counts but are still displayed as smoothed continuous densities to ease readability and comparability across facets. As is clearly visible on the left-hand side, the PAD model with normal likelihood predicts a lot of theoretically impossible negative counts and can neither predict the spike at counts close to zero, nor the heavy right tail.}
    \label{fig:pp-checks-epi}
\end{figure}

The scale of a variable determines not only what information it represents but also how it should ideally be treated within a P model.
For example, if the response variable consists of count data without a known or practically reachable upper bound, we should model this data via an appropriate (unbounded) discrete distribution (e.g., Poisson, or some of its generalizations) to sensibly capture the aleatoric (irreducible) uncertainty in those count responses \citep{vives_count_2006, frome_analysis_1983, winter_poisson_2021}.
What is more, this ensures that the variables' natural boundaries are respected (e.g., lower bound of zero for count data), such that the corresponding model predictions cannot go beyond the data space that is possible in reality (see Figure~\ref{fig:pp-checks-epi} for an illustration). 
As another example, if our response variable is ordinal, that is, it consists of discrete ordered categories without guarantees that the categories can be considered equidistant, we should model such data via an ordinal distribution \citep{mccullagh_regression_1980, liddell_analyzing_2018, burkner_ordinal_2019}. 
The same points hold also for predicting variables even if they are not explicitly modeled with a distribution \citep{burkner_mo_2020, gertheiss_penalized_2009}. 
Failure to consider the variable scales in P models can have detrimental consequences for the validity of the obtained results \citep{gertheiss_penalized_2009, burkner_mo_2020, liddell_analyzing_2018}. 
Equivalently, respecting the intrinsic scales of all quantities included in a P model can help to avoid unreasonable parameter estimates or implausible (or worse, impossible) predictions. 

\subsubsection{Probabilistic Structures}
\label{probabilistic-structures}

Observed data often exhibits specific probabilistic structures that can be inferred from (qualitative) understanding of the data-generating process.
For example, if we collect psychometric data from multiple students in the same class, it is highly unlikely that the data points will be mutually independent (e.g., because students share the same teacher, rooms, peers, etc.).
This situation is prototypical for the application of \emph{multilevel models}, which aim to capture such dependencies \citep{gelman_data_2006, lme4, brms1, brms2}.
Multilevel models treat such dependencies of observations belonging to the same group as equivalent to variation between groups \citep{gelman_data_2006}.
In other words, if there were no variation between groups, there would be no structural dependency of observations within groups (at least none elicited by this grouping structure).

\begin{figure}[t]
\centering
\includegraphics[width=0.99\textwidth]{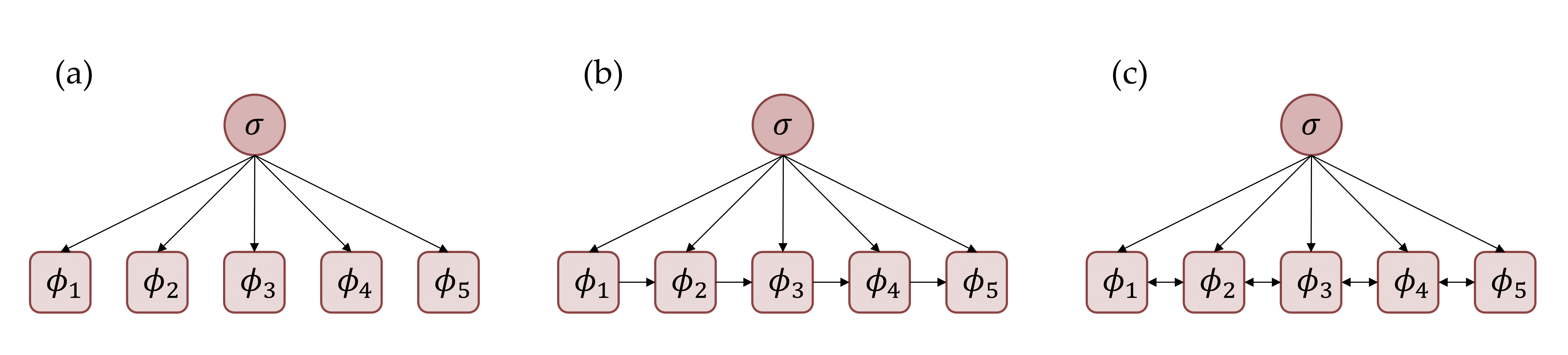}
\caption{Graphs illustrating common probabilistic structures. Rectangles depict nested parameters within a given probabilistic structure. Circles depict the corresponding hyperparameters. (a)
Exchangeable parameters; (b) Conditionally dependent parameters with a directed (e.g., temporal) dependency structure.
(c) Conditionally dependent parameters with bidirectional (e.g., spatial) dependency structure.}
\label{fig:structural-dependence}
\end{figure}

There are three major types of structural dependence between groups that can be expressed as multilevel models: exchangeable, directed, and undirected \citep{rue_GRF_2005, fuglstad_GRF_2019, gao_priors_2021}, illustrated schematically in Figure~\ref{fig:structural-dependence}.

Exchangeable groups are the most common assumption in multilevel models and imply that (before seeing any data) we hold the same prior beliefs about each of the groups but assume they are all drawn from the same population (e.g., students within classes, classes within schools, schools within cities, etc.).
In the most simple case (i.e., two-level structure, univariate and normally distributed parameters), we would specify a univariate normal prior for each group indexed by $i$ and group parameter $\phi_i$ as
\begin{equation}
    \phi_i \sim \text{Normal}(\mu, \, \sigma),
\end{equation}
where $\mu$ and $\sigma$ are the mean and standard deviation parameters shared across groups, respectively.
Typically, we would estimate the across-group parameters from the data along with the group-specific parameters $\phi_i$ themselves.

In directed dependency structures, adjacent groups are assumed to have directed influence on each other in a way that group $i$ can affect group $j$, but not vice versa.
The most common example is temporal autocorrelation where a variable at time $i$ can potentially be influenced by a variable at time $i-1$ \citep{spirtes_causal_2016, gao_priors_2021}.
For a univariate Gaussian random walk, we would formalize this assumption with the following prior
\begin{equation}
    \phi_i \sim \text{Normal}(\phi_{i-1}, \, \sigma).
\end{equation}

In undirected dependency structures, the influence of adjacent groups can go both ways, with spatial autocorrelation being the most common example \citep{besag_spatial_1974, gelfand_car_2003, morris_spatial_2019}. 
For example, in (spatial) conditional autoregressive (CAR) structures \citep{besag_spatial_1974}, we could write down the prior on the group coefficients as 
\begin{equation}
    \phi_i \sim \text{Normal} \left(\frac{1}{|\mathbf{N}_i|} \sum_{j \in \mathbf{N}_i} \phi_{j}, \, \sigma \right),
\end{equation}
where $\mathbf{N}_i$ is the set of groups that are neighbours of group $i$.
Importantly, a shared feature of these dependency structures is that they are agnostic towards the underlying causal mechanisms -- their purpose is purely to accurately represent the inherent probabilistic structure of the observed data \citep{gelfand_car_2003, williams_litter_2017, gao_priors_2021}.

But what if the data-generating process suggests a certain kind of dependency for which we find no empirical support?
For example, shall we retain a grouping term of classes even if the PAD model suggests close to zero variation between groups? 
There are good arguments for both choices. 
On the one hand, excluding such a term implies a simpler model with higher parsimony \citep{bates_parsimonious_2015} (see also Sections \ref{parsimony}), although the increase in parsimony will be quite small due to the \emph{partial pooling} property of multilevel models induced by their hierarchical priors if there is a sufficient number of groups \citep{gelman_data_2006, hodges2001}.
On the other hand, including the term sets a good example for future replications and applications of the same P model, in the same or different contexts.
That is, if someone applies this P model to a new data set, they may very well find the between-group variation under question to be non-zero, thus justifying the inclusion of the corresponding grouping term.

\subsubsection{Physical Constraints}

In the domains of physics and natural sciences, we tend to have strong prior knowledge about the functional P model structure in the form of known hard constraints such as symmetries, invariances, or conservation laws \citep{sanz-sole_port-hamiltonian_2007, raissi_pinn_2019, karniadakis_piml_2021, baddoo_pidmd_2021}.
For example, a harmonic oscillator expressed by the second-order differential equation
\begin{equation}
\label{harmonic-oscillator}
\ddot{x}(t) = k \, x(t),
\end{equation}
with functional solution $x$, second derivative $\ddot{x}$, as well as constant $k$, represents an isolated system that is energy conserving \citep{lurie_mechanics_2002}.

Similar to a harmonic oscillator, most physical hard constraints can be expressed via differential equations whose direct inclusion in a P model is computationally demanding if we do not have access to an analytic solution \citep{cranmer_sbi_2020, lavin_simulation_2021, sunnaker_abc_2013}. Accordingly, building a more flexible, data-driven P model as a surrogate is a computationally attractive choice \citep{lavin_simulation_2021, burkner_pce_2022}.
Still, even for such a surrogate, it remains beneficial to incorporate known physical constraints to eliminate the need to learn them directly from data.
This is likely to increase the model's data efficiency, that is, the amount of data required by the model to achieve a certain predictive goal \citep{raissi_pinn_2019, lavin_simulation_2021}.
The discussion about physics-informed modeling is particularly prominent in core areas of high-dimensional machine learning, such as neural networks that tend to be very data-hungry \citep{raissi_pinn_2019}, but in principle applies to all P models created for representing data with known physical constraints.




\subsection{Parsimony}
\label{parsimony}

\emph{Parsimony} refers to the formal simplicity of a Bayesian model; some might define it as the conceptual or mathematical elegance of the underlying interpretative framework.
Here, we view parsimony as a quantifiable property of a Bayesian model.
We treat it also as a relative quantity -- it is always possible to propose a more complex model (or possibly a simpler one) which is equally consistent with the available data.

Within our PAD framework, we will distinguish two types of parsimony: P-parsimony and A-parsimony.
P-parsimony characterizes the formal simplicity of a P model and should be measurable from the structure of the joint distribution $\joint$.
A-parsimony characterizes the simplicity of an approximator and should be measurable through the interface of A.
The former is directly related to the theoretical appeal of a P model's probabilistic assumptions; the latter is directly associated with the usability of an approximator.

\subsubsection{P-Parsimony}
\label{P-parsimony}

\begin{figure}
\centering
\includegraphics[width=0.99\textwidth]{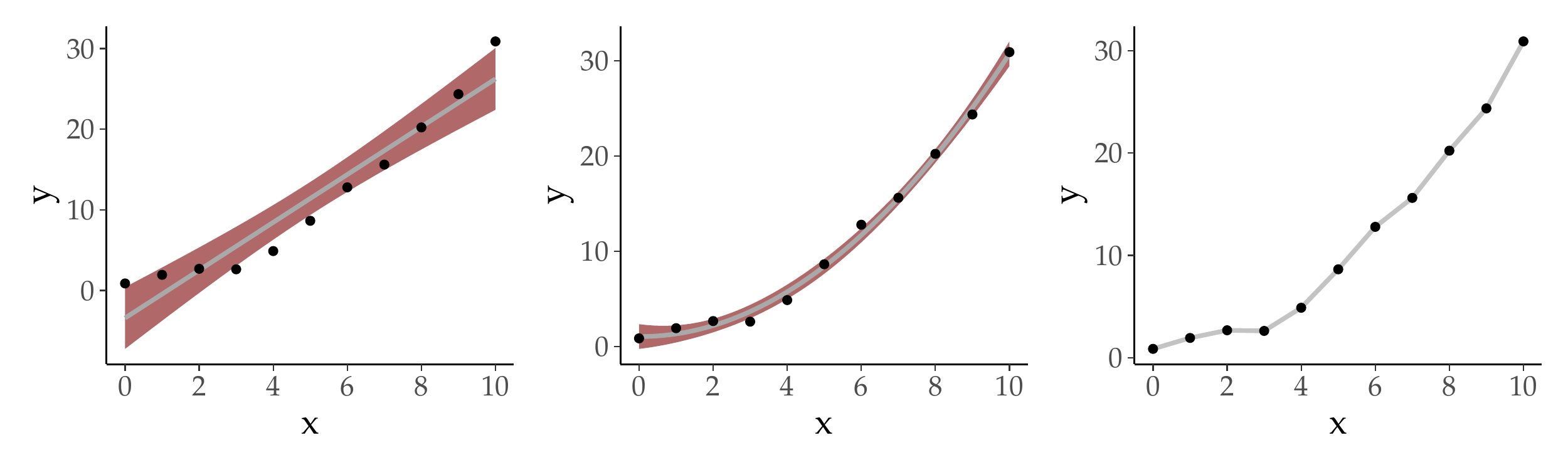}
\caption{P models of different complexity applied to a data set D of $11$ observations following a quadratic relationship in expectation. Left: Most parsimonious, linear model with a $3$-parameter likelihood $y \sim \text{Normal}(\beta_0 + \beta_1 x, \sigma)$. This model is too simple for the data. Center: Slightly less parsimonious, quadratic model with a 4-parameter likelihood $y \sim \text{Normal}(\beta_0 + \beta_1 x + \beta_2 x^2, \sigma)$. This model's complexity is \textit{just right} for the data. Right: Least parsimonious, linear interpolation model between adjacent points that has as many parameters as observations in the data (1 intercept and 10 linear slopes). This model is too complex for the data. Shaded areas indicate 95\% credible intervals of the regression line for models where this uncertainty can be computed.}. 
\label{fig:parsimony-plots}
\end{figure}

In many real-world modeling scenarios, we have limited data and strive for P models that can capture all relevant latent properties with as little data as possible (see Figure~\ref{fig:parsimony-plots} for a simple illustration).
One particular aspect of this goal is captured by the dimensionality of the parameter space, whereby higher parsimony simply means lower parameter dimensionality.
Canonical examples for high parsimony are physical simulators defined by complex (white-box) forward models with intractable likelihoods \cite{cranmer_sbi_2020}.
The latter are informed by strong subject matter knowledge and are thus able to maintain low parameter dimensionality (e.g., consider the harmonic oscillator Equation~\eqref{harmonic-oscillator}, which only requires a single parameter to describe highly non-linear, non-monotonic behavior).
On the other end of the spectrum are neural network models that tend to use simple likelihoods (e.g., Gaussian or categorical), but are characterized by an extremely high parameter dimensionality and large compositions of non-linear transformations, such as GPT-3 featuring $175$ billion parameters \citep{floridiGPT3ItsNature2020}.
In a way, we need to compensate for our lack of \textit{a priori} knowledge (or inability/unwillingness to use it) by applying less parsimonious models that replace more restrictive model structures with a heightened hunger for data.

The motivation for parsimony is related to other utilities as well, since more parsimonious P(A)D models tend to require less data to achieve the same reduction in epistemic uncertainty (parameter recoverability; Section~\ref{parameter-recoverability}) and predictions (predictive performance; Section~\ref{predictive-performance}), and tend to be easier to comprehend in real-world applications (interpretability; Section~\ref{interpretability}). 
Still, we can construct chaotic models -- where minimal changes in the parameters lead to strong changes in the predictions -- that are highly parsimonious, yet uninterpretable and extremely flexible in terms of the function space they can approximate \citep{piantadosi_one_2018}. 
However, most P models applied in current practice do not exhibit such chaotic behavior.

Despite its intimate connection to other utilities, we think that parsimony deserves to be a utility in its own right, harmonized with Occam's razor: Given two models, and other things being equal, one should choose the more parsimonious one \citep{blumer1987occam}.
Increasing the parsimony of a model (or a scientific theory, for that matter) implies making more restrictive assumptions (i.e., reducing the function space that can be theoretically approximated by the model), thus increasing its \emph{falsifiability}: We can more easily create situations where the model is wrong.
Furthermore, in applied settings, sparser models may lead to more efficient data collection and more economical measurement designs (i.e., fewer variables to measure or less acquisition trials in design optimization) \citep{pavone_refmodels_2022}.
Nevertheless, the strive for parsimony may not always be a useful guide to our scientific exploration, if the aesthetics of parsimonious P models make us blind for potentially more appropriate (e.g., in terms of other utilities), but less parsimonious representations. 
For example, the strive for parsimony may be one of the factors that has stalled the scientific progress in the foundations of physics during the past decades \cite{hossenfelder2018lost}.

\paragraph{Effective Number of Parameters.}
There are different ways to measure parsimony, with simply counting the number of parameters\footnote{More precisely, we have to count the minimal number of unconstrained parameters that can be invertably transformed to the space of the original model parameters. 
For example, a simplex parameter vector of length $K$ is equivalent to only $K-1$ unconstrained parameters because the $K$-th one is determined by the sum-to-one constraint.} of a P model being the most straightforward approach. 
For simple models, such as linear regression, this measure of parsimony matches the concept of \emph{degrees of freedom} (DoF) in frequentist statistics. 
In the same way, the DoF concept becomes awkward even for slightly more complex models \citep{janson2015}, the former is not a generally useful measure of parsimony either \citep{piantadosi_one_2018}. The reason for this is that, from a Bayesian perspective, any prior information on a parameter increases a P model's parsimony, such that the \emph{effective number of parameters} ($\enp$), might be substantially smaller than the nominal number of parameters \citep{vehtari_2017_loo}. 
The same mechanism also underlies the difficulty in computing the DoF of test statistics in frequentist multilevel models, because random effects distributions are equivalent to priors \citep{hodges2001}.

There are several $\enp$ measures in the literature \citep{spiegelhalter_enp_1998, watanabe_waic_2010, vehtari_2017_loo, piironen_sparsity_2017}, often defined in the context of information criteria. For the information criterion based on leave-one-out cross-validation (LOO-CV), $\enp$ is measured as the sum of the differences between the pointwise log predictive densities of the full posterior and the pointwise log predictive densities of the LOO posteriors \citep{vehtari_2017_loo}:
\begin{align*}
\label{enp-loo}
    \enp_\text{LOO} &=
    \sum_{n=1}^N \left( \log p(y_n \mid y) - \log p(y_n \mid y_{-n}) \right) \\
    &= \sum_{n=1}^N \left( \log \int p(y_n \mid \theta) \, p(\theta \mid y) \, d \theta - \log \int p(y_n \mid \theta) \, p(\theta \mid y_{-n}) \, d \theta \right). \numberthis
\end{align*}
\newcontent{
The notation $y_{-n}$ indicates that the $n$-th data point in $y$ has been excluded. As more parameters are added to the model, the in-sample predictive performance represented by $\log p(y_n \mid y)$ grows more quickly than the out-of-sample predictive performance represented by $\log p(y_n \mid y_{-n})$ such that the sum of their pointwise differences grows. This provides an intuition why $\enp_\text{LOO}$ can be considered a measure of parsimony. Its concrete interpretation as an effective number of parameters is inspired by the following observation: When using very wide or even completely flat priors over all parameters, $\enp_\text{LOO}$ will roughly coincide with the nominal number of parameters, but becomes smaller than the latter in the presence of prior information \cite{vehtari_2017_loo}.
}

Bayesian LOO-CV can usually be computed efficiently via importance sampling without any model refitting, and so can $\enp_\text{LOO}$ be computed without any actual refitting \citep{vehtari_2017_loo, vehtari_pareto_2021}.
For a large number of observations $N$, $\enp_\text{LOO}$ can be asymptotically approximated by the sum of the full posterior variances over the pointwise log-likelihood values, which is the ENP estimate used in the widely applicable information criterion (WAIC) \citep{watanabe_waic_2010}:
\begin{equation}
\label{enp-waic}
      \enp_\text{LOO} \approx \enp_\text{WAIC} = \sum_{n=1}^N \text{Var}_{p(\theta \mid y)} \left[ \log p(y_n \mid \theta) \right]
\end{equation}
\newcontent{Intuitively, as the number of parameters grows, so does the epistemic uncertainty in the posterior, which leads to an increase in the variance of posterior predictive quantities, such as $\log p(y_n \mid \theta)$.}
The WAIC approximation of LOO-CV performance can be quite unreliable so using $\enp_\text{LOO}$ is highly recommended whenever possible \citep{vehtari_2017_loo}.
What becomes apparent in these equations is that parsimony, at least when measured through these $\enp$s, may depend on the specifically realized data $\data$, and as such needs to be defined over PD models. 
This is specifically true for models with hierarchical priors, where the amount of hierarchical shrinkage (i.e., the influence of the hierarchical priors) is data-dependent \citep{gelman_data_2006}. 
Practically, the posterior integrals in \eqref{enp-loo} and \eqref{enp-waic} for PAD models are efficiently approximated via Monte Carlo estimates based on posterior draws from an approximator \citep{vehtari_2017_loo}.

The huge advantage of these ENP measures is that they do not need to be aware of the internal structure of a P model, but only require its predictive outputs in the form of pointwise log-likelihood values. 
However, the need for the latter has the drawback that ENP measures do not work natively with $\implicit$ models due to their lack of tractable likelihoods; unless one has learned not only the model's posterior but also its likelihood density during training \citep{wiqvist2021sequential}. 
What is more, if the model includes residual dependencies between observations, the pointwise (log-)likelihood may not be available, even if the joint likelihood is analytic \citep{burkner2021nfloo}.

\paragraph{Prior P-Parsimony.}
In the above-described ENP definitions, we integrate over the posterior distribution and so, in this sense, measure \emph{posterior parsimony}.
This naturally raises the question of whether we can define measures of \emph{prior parsimony} as well.
In a Bayesian setting, prior parsimony is automatically embodied in the marginal likelihood (sometimes called Bayesian evidence) \citep{kass1995bayes, mackay2003information, lotfi2022bayesian}, which we already encountered in our discussion on prior predictive performance (see Section~\ref{pp-prior-post}).
As a reminder, we obtain the marginal likelihood by marginalizing the joint P model over its prior
\begin{equation}\label{eq:marg_lik}
    p(y) =  \mathbb{E}_{\prior}\left[\lik\right] = \int \lik \, \prior \, d\theta.
\end{equation}
Accordingly, we can interpret the marginal likelihood as the expected probability of generating data $y$ from a P model when we randomly sample from the prior $\prior$.
Through the prior's role as a weight on the likelihood, the marginal likelihood encodes a probabilistic version of Occam’s razor by penalizing the prior complexity of a P model \citep{kass1995bayes, mackay2003information}.

\begin{figure}[t]
    \centering
    \includegraphics[width=0.99\textwidth]{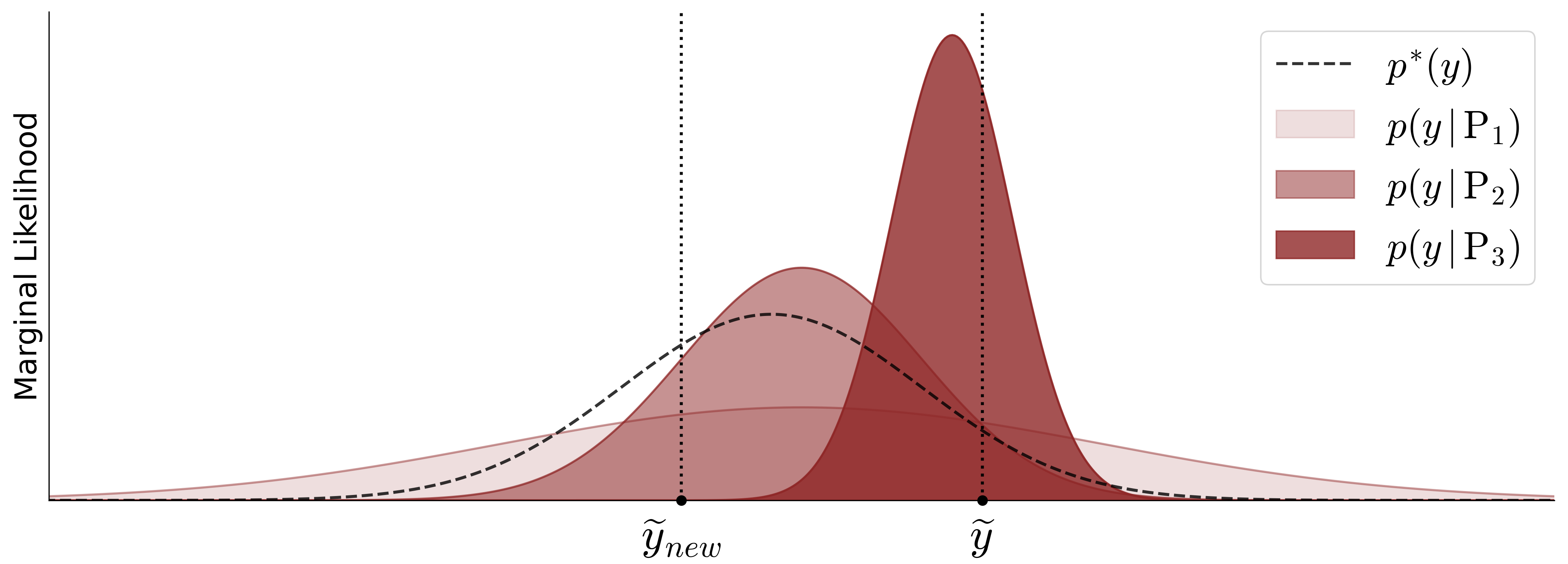}
    \caption{Hypothetical scenario with three P models of descending complexity: $\text{P}_1$, $\text{P}_2$, and $\text{P}_3$. The most complex model $\text{P}_1$ can account for the broadest range of observations at the cost of diminished sharpness of its marginal likelihood; in contrast, the simplest model $\text{P}_3$ has the sharpest marginal likelihood which concentrates onto a narrow range of possible data. Even though the observed data $\data$ is well within the generative scopes of models $\text{P}_1$ and $\text{P}_2$ too, the simplest model $\text{P}_3$ has the highest marginal likelihood at $\tilde{y}$ among the three candidates and is therefore favored from a marginal likelihood perspective. However, the higher relative marginal likelihood of the simplest model $\text{P}_3$ is a poor proxy of its predictive performance for new data sets, as it assigns close to $0$ density to the new data set $\tilde{y}_{\rm new}$, suggestive of overfitting. The model $\text{P}_2$, whose marginal likelihood is closest to the data-generating distribution $p^*$, would have been favored, had $\tilde{y}_{\rm new}$ instead of $\tilde{y}$ been used for computing the associated Bayes factors.}
    \label{fig:ml_evidence}
\end{figure}

However, the marginal likelihood is not an explicit measure of parsimony; rather, it represents an implicit relative quantity which combines prior parsimony with the ability of a P model to fit the data by considering its entire generative scope (see Figure~\ref{fig:ml_evidence}).
Following \cite[][Chapter 28]{mackay2003information}, we can illustrate the above conflation by assuming that the posterior of a P(A)D model is well represented by a (multivariate) Gaussian.
In this case, the marginal likelihood can be approximated as:
\begin{equation}\label{eq:ml_fact}
    p(y) \approx p(y \mid \theta_{\text{MP}}) \times p(\theta_{\text{MP}}) \det (\boldsymbol{H}(\theta_{\text{MP}}) / 2\pi)^{-\frac{1}{2}},
\end{equation}
where $\theta_{\text{MP}}$ is the posterior mode and $\boldsymbol{H}(\theta_{\text{MP}})$ is the Hessian of the likelihood evaluated at $\theta_{\text{MP}}$.
The multiplicand $p(\theta_{\text{MP}}) \det (\boldsymbol{H}(\theta_{\text{MP}}) / 2\pi)^{-\frac{1}{2}}$ is termed an \textit{Occam factor} and represents the factor by which a P(A)D model's parameter space contracts as the prior is updated to the posterior based on the information contained in D.
Thus, under the Gaussian assumption, the magnitude of the Occam factor is an explicit measure of prior complexity (i.e., inverse prior parsimony) related to the information gain a P model can achieve over its generative scope \cite{mackay2003information, lotfi2022bayesian}.
Consequently, a P model with a vague prior will incur a larger penalty by the Occam factor than a different P model with a sharper prior, provided that both models share the same likelihood. 
However, if the Gaussian assumption is inadequate, the approximation of Equation~\eqref{eq:ml_fact} can sustain a large error and may no longer be useful. 
Unfortunately, we are not aware of a more general decomposition of the marginal likelihood into a prediction factor and a parsimony factor, as is the case with $\text{ENP}_{\text{LOO}}$.

A closely related concept is the principle of Minimum Description Length \cite[MDL,][]{rissanen1978modeling, hansen2001model}, which views parsimony through the lens of information theory.
In the MDL framework, a probabilistic model represents a \textit{coding scheme} designed to describe the data $\data$. 
Accordingly, a parsimonious P model provides a concise description of the data in terms of code length (relative to a competing P model).
Note, that MDL is not a unique measure, but rather an umbrella framework for deriving measures of parsimony/complexity in various application contexts (see \cite{hansen2001model} for a comprehensive exposition).
For instance, in a Bayesian context, one can show \cite{hansen2001model} that a canonical measure of description length for model P is given by
\begin{equation}
    \text{DL} = - \log \int \lik\,\prior\,d\theta,
\end{equation}
which we recognize as the negative logarithm of the marginal likelihood introduced in Equation~\eqref{eq:marg_lik}.
In this way, MDL not only highlights the theoretical connection between Bayesian model comparison and information theory but also provides a principled way for deriving new measures of prior parsimony in future basic research.

\newcontent{
\paragraph{Sparsity-inducing priors.} Another perspective on P-parsimony is provided by sparsity-inducing priors, especially global-local shrinkage (GLS) priors \citep{van_erp_shrinkage_2019, bhadra_horseshoe_2020, vanDP2021theoretical}. These priors will shrink redundant coefficients towards values close to zero, inducing sparsity in the posterior\footnote{Shrinkage priors will not shrink coefficients exactly to zero but only close to it. Thus, such coefficients remain in the regression equation but exert a minimal impact on predictions. If desired, exact sparsity can be achieved in a second step via a variable selection procedure \cite{piironen_comparison_2017, catalina_projection_2022, pavone_refmodels_2022}.}. GLS priors can be applied in many model classes, including linear and generalized linear models, non-linear and non-parametric function estimation, time series, as well as deep neural networks \cite{van_erp_shrinkage_2019, ghosh_model_2019, bhadra_horseshoe_2020, schafer_locally_2023}.
Here, we focus our discussion on Gaussian linear models as this case is most intuitive and theoretically best understood.
Given a linear regression model in its simplest form, GLS priors are defined on the $K$ regression coefficients $\beta_k$ as follows:
\begin{equation}
\label{glprior}
      \beta_k \sim \text{Normal}\left( 0, \lambda_k^2 \tau^2\right), \ \ \lambda_k \sim p(\lambda_k), \ \ \tau \sim p(\tau),
\end{equation}
where $\lambda_k$ denotes the local scale parameter unique to each coefficient and $\tau$ denotes the global scale parameter that is shared across all coefficients. The choice of the hyperpriors $p(\lambda_k)$ and $p(\tau)$ determines the specific properties of the GLS prior, leading to, for example, the horseshoe \citep{carvalho_horseshoe_2010, piironen_sparsity_2017} or the R2D2 prior \cite{zhang_R2D2_2020, aguilar_intuitive_2023}; see \cite{van_erp_shrinkage_2019} for a comprehensive overview.

The implied posterior of the coefficients has a highly interesting relationship with the maximum likelihood (ML) estimate $\hat{\beta}_k$ that can be obtained from the same likelihood and data but under the assumption of flat priors on the coefficients. Concretely, and assuming that the ML estimate exists, the posterior mean $\mathbb{E}_{\theta \mid y}(\beta_k)$ can be computed as follows \cite{piironen_sparsity_2017, aguilar_intuitive_2023}:
\begin{equation}
\label{postbmleshrinkage}
\mathbb{E}_{p(\theta \mid y)} [ \beta_k ] = (1- \kappa_k) \hat{\beta}_k,
\end{equation}
with
\begin{equation}
\label{defkappa}
\kappa_k = \frac{1}{1+ a_k \lambda^2_k \tau^2}.
\end{equation}
Here, $a_k$ is some constant that depends on the response's and the $k$-th predictor's scales. Accordingly, the smaller $\lambda_k$ and $\tau$, the stronger the shrinkage of $\beta_k$ to zero, relative to the ML estimate $\hat{\beta}_k$. Conversely, the larger $\lambda_k$ and $\tau$, the closer the posterior mean of $\beta_k$ will be to $\hat{\beta}_k$. Given these properties, $\kappa_k$ are called \textit{shrinkage factors} \cite{piironen_sparsity_2017, aguilar_intuitive_2023}.

The model leading to the ML estimate has $K$ coefficients, which are all counted fully when it comes to determining the number of parameters (see above). Since the posterior mean $\beta_k$ implied by the GLS prior is equal to $(1 - \kappa_k) \hat{\beta}_k$, we see that summing over all $(1 - \kappa_k)$ terms can be considered a measure of the \textit{effective number of coefficients} \cite[ENC,][]{piironen_sparsity_2017}:
\begin{equation}
\label{meffcoefoverall}
\enc_\text{GLS} = \sum_{k=1}^{K} (1-\kappa_k).
\end{equation}
In contrast to the above ENP measures, $\enc_\text{GLS}$ is essentially limited to linear models. What is more, $\enc_\text{GLS}$ only considers regression coefficients, not necessarily all P model parameters (e.g., it ignores the residual standard deviation $\sigma$). These are not the only differences between these measures though. Even though both are derived as generalizations of simply counting parameters, the ENC measures focus on posterior variance (which is explicit in the definition of $\enp_\text{WAIC}$), while $\enc_\text{GLS}$ focuses on the posterior mean. Thus, they consider different aspects of the posterior when measuring parsimony. Studying the relationships between these measures more closely would be an interesting endeavor for future research. 
}



\subsubsection{A-Parsimony}
\label{A-parsimony}

As we discussed in Section~\ref{sec:pa_models} concerning PA models, posterior approximators can range from relatively simple optimization algorithms to high-dimensional parametric models (e.g., neural networks) which themselves can be viewed as standalone P models (e.g., Bayesian neural networks). 
The notion of A-parsimony intends to capture our intuition that these different approximators have varying degrees of complexity.
Here, we propose a very straightforward definition of A-Parsimony: The cardinality of the hyperparameter space $\mathcal{H}$ available for fine-tuning through the implementation interface $\mathcal{I}$ of the underlying mathematical algorithm $\mathcal{A}$.
For instance, the widespread use of MCMC in Bayesian inference is partly because probabilistic programming languages provide relatively simple interfaces, which abstract away a staggering multitude of hyperparameters of complex MCMC samplers \citep[e.g., NUTS,][]{hoffman2014no}).
On the other hand, neural approximators \citep[e.g.,][]{radev_amortized_2020, greenberg2019automatic} inherit the vast hyperparameter spaces of deep neural networks and are thus currently still rather challenging to apply or fine-tune \cite{victoria2021automatic}.

A-parsimony is not only relevant for the usability of approximators, but also plays an important and limiting role in comparison or benchmarking studies assessing the relative performance of different approximators. 
Suppose we wish to compare approximator $\text{A}_1$ having no hyperparameters with approximator $\text{A}_2$ having a single continuous hyperparameter $h \in [0, 1]$, in the context of some P model.  
A comparison of approximators must naturally be based on some metric (or a set of metrics) $q(\text{A}, \text{P})$ which quantifies the approximation quality of A with respect to a given P model (e.g., the distance between corresponding PD and PAD models or the estimation speed of the approximator).
However, even for the simple scenario outlined above, it is not clear how to systematically carry out such a comparison due to the presence of hyperparameters.
One approach would be to approximate the average approximation quality given by $\int_0^1 q(\text{A}_2(h), \text{P})\,p(h)dh$ of $A_2$ and compare it to $q(\text{A}_1, \text{P})$.
Another approach would be to seek the best approximation quality given by $\max_{h \in [0, 1]} q(\text{A}_2(h), \text{P})$ and compare it to that of $q(\text{A}_1, \text{P})$.
Needless to say, the difficulty of ranking and benchmarking approximators with large hyperparameter spaces drastically increases, which makes A-parsimony a key limiting factor as well as a desirable utility to improve upon.

Finally, A-parsimony is related to robustness (see Section \ref{robustness}) and convergence (see Section \ref{convergence}), as the presence of multiple hyperparameters raises the question of how to choose hyperparameter settings which i) lead to stable results and ii) generalize to various applications of a PA(D) model.
For some approximator classes (most notably, MCMC) and P models (e.g., linear models), empirical guidelines and theoretical considerations may suggest relatively robust default choices.
For newer approximator classes (e.g., neural density estimators) or more exotic applications, some form of sensitivity analysis or hyperparameter search might be necessary to ensure sufficient robustness or generalizability.

\subsection{Interpretability}
\label{interpretability}

\emph{Interpretability} of a P(A)(D) model can be qualitatively defined as ``the degree to which a human can understand the cause of a [model-based] decision'' \citep{miller_explanation_2019} or as ``the degree to which a human can consistently predict the model’s result'' \citep{kim_examples_2016}.
A more precise, perhaps even mathematical, definition is difficult to provide given the context and expertise-dependent nature of interpretability, but there is progress in this direction \citep{doshi-velez_interpretable_2017}. 
In any case, achieving interpretability will help us understand \emph{why} a P(A)(D) model behaves the way it does (e.g., in terms of predictive performance; see Section~\ref{predictive-performance}).
Such understanding can have not only profound epistemological, but also far-reaching ethical and social implications \citep{eu_regulation_2016, doshi-velez_interpretable_2017, molnar_interpretable_2020}.

According to \citep{molnar_interpretable_2020}, we can distinguish between intrinsic and post-hoc interpretability.
The former is related to the intelligibility of the P(A)(D) model itself (i.e., its structure and parameters), whereas the latter is related to the \emph{explainability} of the PAD model's results using auxiliary methods, such as permutation feature importance for neural networks \cite{yang2009feature} or random forests \cite{janitza2013auc}.
However, there is a conceptual ambiguity regarding the term in the recent literature.
Some accounts use explainability as a synonym for interpretability in general \citep{molnar_interpretable_2020}, while others use explainability to refer solely to post-hoc interpretability \citep{burkart_survey_2021}.
In our PAD model taxonomy, we view only intrinsic interpretability as a utility of the P(A)(D) model.
Differently, post-hoc interpretability is a utility of an \emph{explanator} that is applied to the original PAD model's results -- in fact, the explanator may just be another, more interpretable P(A)(D) model that is used as a surrogate \citep{burkart_survey_2021}. Accordingly, the following discussion focuses only on intrinsic interpretability, to which we hitherto refer simply as \textit{interpretability}.

P model interpretability relates to the general meaning of its parameters, so it makes sense to differentiate between the interpretability of $\implicit$ and $\explicit$ models since the two model classes often put different demands on the epistemic value of their parameters. 
Further, as we will see below, there are P models whose interpretability can be influenced by both data D and approximator A. 
As such, it can be necessary to further distinguish the interpretability of P, PD, and PAD models.

\subsubsection{Interpretability of $\implicit$ models}

In $\implicit$ models, most parameters correspond to real-world quantities or emergent properties, whose meaning can be understood independently of the $\implicit$ model that is used to estimate them (see Section \ref{sec:p_models}).
For example, in a harmonic oscillator \citep{lurie_mechanics_2002}, the object's mass that serves as a parameter carries a meaning independent of the differential equation that describes the oscillator's behavior.
As such, while the transformations performed to generate data from an $\implicit$ model are highly non-linear and often not analytically tractable \citep{cranmer_sbi_2020}, the interpretability of $\implicit$ models tends to be high (at least in the eyes of domain experts in the field).

However, even for domain experts, it can be exceptionally challenging to predict the generative behavior of a high-dimensional $\implicit$ model given a particular parameter configuration.
This can be the case, even when a P model has a small number of readily interpretable parameters.
Consider, for instance, the prototypical \textit{logistic map} equation \cite{may1976simple} given by
\begin{equation}
    y_{t+1} = \rho \, y_t \, (1 - y_t)
\end{equation}
and having only a single parameter $\rho \in [0, 4]$ which can be interpreted as \textit{growth rate} in population dynamics modeling \cite{storch2017revisiting}.
Despite its beguilingly simple form, the logistic map is known to develop chaotic behavior as the parameter $\rho$ varies in the range from approximately $\rho \approx 3.56995$ to $\rho \approx 3.82843$. 
The model's generative behavior in this range is characterized by a periodic phase, intercepted by bursts of aperiodic fluctuations.
And even though such behavior can be generally abstracted and described for a single parameter, for instance, with the help of bifurcation diagrams \cite{gilmore1995structure}, it can quickly become less amenable to high-level descriptions when it results from the interaction of two or more parameters \cite{barrientos2017chaotic}.
Unsurprisingly, Bayesian analysis of PD or PAD models based on an underlying chaotic $\implicit$ model has long been recognized as a challenging endeavor \cite{berliner1991likelihood}, requiring sophisticated approximators with surrogate likelihoods \cite{springer2021efficient}.

As alluded to above, the interpretability of high-dimensional $\implicit$ models will often depend on whether we focus on individual parameters and their functional role for data generation in isolation (i.e., first-order interpretability) or try to understand interactions between parameters as well as their joint contribution to the generation of $y$ (i.e., higher-order interpretability).
Accordingly, even for complex $\implicit$ models with dozens of parameters, we may still retain relatively high first-order interpretability through the theoretical embedding of each individual parameter, but higher-order interpretability may suffer, since multiple parameters can act similarly on $y$ and interact in surprising ways due to non-linearity.
For instance, the compartmental model of the early COVID-19 pandemics in Germany set up by \cite{radev2021outbreakflow} has $34$ free parameters, each of which has a direct isolated interpretation, for instance, infection rate, number of initially exposed people, weekly modulation, or probability of detection.
However, the exact interplay between these parameters in determining the actual reported number of daily cases might not be immediately obvious from the understanding of individual parameters alone or from the model equations themselves.

Finally, the higher-order interpretability of $\implicit$ models may change once they have been connected to data due to dependencies between parameters.
Oftentimes, we choose a prior $\prior$ which factorizes into independent components, reflecting our assumption of disentanglement or independent generative factors of variation.
However, the resulting PD or PAD models will rarely conserve independence in their joint posteriors (e.g., due to loss of information or an inherent lack of disentanglement in the inverse model).
A canonical example would be a strong posterior correlation between two parameters with initially independent priors, indicating that the parameters do not fulfil orthogonal functional roles for generating the data.

\subsubsection{Interpretability of $\explicit$ models}

In $\explicit$ models, the parameters do not need to correspond to real-world quantities or mechanisms.
Rather, their meaning can often only be understood within the $\explicit$ model they are part of \citep{gelman_prior_2017}.
The archetypal $\explicit$ model is linear regression, where a regression coefficient $\beta$ describes the linear relationship between a predictor variable and the response whilst holding all other predictors constant.
As such, $\beta$ has a clear meaning to an analyst with some statistical knowledge, provided that the measurement scales of predictor and response variables make sense for the task at hand.
However, the requirement to hold all other predictors constant becomes impossible to fulfil if the predictors cannot be varied independently from each other, for example, because they are correlated in purely observational data or because some of them constitute interactions between already included predictors.
As such, even for as few as four or five predictors, interpretability of their joint contribution becomes highly challenging unless predictors are mutually independent \citep{molnar_interpretable_2020}.

The use of non-linear, monotonic transformations in $\explicit$ models, such as link functions in generalized linear models \citep{nelder_glm_1972} or non-linear activation functions in neural networks \citep{sharma_activation_2017} further complicates the interpretability of an originally linear predictor structure. 
For example, when using the logarithmic link (equivalently, the exponential response/activation function), the originally additive relationships become multiplicative, resulting in exponential growth, which is much harder to comprehend for humans \citep{wagenaar_misperception_1975}.
This then reduces the interpretability of the $\explicit$ model's parameters from both their signs and magnitudes to only their signs.
If one were to apply non-monotonic transformations, the interpretability of the parameters' signs would be lost as well.
In addition to non-linear transformations of the whole linear predictor term, every structural deviation from a (latent) linear structure further reduces interpretability. 
For example, interactions, polynomial terms, hierarchical structure \citep{gelman_data_2006}, Gaussian processes \citep{williams_gps_1996, rasmussen_gps_2003}, or splines \citep{friedman_splines_1991,wood_thin_2003} all make interpretation of a $\explicit$ model's parameters harder, if not impossible in some cases.

The interpretability of a $\explicit$ model may also be affected by the data utilized for parameter estimation.
Accordingly, PD models may differ in their interpretability even if their underlying $\explicit$ model is the same.
For example, when employing shrinkage priors for high-dimensional linear $\explicit$ models with lots of irrelevant predictors, the posterior of most regression coefficients will shrink to values very close to zero, effectively eliminating the corresponding predictors from the regression equation \citep{piironen_sparsity_2017, zhang_R2D2_2020} (see also Section~\ref{P-parsimony}).
If only a few coefficients are substantially different from zero, the interpretability of the resulting PD model would be much higher than that of the original $\explicit$ model.

Finally, an approximator A may create a situation where a PA(D) model's interpretability deviates from that of the underlying P(D) model. 
However, that may only happen if the posterior approximation $\posta$ is qualitatively different from its analytic counterpart $\post$ due to an incomplete posterior exploration.
A common case arises when the analytic posterior is multi-modal but the approximator collapses to a single mode \citep{garipov_loss_2018}.
Notably, mode collapse represents a case where the interpretability of the PAD model may be higher than that of the underlying PD model, at the cost of other utilities, such as predictive performance (see Section \ref{predictive-performance}).
An example of an $\explicit$ model class that produces highly multi-modal posteriors are artificial neural networks \citep{garipov_loss_2018, draxler_essentially_2018, izmailov_BNN_2021}.
While the interpretability of the underlying $\explicit$ model is usually low \citep{burkart_survey_2021, zhang_visual_2018, zhang_interpretable_2018}, some of their PAD models can exhibit much higher interpretability if they are steered in the right direction \citep{zhang_visual_2018, zhang_interpretable_2018}.

\newcontent{The above-described notions of interpretability are largely qualitative. While some attempts at a quantitative treatment have been made \citep{doshi-velez_interpretable_2017}, we are not aware of any sufficiently general definition that allows for a more objective, quantitative comparison between P(AD) models with respect to their interpretability. Thus, we hope that our PAD model taxonomy may inspire more focused research on the quantification of interpretability.}



\subsection{Convergence}
\label{convergence}

%
\newcontent{
\emph{Convergence} is a utility of PA and PAD models which rely on complex approximators, such as MCMC, variational inference, or neural density estimators. As explained in Section \ref{sec:pa_models}, approximators provide certain guarantees under specific assumptions, such as infinite draws in the case of MCMC \citep{BDA3} or infinite training and representational capacity in the case of neural density estimators \citep{radev_amortized_2020, radev_amortized_2021, schmitt_bayesflow_2023}. In practice, however, modelers cannot wait a lifetime of infinity for approximators' promises to come true; for the time being, we need to work with finite posterior draws and non-convex optimization objectives teeming with local optima. 

Thus, our convergence utility pertains to the relative distance between a particular (finitely instantiated) PA(D) model and the optimal PA(D) model attainable under perfect conditions for A. 
%
For the above definition to be useful, we need a proxy measure of how close the current approximation is to the optimal approximator outcome. We call such measures \textit{convergence diagnostics} and they are indispensable for ascertaining the validity of PA(D) models. Ideally, good convergence diagnostics should also indicate that the approximation is close to the analytic posterior, but only within the space of distributions the approximator can reach. Accordingly, the relation between convergence and analytic posterior approximation is only indirect for approximators that may be asymptotically biased \citep{yaoYesDidIt2018, dhaka_vi-sgd_2020}. Below, we briefly detail common convergence diagnostics for different types of approximators.
}

\subsubsection{Convergence Diagnostics for Markov Chain Monte Carlo}
\label{convergence-MCMC}

\begin{figure}
\centering
\includegraphics[width=0.99\textwidth]{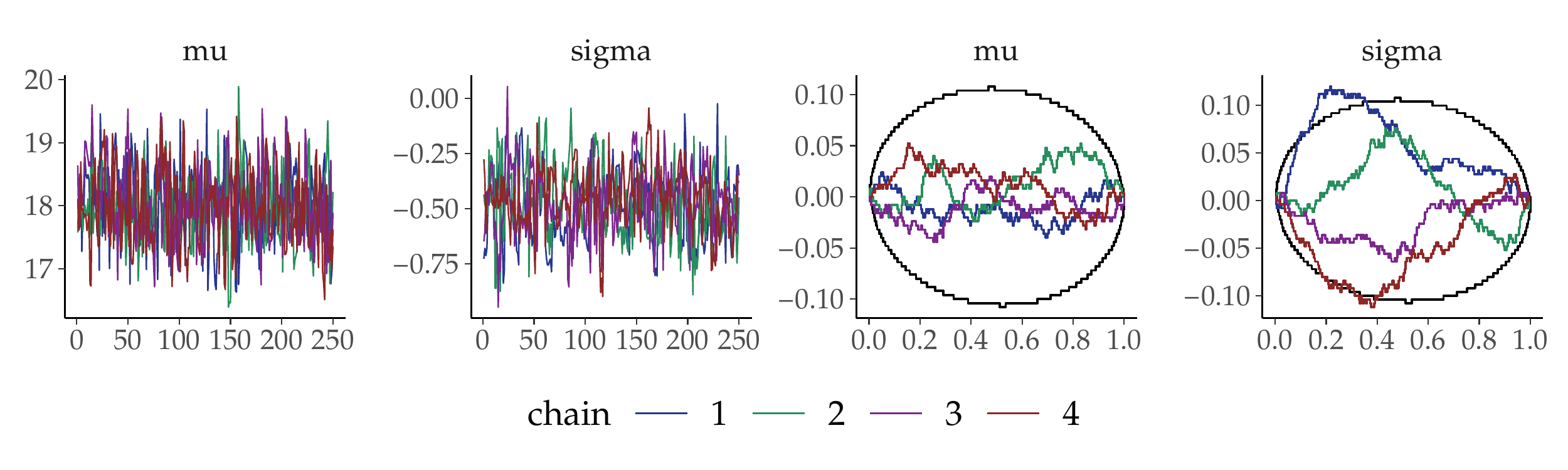}
\caption{Graphical convergence checks of two parameters $\mu$ and $\sigma$. Left: Traditional trace plots. Right: ECDF difference plots with 99\%-confidence envelopes \citep{sailynoja_graphical_2022}. Both kinds of plots use the same posterior draws, but only the rightmost ECDF difference plot highlights that Chains 1 and 4 have some mixing problems for $\sigma$. Example draws obtained from the \texttt{bayesplot} R package \cite{gabry_visualization_2019}.} 
\label{fig:convergence-plots}
\end{figure}

Convergence diagnostics are fundamentally important for posterior approximators that rely on MCMC since these approximators can be arbitrarily bad before full convergence \citep{lambert_Rstar_2022}.
Thus, all model-based inference relies on the quality of the approximation being close enough to the analytic posterior with respect to some minimally required precision. For a quick graphical check, trace plots or ECDF difference plots \citep{sailynoja_graphical_2022} can be used, as illustrated in Figure~\ref{fig:convergence-plots}.

In terms of numerical approaches, three related classes of MCMC convergence diagnostics are applied in today's practice, namely scale reduction factor $\widehat{R}$, effective sample size (ESS) and Monte Carlo standard error (MCSE) \citep{gelman_rhat_1992, cowles_mcmc_1996, robert_mc_1999, flegalMarkovChainMonte2008,  BDA3, dossMarkovChainMonte2014, vehtari_rhat_2021}.
They all provide convergence measures for univariate quantities of interest $\psi = \psi(\theta)$ that are functions of the P model's parameters $\theta$ (see also Section \ref{parameter-recoverability}).
There is not a single ``global'' $\widehat{R}$, ESS, or MCSE measure for $\psi$, but one for each summary statistic $T(\psi)$ of $\psi$, where $T$ can be any posterior expectation or quantile \citep{vehtari_rhat_2021}. 
As such, for example, a set of $S$ posterior draws $\psi^{(s)}$ might yield a very precise estimate for the posterior mean of $\psi$, while at the same time, the estimates of some tail quantiles of $\psi$ (e.g., $5\%$ and $95\%$ quantiles) have much less precision \citep{vehtari_rhat_2021}. 
Accordingly, each of these convergence measures is a function of the quantity of interest $\psi$ and the summary statistic $T$, computed from the $S$ posterior draws $\psi^{(s)}$.

Broadly speaking, the scale reduction factor $\widehat{R}$ compares the between-chain variance $B = B(f_T(\psi))$ to the within-chain variance $W = W(f_T(\psi))$:
\begin{equation}
   \widehat{R}_T(\psi) := \sqrt{\frac{B(f_T(\psi)) + W(f_T(\psi))}{W(f_T(\psi))}},
\end{equation}
where the dependence of $B$ and $W$ on $T$ is realized by an appropriate transformation $f_T(\psi)$ that is applied to each posterior draw $\psi^{(s)}$ before the variances are computed, usually on split chains \citep{BDA3, vehtari_rhat_2021, posterior}. We can conclude that convergence has been reached if $\widehat{R} \approx 1$, that is, if the within-chain variance dominates the between-chain variance.

The ESS estimates the number of independent draws that contain the same amount of information about $T(\psi)$ as the $S$ dependent posterior draws obtained via an MCMC approximator.
As a result, we usually see $\text{ESS} < S$, although the opposite can also happen in case of antithetic (negatively auto-correlated) chains \citep{vehtari_rhat_2021}. We can obtain the ESS from all autocorrelations $\rho_t = \rho_{t}(f_T(\psi))$ of lag $t$ of the chains as
\begin{equation}
  \text{ESS}_T(\psi) := \frac{S}{1 + 2 \sum_{t=1}^\infty \rho_{t}(f_T(\psi))},
\end{equation}
where, in practice, we would truncate the infinite sum at some finite value \citep{geyer_ess_1992}. In modern versions of ESS, $\rho_t$ implicitly depends also on $\widehat{R}$ to take variation across chains into account \citep{BDA3, vehtari_rhat_2021}. In case of independent draws, we have $p_t = 0$ such that $\text{ESS} = S$.

The MCSE describes how much (reducible) uncertainty in $T(\psi)$ remains due to the fact that we only have a finite set of dependent MCMC draws for estimation \citep{flegalMarkovChainMonte2008, dossMarkovChainMonte2014, BDA3, vehtari_rhat_2021}.
If $T$ represents an expectation, we can write down the corresponding MCSE schematically as an overall variance $V = V(f_T(\psi))$ across the $S$ draws divided by the corresponding ESS \citep{flegalMarkovChainMonte2008, vehtari_rhat_2021}:
\begin{equation}
   \text{MCSE}_T(\psi) := \sqrt{\frac{V(f_T(\psi))}{\text{ESS}_T(\psi)}}.
\end{equation}
MCSE estimates for quantiles need to be computed a little differently and are provided in \citep{vehtari_rhat_2021}.

Ideally, we should define convergence of MCMC as reaching or undercutting the maximal MCSE that we find minimally acceptable for the given summary of interest $T(\psi)$. However, The MCSE is scale-dependent as it has the same scale as $T(\psi)$, which requires an understanding of how much of an error is acceptable for a certain quantity, in the context of a particular model and research question. This inherently makes MCSE harder to use in practice and hence the scale-free alternatives $\widehat{R}$ and ESS are often preferred \citep{vehtari_rhat_2021}.

All of the above measures are univariate in the sense that they only concern a univariate $T$ applied to a univariate $\psi$. 
Recently, a more comprehensive measure, called $R^*$ \citep{lambert_Rstar_2022}, has been developed that measures convergence in a multivariate way across multiple model parameters or quantities of interest. 
It is able to detect non-convergence in the joint posterior that may be overlooked by only investigating convergence of a small, non-exhaustive set of univariate quantities \citep{lambert_Rstar_2022}. 
This is achieved by training an expressive machine learning model (i.e., random forest) to predict chain indices from posterior draws.
If the predictive performance of the machine learning model on (unseen) test draws does not exceed chance level, we can assume that the MCMC chains have converged.

In addition to all these sampler-agnostic convergence metrics, there are also few sampler-specific metrics. 
Most notably, this concerns \emph{divergent transitions} in Hamiltonian Monte-Carlo (HMC) \citep{betancourt_hmc_2017}, where every occurring divergent transition in the Markov chain may bias the MCMC results and indicate difficulties of the sampler with exploring the target posterior. 
Divergent transitions tend to occur in regions of high curvature of the explored posterior; regions that most other MCMC samplers struggle to explore as well, only that they fail more silently compared to HMC \citep{betancourt_hmc_2017}.

\subsubsection{Convergence Diagnostics for Optimization-Based Algorithms}
\label{convergence-optimization}

Many classes of posterior approximators are based on optimization algorithms. 
The simplest of such approximators aim to find a single point estimate to approximate the analytic posterior, namely the posterior mode, also known as maximum a posteriori (MAP) estimate \citep{BDA3, mackay2003information}. 
Variational inference (VI) approximators also use optimization, but instead of finding the MAP, they aim to find a parametric distribution (e.g., a multivariate Gaussian) that approximates the analytic posterior as closely as possible \citep{fox_vbi_2012, ranganath_bbvi_2014, blei_vi_2017, welandawe_vbi_2022}. 
The optimization then targets the parameters of this parametric distribution (e.g., the means and standard deviations in Gaussian mean-field VI). 
Expectation propagation \citep[EP,][]{opper_ep_2000, minka_ep_2013, vehtari_ep_2020} and integrated Laplace approximation \citep[INLA,][]{rue_inla_2009, lindgren_inla_2015, rue_inla_2017} work in a conceptually similar fashion, but the structure of their parametric approximators and their target distributions are different (e.g., for INLA, the conditional posteriors of the parameters, instead of their joint posterior). 
Again, highly similar in terms of their use of optimization, neural approximators (e.g., invertible neural networks; \citep{ardizzone_inn_2018, radev_amortized_2020}), use optimization to find the neural network parameters that yield the best posterior approximation within the generative scope of the network \citep{papamakarios_nde_2017, lueckmann_nde_2017, greenberg2019automatic, papamakarios2021normalizing, radev_amortized_2020, schmitt_bayesflow_2023} (but see Section~\ref{convergence-amortized} for specifics in diagnosing convergence of amortized neural approximators).

Regardless of how optimization is applied for posterior approximation, the aim is always to find a single point in a potentially high dimensional space that leads to the best approximation of the analytic posterior within the set of realizable approximations. 
Accordingly, all traditional convergence criteria for iterative point optimization apply. 
That is, for non-stochastic optimization algorithms (e.g., gradient-decent or L-BFGS; \citep{nocedal1999numerical}), small absolute or relative changes in the point estimate, small absolute or relative changes in the target function, or small absolute or relative closeness of the target function's gradient to zero (if the gradient is available) \citep{nocedal1999numerical, stan_2022}, would indicate convergence.
For stochastic optimization algorithms (e.g., stochastic gradient-decent or more sophisticated versions, such as Adam; \citep{nocedal1999numerical, kingma_adam_2017}), measuring convergence becomes less straightforward due to the stochasticity in the objective's trajectories. 
If the step size is held constant, they yield a Markov chain around the target point, once the algorithm comes close enough, instead of converging directly to the target \citep{raginsky_sgd_2017, erdogdu_sgd_2018}. 
The latter implies that MCMC convergence diagnostics, in particular $\widehat{R}$, can be applied to diagnose convergence of stochastic optimization algorithms \citep{dhaka_vi-sgd_2020}.

\subsubsection{Convergence Diagnostics for Sequential Monte Carlo}

Sequential Monte Carlo (SMC; aka particle filtering) comprises a heterogeneous class of posterior approximators for PD models whose underlying P models can be expressed in the form of a sequence of conditional distributions (i.e., time series P models) \citep{doucet_smc_2001, del_moral_smc_2006}. Most SMC samplers can be shown to provide asymptotically correct inference as the number of draws (particles) approaches infinity \citep{dai_smc_2020}. 
However, empirical convergence diagnostics in the pre-asymptotic regime appear to be relatively scarce still \citep{cusumano_smc_2017, lee_smc_2018, dai_smc_2020}. 
Perhaps this is because SMC approximators consist of multiple iteratively applied components \citep{dai_smc_2020}, each with their own pre-asymptotic behavior requiring their own local convergence diagnostics: To assess the convergence of importance sampled (IS) particles at a given step, ESS estimates for weighted samples \citep{kong_smc_1994, zhou_smc_2016} or variance measures driven by the number of siblings per particle (i.e., the number of particles with the same ancestor node at step zero) \citep{lee_smc_2018} can be applied. 
The trustworthiness of the IS weights themselves could be diagnosed via the Pareto-$k$-diagnostic of Pareto-smoothed importance sampling (PSIS; \citep{vehtari_pareto_2021, burkner_lfo_2020}), although we are not aware this has been tried so far in the context of SMC (for a closely related application, see \citep{burkner_lfo_2020}). 
Convergence of MCMC kernels that are part of many SMC algorithms \citep{dai_smc_2020} could be assessed via MCMC convergence diagnostics (see Section~\ref{convergence-MCMC}). 
While each of these diagnostics may be locally informative for a given SMC component at a given step, whether and how they convey global convergence to the target joint posterior remains to be studied further.

\subsubsection{Convergence Diagnostics for Approximate Bayesian Computation}

The standard ABC rejection algorithm \cite{rubin_bayesian_1984, diggle1984monte, tavare1997inferring, pritchard1999population} requires a distance function which quantifies the difference between simulated data $y$ (generated from a P model with a particular parameter configuration $\theta$) and observed data $\data$.
Further, it needs a tunable tolerance level $\epsilon$ according to which the algorithm rejects a fraction of $1 - \epsilon$ simulated parameter values.
The algorithm then keeps the remaining parameter values as random draws from an approximate posterior $p_{\epsilon}(\theta \mid y)$.

The ESS of standard ABC rejection samplers is thus typically equal to $(1 - \epsilon)\,S$, with $S$ denoting the total simulation budget, since vanilla ABC samplers perform \textit{independent} sampling.
However, this does not mean that their sampling efficiency is particularly appealing, especially for high-dimensional P models.
That is because ABC samplers notoriously suffer from the curse of dimensionality: Most simulated data sets from a high-dimensional P model will be rejected and so it becomes challenging to obtain enough random draws from $p_{\epsilon}(\theta \mid y)$ for a reasonable reduction of the MCSE.

More sophisticated ABC algorithms, such as ABC-SMC \citep{sisson2007sequential, klinger2018pyabc} or ABC-MCMC \citep{marjoram2003markov, turner2014generalized} alleviate some of these issues and inherit the convergence diagnostics of SMC and MCMC.
However, whenever hand-crafting of distance functions and summary statistics of the data $H(\data)$ (i.e., dimensionality reduction) is involved, ABC algorithms can converge \textit{at best} to $p_A(\theta \mid H(\data))$.
This issue does not exist whenever $H(\data)$ is a sufficient summary statistic, but it can potentially lead to overestimation of $V(f_T(\psi))$ if $H(\data)$ results in considerable loss of information about the parameters $\theta$.

Recent work on ABC focuses on building robust ABC approximators and exploring the possibility of utilizing hand-crafted summary statistics as a key element of misspecification analysis and error correction \cite{frazier2021robust, martin2021approximating}.
A related line of work suggests comparing posterior moments recovered by differently configured ABC approximators as an empirical diagnostic \cite{frazier2020model}.
It remains an interesting open question whether similar ``ensemble approaches'' can be generalized to other approximators for simulation-based inference, such as amortized neural surrogates \citep{radev_amortized_2020, greenberg2019automatic, papamakarios2019sequential} or ABC with learned summary statistics \citep{chen2020neural, jiang2017learning}.

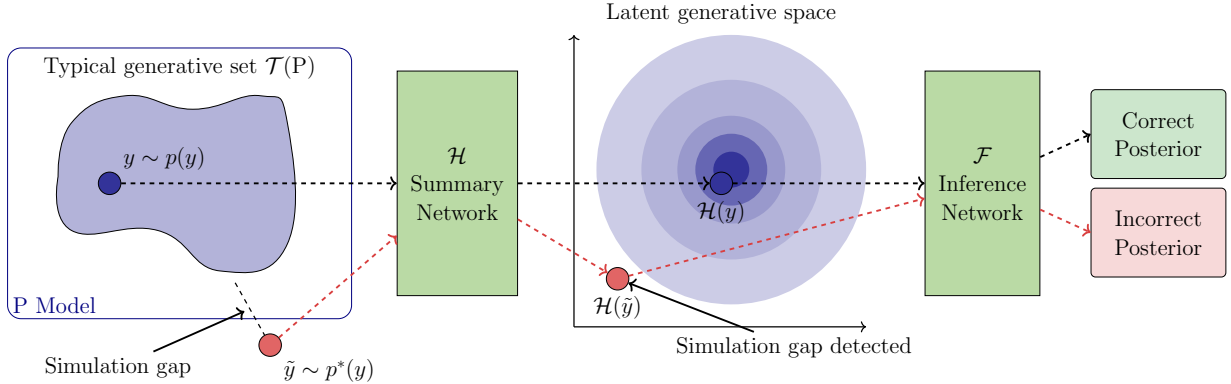
\begin{figure}
    \centering
    \begin{adjustbox}{width=0.99\textwidth}
    \input{graphics/simulation_gaps.tex}
    \end{adjustbox}
    \caption{Detecting model misspecification (i.e., simulation gaps) with amortized neural approximators \cite{schmitt_bayesflow_2023}. A summary (aka embedding) network $\mathcal{H}$ transforms the typical generative set $\mathcal{T}{(\text{P})}$ of a $\text{P}$ model (i.e., the finite set of ``in-distribution'' data simulations that a P model typically generates) into the typical set of a simple distribution (e.g., multivariate Gaussian). Discrepancies between the model-implied data distribution $p(y)$ and the true data distribution $p^*(y)$ manifest themselves as detectable anomalies, causing potential posterior errors by the inference network $\mathcal{F}$. We can detect these anomalies via standard our-of-distribution (OOD) detection techniques.}
    \label{fig:mms}
\end{figure}

\subsubsection{Convergence Diagnostics for Amortized Approximators}
\label{convergence-amortized}

In contrast to MCMC-based approximators, there are no standard convergence diagnostics for amortized approximators yet, as the latter are grounded in diverse, and still fast-evolving theoretical frameworks.
However, whenever we employ neural posterior approximators, we can resort to convergence checks used commonly in deep learning applications \cite{goodfellow2016deep}; see also Section~\ref{convergence-optimization}.

Since most modern neural architectures are trained using some form of stochastic gradient-based optimization, optional stopping (i.e., discontinuation of training once the cost function does not improve over some tolerance period) and other convergence heuristics can be used to determine when a neural approximator has reached a stable local minimum of its cost function.
That said, due to the nature of non-convex optimization, simply assessing local convergence is not enough for trusting PA(D) models coupled with amortized neural approximators.

Moreover, amortized neural approximators require simulations to be faithful proxies of reality and might yield arbitrarily bad posterior approximations when confronted with data that are atypical under the assumed P model \cite{schmitt_bayesflow_2023}.
The latter case is also known as a simulation gap and it occurs when a P model does not accurately represent the behavior of the modeled real-world system (i.e., when the model is misspecified).
Consequently, amortized approximators must be able to detect simulation gaps and potential posterior errors, so that they can warn users about suspicious input data and resulting inference.

Generally, there are two broad types of empirical convergence diagnostics we need to utilize in the context of amortized neural approximators: those of a PA model and those of a PAD model.
Model-agnostic tools, such as different variants of SBC (see Section~\ref{calibration}), are only applicable to PA models, as they assume that we have access to the actual data-generating parameters.
On the other hand, if the neural approximator is a generative neural network \citep{radev2020bayesflow, greenberg2019automatic}, the latent space can be used as a source of convergence information for the PA model.
For instance, flow-based networks \citep{kingma2018glow, papamakarios2021normalizing} are trained to transform an intractable posterior into a simple base distribution from which random draws can be easily obtained.
Thus, convergence of the PA model can generally be determined by a divergence between the prescribed and the learned base distribution.

Unfortunately, neither of the above PA diagnostics can tell us whether the corresponding PAD model will be able to yield faithful estimation due to potential discrepancies between the simulation model and reality (see above).
Thus, further diagnostics are necessary to promote the trustworthiness of amortized posteriors.
One such diagnostic is the maximum mean discrepancy \citep[MMD,][]{mmd} between summary statistics of simulated and real data which tells us whether the observed data belongs to the typical generative set of the simulator or not \cite[][cf. Figure~\ref{fig:mms}]{schmitt_bayesflow_2023}.
As the field of simulation-based inference is still in its infancy, we expect a rapid development of convergence diagnostics for amortized approximators in the future.

\subsection{Estimation Speed}
\label{speed}

For non-amortized approximators (cf. Figure~\ref{fig:non-amortized}), we can define \emph{Estimation Speed} as the time from the start of running the approximator A of a PAD model (or a particular instance of a PA model) until convergence, defined by approximator-specific convergence diagnostics (see Section \ref{convergence}). In certain cases (see Section~\ref{hierarchies_tradeoffs}), it may be sensible to define estimation speed less strictly as the time until termination of the approximator run after which useful results are obtained, without necessarily having achieved convergence. For a given PAD model, estimation speed tends to vary by several orders of magnitude across different classes of approximators. For example, MAP estimators, VI, or other (non-amortized) optimization-based approximators, will usually require a fraction of what sampling-based approximators such as MCMC or SMC need \citep{rue_inla_2009, brms1}.
When considering estimation speed in isolation, there is no doubt that ``faster is better''.
However, to obtain faster approximators, we often have to give up accuracy or asymptotic guarantees of the resulting posterior approximation \citep{fox_vbi_2012, yaoYesDidIt2018}.
Thus, increasing speed by changing the approximator class may have an adverse effect on other utilities of PA(D) models, specifically on parameter recoverability and predictive performance (see Sections \ref{parameter-recoverability} and \ref{predictive-performance}).

Within a given class of approximators, hyperparameter choices can greatly affect the estimation speed as well \citep{hoffman2014no, yaoYesDidIt2018, radev_amortized_2020}.
As an example, consider static HMC where the number of leapfrog steps per Markov transition has to be chosen \textit{a priori} \citep{hoffman2014no, betancourt_hmc_2017}.
On the one hand, if the number of leapfrog steps is too small, MCMC draws will be highly auto-correlated and thus more draws are required to achieve convergence. On the other hand, if the number of leapfrog steps is too large, a lot of computation time is wasted by unnecessary leapfrog steps; or auto-correlation might even get worse again when the HMC sampler eventually makes a so-called ``U-turn'' to come back to its starting point \citep{hoffman2014no}.
Such problems due to hyperparameter choice can be mitigated by automatically tuning hyperparameters in a ``warm-up'' phase or adapting them on the fly, conditional on the local geometry of the approximated analytic posterior.
For example, when using the No-U-turn sampler \citep[NUTS,][]{hoffman2014no}, a generalization of HMC, the number of leapfrog steps is adaptive.
It removes the requirement for the user to choose this hyperparameter manually and may even have better estimation speed than optimally tuned, static HMC \citep{hoffman2014no}.

The above definition of estimation speed is straightforward but can be misleading in practice if convergence is not achieved (or unachievable) within a given run of A until its termination, such that A has to be restarted \citep{gelman2020workflow}.
A typical reason is sub-optimal choices of A's hyperparameters, for example, if the leapfrog step size in HMC is too large leading to divergent transitions whose occurrence implies irrecoverable non-convergence for the current approximator run \cite{betancourt_hmc_2017}.
Thus, estimation speed in practice may be strongly affected by an approximator's ability to run reliably out of the box without much tuning.
Tuning demand can be reduced by adapting hyperparameters automatically on the fly or by having only a small number of sensitive hyperparameters (see Section~\ref{robustness}).
The latter property can further be understood as determining approximator parsimony (see Section~\ref{A-parsimony}). 

A manually but skilfully tuned approximator A$_1$ might beat an auto-tuned approximator A$_2$ in terms of estimation speed when considering only the final, converging run. 
However, the overall time (including failed runs) it can take to get A$_1$ to this optimal state may easily more than offset its final speed advantage. 
As a result, in an honest comparison of practical estimation time, it may be A$_2$ that comes out ahead by a substantial margin. 
Along similar lines, the particular P model implementation may be more or less favourable for different approximators, which can also strongly affect estimation times \citep{betancourt_hmc_2017, vehtari_rhat_2021, beraha_mcmc-comp_2021, devalpine_nimble_2021}.

Sometimes, the reason for an approximator's termination without convergence may also lie in the computational environment, for example, time or memory constraints on a computing cluster that limit the resources available for a single job. For example, if the estimation of a PA(D) model takes longer than expected, estimation might be terminated prematurely, in the worst case leaving no intermediate result to restart from. In this scenario, even if the second run would then be successful, we still had to deal with at least a doubled estimation time. Accordingly, both the predictability of the expected resource requirements and the small variance in resources between repeated runs of the same PA(D) model can imply substantial practical speed improvements.

\subsubsection{Sampling Efficiency}
\label{sampling-efficiency}

For sampling-based approximators, convergence in terms of reaching a given MCSE value (for given quantities of interest and summary statistics), is strongly application-dependent, and so is the estimation speed associated with it (see also Section~\ref{convergence-MCMC}).
For more general comparisons of sampling-based approximators, the concept of \emph{sampling efficiency} is easier to handle and we define it as the average ESS per unit time (for a given quantity of interest $\psi$ and summary statistic $T$):
\begin{equation}
    \textrm{Eff}_T(\psi) = \frac{\textrm{ESS}_T(\psi) }{t_{\rm end} - t_{\rm start}},
\end{equation}
where $t_{\rm start}$ and $t_{\rm end}$ are the start and end times of the approximator run, respectively.
While most approximators, even optimization-based ones, \emph{can} be used to obtain posterior draws upon convergence \citep{rue_inla_2009, fox_vbi_2012}, we restrict the class of sampling-based approximators to those that return only posterior draws as their immediate endpoint instead of the parameters of (closed-form) density functions.
MCMC, SMC, and rejection sampling are the most important members of this class \citep{rubin_bayesian_1984, BDA3, dai_smc_2020}.

Within a class of sampling-based approximators, say MCMC, the same convergence diagnostics, in particular the same ESS, can be applied to all competitors, which simplifies comparisons \citep{beraha_mcmc-comp_2021, devalpine_nimble_2021}. Here it is important to not only use the same \emph{implementation} for these diagnostics across all approximators, but also to ensure that this implementation follows the current state-of-the-art of diagnostic development \citep{vehtari_ess_2021}. Otherwise, comparisons may be biased by outdated diagnostics.
Additionally, one needs to verify empirically that the obtained stationary distribution for a given PAD model is the same for all the compared approximators. Otherwise, sampling efficiency will be misleading since at least one approximator would not have estimated the analytic posterior of the underlying PD model well enough.
Comparisons between approximators belonging to different sampling-based classes may require even more care to ensure that ESS diagnostics across classes are comparable, for example, when comparing MCMC with SMC approximators.

Whenever we are performing sampling efficiency comparisons for PA instead of PAD models, we not only have, in principle, an infinite number of data sets as a basis for comparison but can utilize SBC to falsify the correctness of the achieved stationary distributions (see Section \ref{calibration}). 
However, when we investigate sampling efficiency on a set of simulated data sets, different data-generation scenarios should be studied, since well-specified P models may have different efficiency than misspecified P models.

When studying sampling efficiency, the same practical caveats apply as for estimation speed in general. 
For instance, to achieve a comparison that is ecologically valid for real-world situations, we have to investigate practical sampling efficiency that considers both optimized and non-optimized P model implementations, as well as failed or prematurely terminated approximator runs.



\subsubsection{Estimation Speed of Amortized Approximators}
\label{speed-amortized}

Amortized approximators, such as the pre-paid estimation method \cite{mestdagh2019prepaid} or neural density estimation methods \cite{radev2020bayesflow, greenberg2019automatic}, require a slightly modified view on estimation speed, since they tend to split inference into two phases (cf. Figure \ref{fig:amortized}).
Convergence in the context of amortized neural approximators typically happens before any posterior draws have been obtained \citep{von2022mental, gonccalves2020training}, so the primary computational load falls into the upfront simulation-based training phase.
In contrast, the computational cost of applying a pre-trained amortized approximator to obtain thousands of posterior draws or perform density estimation on real data is typically negligible and only a matter of seconds even for high-dimensional posteriors \cite{radev2020bayesflow}.
In this way, amortized approximators can be extremely useful for studying the (global) information gain (see Section \ref{information_gain}) or calibration (see Section \ref{calibration}) as part of the parameter recoverability utility of PA models, since these demand inference on many, potentially thousands of data sets simulated from the underlying P model.

Due to the properties of amortized approximators, we can modify the definition of estimation speed as the time until convergence of the training phase plus the time for obtaining a sufficient number of posterior draws on real data to reduce the MCSE beyond a pre-defined threshold.
In this context, estimation speed will greatly depend on the \textit{simulation time}, that is, the computational cost of performing a sufficient number of model simulations.
Some amortized methods make it possible to further subdivide estimation speed into three parts: simulation time, training time, and inference time\footnote{Other amortized approximators, such as the pre-paid method \cite{mestdagh2019prepaid}, only entail a simulation phase and an inference phase.}, for instance, when using BayesFlow in an offline training regime \cite{radev_amortized_2020} or when applying sequential neural estimators \cite{greenberg2019automatic} with the prior as a sole proposal distribution throughout training.

In any case, estimation speed for amortized approximators will be dominated by the time spent before obtaining posterior draws. Accordingly, it is often important to determine the break-even point between an amortized and a non-amortized method, that is, after how many observed data sets does the training effort amortize in terms of ESS per unit time?
Naturally, this break-even point will heavily depend on the modeling context.
For some P models, the break-even point between neural estimation and ABC can occur after as few as $5$ or even fewer observed data sets \cite{radev_amortized_2021}, but it can also occur only after as many as dozens when comparing different neural approximators \cite{radev2020bayesflow}.
In addition, amortized neural samplers can often yield independent posterior draws upon convergence \citep{radev2020bayesflow, gonccalves2020training, greenberg2019automatic}, so their sampling efficiency during the inference phase (cf. Figure \ref{fig:amortized}) will be generally superior to non-amortized approximators (i.e., stateful samplers) yielding dependent draws.

Importantly, comparisons between amortized and non-amortized approximators (but also comparisons within the same A class) should take implementation factors into account.
For instance, the estimation speed of neural approximators will be greatly enhanced by using GPU parallelization and even standard ABC rejection samplers can be quite efficient when run on a computing cluster with hundreds of nodes \cite{klinger2018pyabc}.
For simulation-based inference, the implementation of the simulation model presents a further potential bottleneck, which can be alleviated via parallelization, model reformulation reducing the algorithmic complexity of the simulator, or calibration of large-scale simulators via simpler surrogates \cite{lunderman2020estimating}.
In addition, recent hybrid methods employ a mixture of amortized and non-amortized components, such as amortized likelihood ratio approximators within non-amortized MCMC \cite{hermans2020likelihood} or neural likelihood surrogates \cite{papamakarios2019sequential}.
These hybrid methods blur the distinction between amortized and non-amortized methods and render the definition of estimation speed even more challenging.
The dependence on these various implementation factors should make us wary of comparisons between approximators in terms of estimation speed and appreciate the challenges of building scalable PA(D) models.

\subsection{Robustness}
\label{robustness}

A common question that arises when we discuss substantive conclusions derived from model-based inference is how fragile these conclusions are with respect to crucial aspects of P, A, or D.
Can we ``break'' the analysis by a barely perceptible change in the data or by using a slightly different approximator? 
Or are the main results of the analysis largely impervious to such seemingly unsubstantial changes?
The \emph{Robustness} utility attempts to answer such questions by measuring how much a PA, PD, or PAD model's implications change as we (systematically) perturb some of its components.

In the above definition, we use the term ``component'' very generally.
It can refer to (structural) aspects of the P model, most notably, to priors or their hyperparameters \citep{depaoli_sens_2020, kallioinen_2021_sens, perez_sens_2006, berger_overview_1994, evans_weak_2011, evans_checking_2006, roos_sens_2015} or to aspects of the likelihood function \citep{berger_overview_1994, bonat_2013_unit}.
It can also refer to the choice of data D, for example, the percentage of left-out observations \citep{kallioinen_2021_sens, ohagan_fractional_1995, vehtari_pareto_2021, vehtari_2017_loo}, or to hyperparameters of the posterior approximator A \citep{hoffman2014no, radev_amortized_2020, doucet_smc_2001, yaoYesDidIt2018}.
Thus, the term essentially refers to any aspect in which a P(A)(D) model can be sensibly modified.

More formally, we want to investigate the sensitivity (inverse robustness) of the posterior of some quantity $\psi = \psi(\alpha)$ with respect to some variable $\alpha$ which exerts a potential influence on a component of interest.
If we are only interested in investigating the sensitivity of a specific (point) summary of the posterior, we convey this by writing $T(\psi(\alpha))$ for an arbitrary summary statistic $T$.

For example, we can study likelihood or prior sensitivity by power-scaling the respective components of the P model, that is, replacing the joint model $\lik \, \prior$ with $\lik^\alpha \, \prior$ or $\lik \, \prior^\alpha$, respectively \citep{ohagan_fractional_1995, ibrahim_power_2000, bissiri_general_2016, kallioinen_2021_sens}. 
Of course, one can also choose to power-scale only parts of the likelihood or parts of the joint prior. 
Although it is just one of many ways to systematically perturb a P model, power-scaling is a popular approach due to its simplicity and natural integration with existing workflows \citep{bissiri_general_2016, kallioinen_2021_sens}.

Regardless of the exact perturbation method, we can define (local) sensitivity as a measurable distance (represented by a function $f$) between the results of the current P(A)(D) model at value $\alpha_0$ and an alternative value $\alpha_1$ that implies a different P(A)(D) model, diverging from the original one only in the choice of $\alpha$ \citep{kallioinen_2021_sens, roos_sens_2015}:
\begin{equation}
   \text{Sen}_{\alpha}(T(\psi), \alpha_0, \alpha_1) := f(T(\psi(\alpha_0)), T(\psi(\alpha_1)).
\end{equation}
For example, if $T$ were a posterior expectation or a quantile of some univariate quantity $\psi$ and $\alpha$ were a hyperparameter of the prior $p(\psi) = p(\psi \mid \alpha)$, then $f$ could simply be the absolute difference between these expectations or quantiles as implied by $\alpha_0$ and $\alpha_1$, respectively. 
This definition can further be generalized to sets of alternative $\alpha$ values in the neighborhood of $\alpha_0$ \citep{roos_sens_2015}.

If $\alpha$ can be perturbed continuously (e.g., using power-scaling) and if $T(\psi(\alpha))$ is differentiable at $\alpha_0$, we may also define sensitivity as a function of the derivative of $T(\psi(\alpha))$ evaluated at $\alpha_0$ \citep{kallioinen_2021_sens, perez_sens_2006}:
\begin{equation}
\text{Sen}_{\alpha}(\nabla T(\psi), \alpha_0) := f \left( \left. \frac{d \, T(\psi(\alpha))}{d \, \alpha} \right|_{\alpha=\alpha_0} \right).
\end{equation}
The latter definition has the advantage that no further value $\alpha_1$ has to be chosen, but it has a smaller range of applicability and potentially more difficult interpretation.
In both of the above definitions, we can always choose $f$ such that the sensitivity is non-negative with a value of $0$ indicating complete insensitivity.

For complex models, small amounts of sensitivity are almost always expected but may not practically matter. Accordingly, we would say that $T(\psi(\alpha)$ is \emph{practically sensitive} with respect to $\alpha$ if
\begin{equation}
   \text{Sen}_{\alpha}(T(\psi), \alpha_0, \alpha_1) > \delta \qquad (\text{or }  \text{Sen}_{\alpha}(\nabla T(\psi), \alpha_0) > \delta)
 \end{equation}
for some chosen threshold $\delta$ that depends on the sensitivity measure and the modeling context \citep{kallioinen_2021_sens}. 
For example, let $T$ denote a posterior mean and $\psi$ denote a standardized effect size that we would deem sensitive if a change exceeds $\delta = 0.2$ standard deviation units.
Then, the results would be practically sensitive to the given perturbation, if changing $\alpha$ from $\alpha_0$ to $\alpha_1$ implied $|T(\psi(\alpha_0)) - T(\psi(\alpha_1))| > 0.2$.

When evaluating practical sensitivity related to hyperparameter choice within a class of posterior approximators, robustness is highly desirable, since approximators should ideally converge to the same target (see Section \ref{convergence}).
What is more, as PA models grow in complexity, analyses based on a single approximator may gain trustworthiness by some form of multiverse analysis employing multiple approximators \cite{wagenmakers2022one}.
However, it is currently unclear how to systematically weigh the relative contribution of different approximators when trying to aggregate results from multiverse analysis, since some approximators might yield very poor posterior approximations and thus skew any substantial conclusion.

Differently, when it comes to perturbations in P model assumptions or the observed data, neither practical sensitivity nor insensitivity is desirable \textit{per se}.
Rather, we would like results to be practically robust to perturbations if (a) the perturbations affect only nuisance components of the P(A)D models that are equally justifiable within the given context, or (b) if the perturbations are so small that they could very well have occurred due to uncontrolled or uncontrollable influences. 
In contrast, when different P model assumptions represent competing substantive theories of interest, we want the corresponding P(A)D models to be sensitive to these assumptions. 

Examples for (a) include different choices of non-equivalent likelihood families that have an overall similar complexity (e.g., Log-Normal vs. Gamma distribution for continuous positive data) or different P(A)D models that are capable of similar predictive performance (see Section \ref{predictive-performance}), in case the latter is not already the quantity of interest itself.
Examples for (b) include adding small amounts of noise to the data \citep{madry_adverserial_2017}, leaving out a small subset of the data \citep{vehtari_pareto_2021,vehtari_2017_loo}, or slightly changing prior hyperparameters when the goal is to specify weakly informative priors \citep{kallioinen_2021_sens}.
As the magnitude of the perturbations increases, we expect results to become practically sensitive to these perturbations and observed insensitivity would then be a reason for concern.
For example, if drastically increasing the amount of data would not reduce the posterior standard deviation of $\psi$, this would be an indication of empirical non-identifiability (\citep{gelman_prior_2017}; see also Section \ref{information_gain}) or an error in our model code \citep{gelman2020workflow}.

Another type of sensitivity, arising in modeling dynamic systems, is sensitivity to initial conditions \cite{lorenz1963deterministic, schuster2006deterministic}, which in our taxonomy can be understood as part of $\implicit$ models.
Sensitivity to initial conditions, popularly known as the \textit{butterfly effect}, implies that an arbitrarily small change in initial conditions can result in considerably different subsequent system states (or observed trajectories).
Moreover, this type of sensitivity can be considered as a hallmark of deterministic chaos \cite{schuster2006deterministic}.
In the context of dynamic models, the so-called Lyapunov exponent can measure a model's sensitivity to initial conditions \cite{benettin1980lyapunov}.
Lyapunov exponents characterize the rate of exponential divergence from perturbed initial conditions and the maximal Lyapunov exponent can be used to summarize the overall sensitivity of a model into a single number \cite{kantz1994robust}.
For more details on dynamic systems and deterministic chaos, we refer the interested readers to \citep{schuster2006deterministic, boccaletti2000control}.

\begin{table*}
\centering
\caption{Applicability of the ten utility dimensions for each model class of the PAD-taxonomy.}
\newcontent{
\begin{tabular}{lllll}
\toprule
 & P & PA & PD & PAD \\
\midrule
Causal Consistency & \cmark & \cmark & \cmark & \cmark \\
Parameter Recoverability & \cmark & \cmark & \xmark & \xmark \\
Predictive Performance & \xmark & \xmark & \cmark & \cmark \\
Fairness & \xmark & \xmark & \cmark & \cmark \\
Structural Faithfulness & \cmark & \cmark & \cmark & \cmark \\
Parsimony & \cmark & \cmark & \cmark & \cmark \\
Interpretability & \cmark & \cmark & \cmark & \cmark \\
Convergence & \xmark & \cmark & \xmark & \cmark \\
Estimation Speed & \xmark & \cmark & \xmark & \cmark \\
Robustness & \xmark & \xmark & \cmark & \cmark \\
\bottomrule
\end{tabular}
}
\label{table:applicability}
\end{table*}

\newcontent{
\subsection{Intermediate Summary II}
\label{intermediate_summary2}

In the preceding sections, we have proposed to focus on ten general utility dimensions pertaining to the different Bayesian model classes within our PAD model taxonomy.
\autoref{table:applicability} summarizes the applicability of each utility dimension to each model class; the justification for this classification can be gathered from the treatment of individual utilities. 
Throughout, we have tried to elucidate the rather diverse facets of these utilities compactly, while providing a comprehensive list of references for the interested readers.
We believe that considerations regarding these utilities are already \textit{implicit} in much of the applied Bayesian literature. 
However, besides disambiguation of core concepts, a major goal of collecting these utilities has been to stimulate their \textit{explicit} consideration in Bayesian workflows.
In the following section, we will highlight the interactions between some of these utilities and discuss their relative importance in terms of application-dependent utility hierarchies.
}

\section{Utility Hierarchies and Trade-offs}
\label{hierarchies_tradeoffs}

\newcontent{For comprehensive Bayesian model comparison across all utility dimensions, we need a thorough understanding of how the latter relate to each other.
Thus, we begin by highlighting important connections between selected utilities. 
Some connections have been discussed already in different places in Section \ref{good_bayesian_model}, but we summarize some of them here again for the reader's convenience.
}

\newcontent{
\subsection{Interplay Between Utilities}

\textit{Predictive Performance and Parameter Recoverability.} 
Any modeling goal can be summarized by a set of quantities whose inferred model-based approximations are then used in subsequent decision-making. 
These quantities of interest may be manifest (observable) or latent (unobservable), leading either to a focus on predictive performance (observable quantities) or parameter recoverability (latent quantities). 
Statistically, these two utilities can be evaluated similarly by comparing model-based approximations with their real-world or ground-truth counterparts. 
However, despite the statistical similarity in evaluation, prioritizing either of the two leads to substantially different modeling workflows, as detailed further below.

\textit{Parameter Recoverability and Convergence.} 
Both parameter recoverability and convergence aim to quantify the difference between model-based results and some ideal theoretical target that we want to approximate as best as possible. 
However, the two utilities differ in what the \textit{ideal target} is. 
For parameter recoverability, it is a known ground-truth from which the data D was implicitly or explicitly generated. 
For convergence, it is the best possible posterior approximation that a particular approximator A can achieve for a given PD model. 
Indeed, these utilities are different for two main reasons.
First, even the analytic posterior of a PD model may perform very poorly in terms of parameter recovery, for example, when D contains too little information relative to the complexity of P. 
Second, for biased approximators, the ideal convergence target is not even the analytic posterior itself but rather the ``closest possible''  distribution within the scope of A.

\textit{Convergence and Estimation Speed.} 
In most cases, convergence determines the definition of estimation speed by defining the latter as the time from the start of running the approximator to convergence. While this definition usually works well, there are some scenarios where it falls short. First, sometimes we may want to define estimation speed less strictly as the time until
termination of the approximator run after which useful results can be obtained. This definition can be sensible especially when the goal is to achieve good predictive performance. If that goal was already achieved with a formally non-convergent model, there may be no need to bother with convergence anymore. Second, when using amortized approximators, we split the approximation process into a time-intensive training step and a subsequent inference step, which is then almost instant. In this context, convergence can be easily defined only for the training step, while estimation speed has clear definitions both for the training and inference step.

\textit{Causal Consistency and Fairness.}
Fairness is not detached from causal considerations.
For example, measurement fairness aims to ensure that the items in a test battery measuring a psychometric construct relate to this construct in the same way for all groups differing in protected attributes. 
In the associated causal graph, the causal effect of the latent construct on the measured items should be conditionally independent of the protected attributes given the unprotected attributes.
However, fairness considerations reach beyond causality, as they carry important ethical, political, and societal aspects that causal modeling alone cannot appropriately account for (i.e., questions of fact vs. questions of value).

\textit{Causal Consistency and Structural Faithfulness.} 
Both causality and structural faithfulness aim to represent an empirical phenomenon or the characteristics and constraints of an assumed stochastic process as closely as possible via an adequate P model. 
However, they differ in what aspects of the process they try to encompass. 
Causal consistency focuses on the specification of conditional independence between (sets of) variables given (sets of) other variables. 
This structure is then reflected in the P model’s probabilistic factorization without referring to specific distributions or functional forms. Differently, structural faithfulness deals with the details of the P model itself, for example, what functional forms connect the conditionally dependent variables or which probability distribution they follow. 

\textit{Structural Faithfulness and Parsimony.}
Changing structural faithfulness affects parsimony although in different directions depending on the kind of P models being considered and how structural faithfulness is increased. For example, when adding physical constraints such as symmetries, parsimony is increased along with structural faithfulness as the model does not have to learn the constraints from data. In contrast, when modeling additional probabilistic structures (e.g., of time and space), parsimony is usually decreased as new parameters have to be added to the P model to account for these structures. 

\textit{Parsimony and Interpretability.}
Higher parsimony is often associated with higher interpretability. However, there are notable expectations. For example, a highly regularized P model (e.g., through continuous shrinkage priors) may be considered highly parsimonious in terms of the effective number of parameters, but at the same time remain largely opaque because the parameter dimensionality itself remains high. As another example, low dimensional yet highly non-linear systems may feature directly interpretable parameters (e.g., growth rate in logistic map models), but their joint influence on the system's behavior may be very hard to understand without the aid of model simulations. 

\subsection{Utility Trees}
\label{utility_trees}

Having highlighted some key connections between utilities, we will now move on to evaluating the utilities' relative importance depending on the goal of inference. We will differentiate between a utility \emph{hierarchy}, where one utility is strictly more important than another, and a utility \emph{trade-off}, where we can achieve a gain of one utility at the cost of a loss of another. 
Utility hierarchies, utility trade-offs, as well as the relative importance of different utilities, are inevitably application-specific and contingent on the particular modeling goals. 

The first branching point in ranking the different utilities is whether we are interested in observable or latent quantities for subsequent decision-making, leading to what we call observable and latent inferential goals. 
This distinction is equivalent to focusing on either predictive performance or parameter recoverability as a primary utility. Related binary perspectives have been put forward, for example, as the difference between two ``statistical cultures'' \cite{breiman_statistical_2001} or the prediction-explanation dilemma \cite{yarkoni2017choosing, shmueli2010explain}. 
However, the distinction between the two goals has not been discussed in the context of explicit model utilities. 
Below, we present two utility trees defining hierarchies and trade-offs for the two kinds of inferential goals.

The way we would like to see these utility trees being used in practice is that analysts (a) build and improve their models in a way that respects utility hierarchies and (b) talk explicitly about the utility trade-offs they have been making in the process. 
This should enable users of the models and consumers of statistical inference to understand which model-building decisions have been made, why they have been made, and how they affect the trustworthiness of model-based decisions.
}

\subsection{Utility Tree for Observable Inferential Goals}

\begin{figure}
\centering
\includegraphics[width=0.99\linewidth]{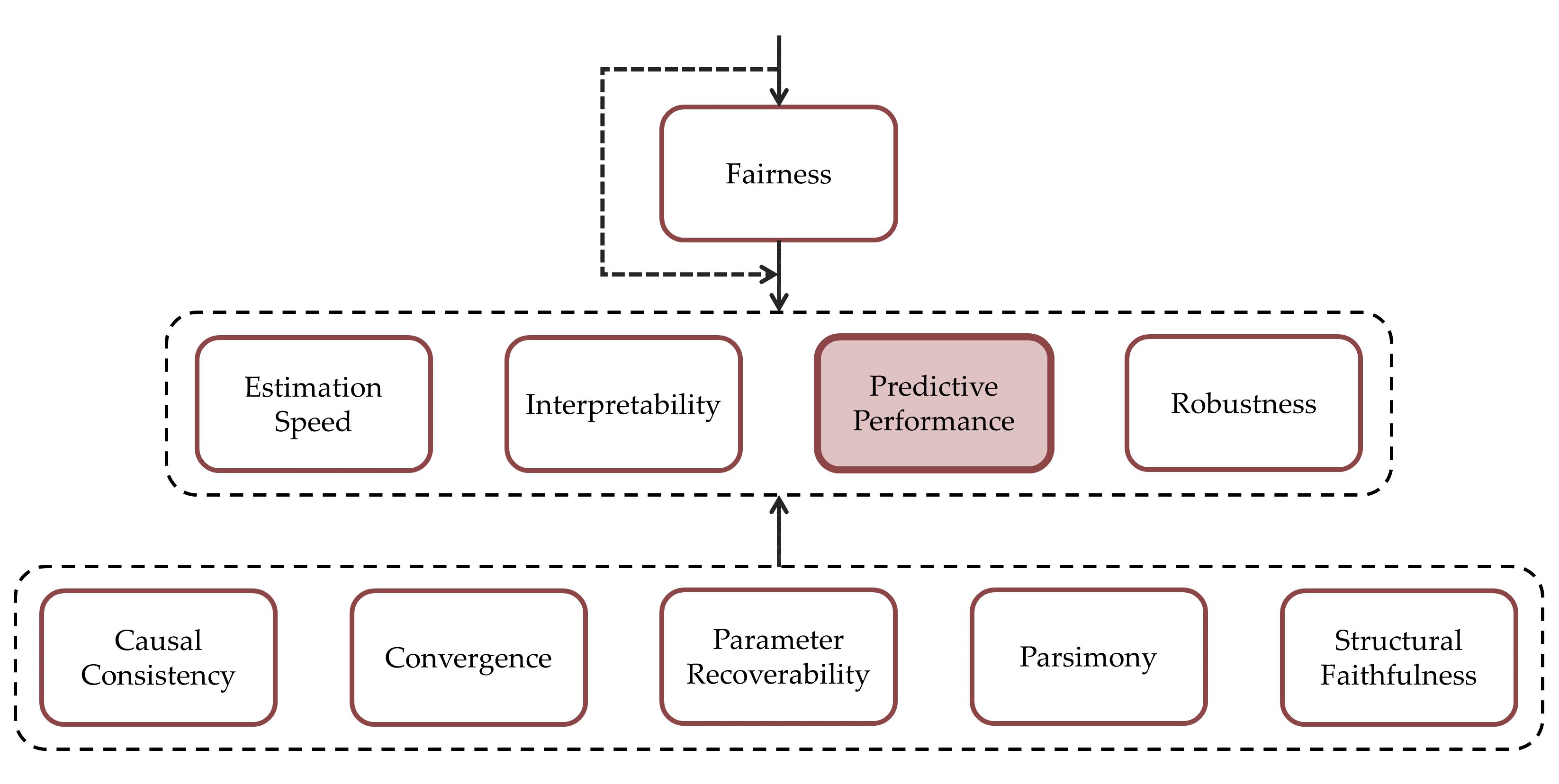} 
\caption{Utility tree for model-based inference of observable quantities (prediction).}
\label{fig:prediction_utility_tree}
\end{figure}

The workflow for observable inferential goals centers around the predictive performance of PAD models as a central utility. 
This is the prevalent perspective on modeling in machine learning research. Given its practical nature, this perspective would require a practically usable representation of a posterior distribution from which predictions can subsequently be obtained, hence the focus on PAD models. 
Below, we discuss our proposed model utility tree for observable inferential goals (see Figure~\ref{fig:prediction_utility_tree}).

\subsubsection{Primary Utilities}

\emph{Fairness}: The fact that predictive performance is the central utility under this perspective does not mean that it would be the sole or even the most important utility to consider. 
Rather, on the top of the utility hierarchy, we need to check whether the predictive goals concern certain aspects related to fairness. If they do, our PAD model needs to satisfy the fairness criteria agreed upon in the corresponding domain, otherwise, it would be considered invalid from an ethical and/or legal perspective (see Section~\ref{fairness}), regardless of how good its predictive performance is. If fairness concerns do not apply to the particular PAD model, the fairness utility can be circumvented. 

\subsubsection{Secondary Utilities}
\label{pp-secondary-utilities}

The secondary level of our predictive utility tree includes (in addition to predictive performance itself), estimation speed, interpretability, and robustness (in alphabetical order) as utilities across which trade-offs can be made. Notably, we do not require convergence of the PAD here and view estimation speed simply until termination of the approximator, regardless of whether or not convergence had been reached (see Section~\ref{speed}).
While the three additional central utilities may exhibit trade-offs among each other, increasing them may in particular justify (some) reduction in predictive performance. 

\emph{Estimation speed}: If achieving high predictive performance requires either a very high dimensional parameter space (e.g., as in a neural network) or the repeated evaluation of a complex simulator (e.g., in a differential equation-based mechanistic $\implicit$ model), then estimation speed will exhibit a trade-off with predictive performance. 
In other words, PAD models with easy-to-evaluate or easy-to-simulate likelihoods may obtain faster posterior approximation, but may also yield worse predictive performance. 
Whenever estimation speed becomes prohibitive for the practical application of a PAD model, the context may justify the use of another P model, even if the latter sacrifices (some) predictive performance. The same logic can justify using approximators that speed up the approximation of a PAD model, even at the expense of losing (some) predictive performance (e.g., using VI instead of MCMC-based approximators; see \cite{blei2017variational} or Section~\ref{sec:pa_models}).

\emph{Interpretability}: Interpretability is often higher in P models with lower parameter dimensionality and linear structure (see Section~\ref{interpretability}). 
However, depending on the complexity of the real data generator, low-dimensional and/or linear(ish) models may have worse predictive performance than higher-dimensional and/or more non-linear models. Yet, even if predictive performance is the main goal, it may still be legitimate (or even legally required) to use a more interpretable PAD model, even at the cost of some predictive performance.
 
\emph{Robustness}: A PAD model yielding high predictive performance on some test data may yield surprisingly poor predictions on slightly modified data (e.g., adversarial attacks on deep neural networks, see \cite{akhtar2018threat}). 
In a similar vein, a PAD model that is well predicting given some reasonable initial values of an approximator A may deliver worse predictions for some other (equally reasonable) initial values \citep{lee_initial_1991}. 
Both of these sensitivity types are not desirable and it can be legitimate to sacrifice (some) predictive performance for an increase in robustness against small perturbations.

\subsubsection{Tertiary Utilities}
\label{pp-tertiary-utilities}

At the third level of the predictive performance tree, we find supporting utilities that may serve as proxies for the central utilities. 
Specifically, these include causal consistency, convergence, parameter recoverability, parsimony, and structural faithfulness (in alphabetical order). Tertiary utilities are often easier to evaluate and available ``early'' in the model-building workflow, for example, when they are only requiring a P model instead of a PAD model. Using these utilities can thus help speed up the model-building process. However, these supporting utilities should only guide final modeling choices whenever multiple models are equally justifiable with respect to primary and secondary utilities.

\emph{Causal consistency}: If the quantities of interest are purely predictive, enforcing a P model to satisfy causal consistency (or even thinking about a causal graph in the first place) is not required and may even have detrimental effects on predictions \cite{mcelreath2020statistical}.
In other words, for predictions, it would usually be entirely irrelevant how the association between two variables came to happen, as long as the input variables are predictive of the outcome variables. 
However, causal consistency can still be a supporting utility to reduce the \textit{a priori} admissible model space by ruling out variables or interactions, as well as related P model terms, for which a causal graph implies a lack of relation to the target variables (i.e., no path between covariates and target, regardless of path direction). 
Considering a genetic association study as an example, we can rule out gene areas that only encode genes whose effects are known and understood to have no plausible relationship to the phenotypes being predicted \citep{wainberg_genes_2022}.
    
\emph{Convergence}: Convergence is not strictly required in the predictive utility tree, since a non-converged PAD model may still exhibit satisfactory predictive performance. 
That said, achieving convergence will likely imply an improvement in central utilities as well. 
Not only is this true for predictive performance itself, but also for robustness. 
For instance, a non-converged PAD model may vary arbitrarily from another non-converged PAD model, whereas we can expect them to be (more) similar upon convergence, at least for approximators that have the potential to explore the full posterior (see Section~\ref{sec:pa_models}). 
Accordingly, before studying central utilities that may be costly to evaluate, we can use convergence as a shortcut to rule out PAD models with low potential to score high in those central utilities.

\emph{Parameter recoverability}: 
Parameter recoverability can indirectly enhance predictive performance since nearly non-identifiable P models and poorly calibrated PA models can be discarded early in a model-building workflow.
Such models can neither yield good predictions, nor trustworthy uncertainty representation, as some information gain is necessary to achieve posterior predictions that are different from prior predictions (see Section~\ref{information_gain}).
However, strict parameter recoverability still plays a secondary role for the central goal of achieving good predictions.
For instance, highly over-parameterized P models may achieve zero Bayesian surprise (i.e., no difference between prior and posterior) for a large fraction of their parameters, but still yield reasonable predictions based on the few identifiable parameters \cite{radev2021outbreakflow}. 
    
\emph{Parsimony}: In addition to having aesthetical value in itself (see Section~\ref{parsimony}), parsimony as a supporting utility bears close relations to estimation speed, interpretability and predictive performance: More parsimonious models tend to be (a) faster to estimate, at least when comparing P models that are nested (i.e., one model is a special case of the other), (b) more interpretable, as fewer parameters have to be considered simultaneously, and (c) less prone to overfitting (although they may be prone to underfitting). 
Some forms of parsimony (both plain number of parameters and \textit{a priori} effective number of parameters; see Section~\ref{P-parsimony}) are available before running any approximator. Correspondingly, we can utilize parsimony as an \textit{a priori} proxy for the central utilities.

\emph{Structural faithfulness}: Structural faithfulness comprises several P model characteristics that we ideally know and understand before running any approximator: variable scales, probabilistic symmetries, and physical constraints (see Section~\ref{structural_faithfulness}). Structural faithfulness is related to multiple central utilities. Most importantly this concerns predictive performance, as structurally faithful models are more likely to predict more accurately while requiring less data and showing better uncertainty calibration \citep{raissi_pinn_2019, williams_litter_2017}. But structural faithfulness is also related to estimation speed (for the better or worse; \citep{gelman_data_2006, brms1}) as well as robustness, for example, small perturbations of the training data \citep{madry_adverserial_2017}. Accordingly, we can also treat structural faithfulness as a proxy for central utilities to reduce the \textit{a priori} considered model space to (sufficiently) structurally faithful models.

\subsection{Utility Tree for Latent Inferential Goals}

\begin{figure}[t]
\centering
\includegraphics[width=0.8\linewidth]{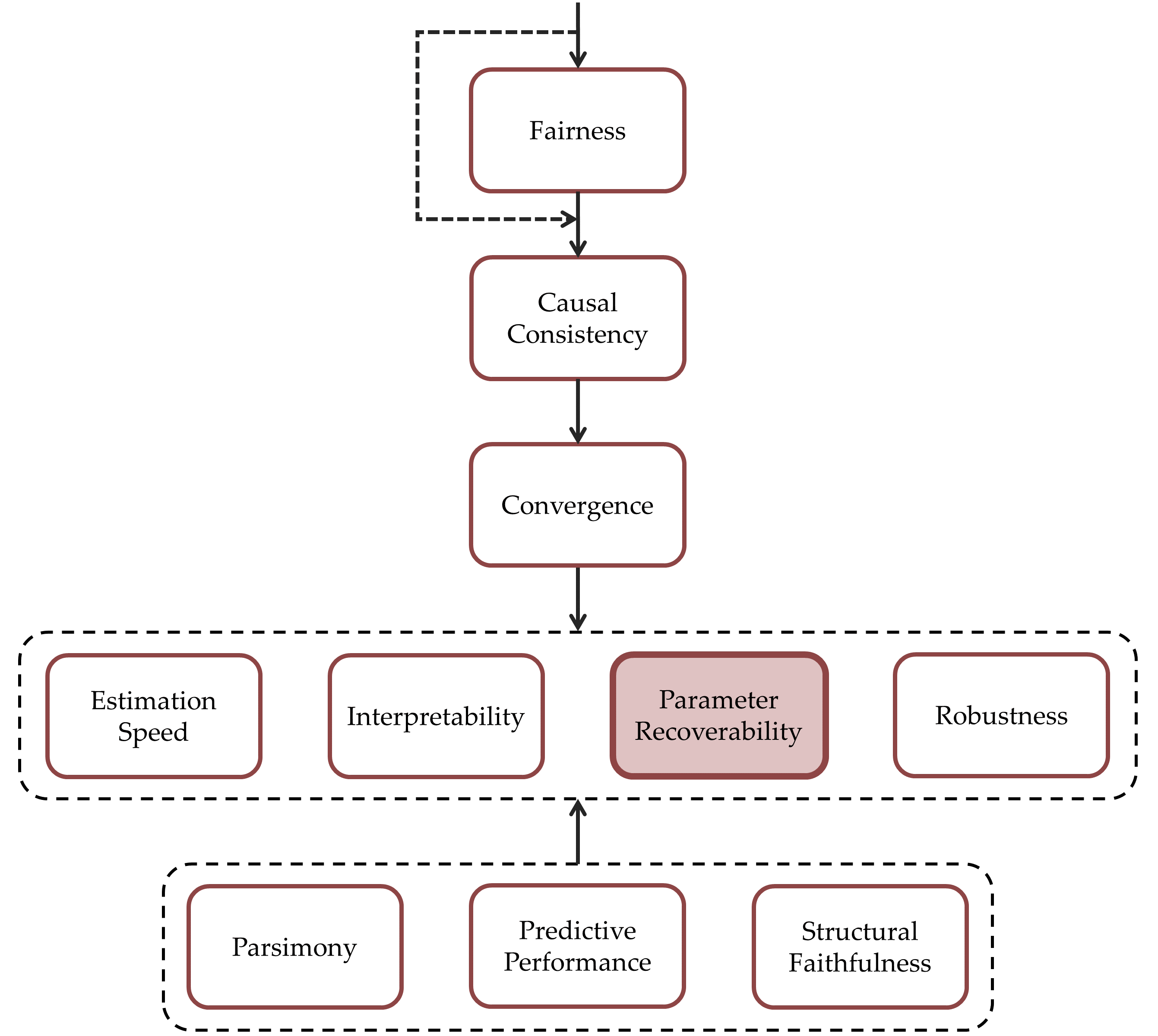} 
\caption{Utility tree for model-based inference of latent quantities (e.g., parameter estimation).}
\label{fig:parameter_utility_tree}
\end{figure}

The utility tree for latent inferential goals centers around the parameter recoverability of P or PD models as the central utility. 
Modeling goals following this perspective are almost entirely of theoretical, epistemic nature, and so the approximator is not itself part of the modeling goal. Yet, in practice, we will still almost always rely on PA and PAD models for practical evaluation, hence the indispensable role of the approximator.
Below, we discuss our proposed model utility tree for latent inferential goals (see Figure~\ref{fig:parameter_utility_tree}).

\subsubsection{Primary Utilities}

\emph{Fairness}: Once again, on top of the hierarchy, we find fairness for ethical and/or legal reasons whenever the modeling context has fairness-related implications (see Section~\ref{fairness}). 
First, at an individual (person-specific) level, we need to ensure that estimated latent parameters are fair with regard to individuals of protected groups. 
Second, at a more general (person-unspecific) level, we need to keep in mind how our inferences about latent parameters might trigger political decisions or societal processes affecting protected groups. 

\emph{Causal consistency}: Next, we need to ensure that the P model is causally consistent with the assumed, and theoretically justified, causal graph (see Section~\ref{sec:causality}). 
We argue that thinking about causal consistency is required for any latent modeling goal. 
Even if studies do not engage in (sufficient) causal analysis, and then correctly state that their results cannot be interpreted causally, there remains the (perhaps implicit) wish that a causal claim would be possible. 
What is more, even a pure measurement goal (e.g., estimating intelligence or personality traits without the need to relate latent parameters to each other) would need a causal model to decide and justify which observable variables to use for the estimation of the latent variables \citep{kaplan_structural_2008}. 
Finally, even if one might find a latent inferential goal that would be honestly satisfied with association only, causal analysis and discussion would still be required to prevent people from interpreting results causally.

\emph{Convergence}: Convergence of PAD models is a prerequisite for any practically trustworthy result of a latent inferential goal because we have no external validation criterion available during inference on real data as we would have when considering observable inferential goals. 
In fact, before convergence, posterior approximations may be almost arbitrarily incorrect, regardless of the kind of approximator being used (see Section~\ref{convergence}). Specifically, for asymptotically biased approximators (e.g., VI), the approximated posterior upon convergence may still be a bad representation of the analytic posterior if the expressive scope of the approximator is limited.
But even in such cases, a converged approximator is more likely to be closer to the analytic posterior than an arbitrary, non-converged approximator, and so the former is to be treated as more trustworthy.

\subsubsection{Secondary Utilities}

When it comes to inferential goals, parameter recoverability is the central utility of the secondary hierarchy. 
However, except for the identification sub-utility, it cannot be studied directly on real data because knowledge of the ground-truth is missing (see Section~\ref{parameter-recoverability}). 
As a result, many studies on parameter recoverability occur in the form of simulations or, if possible, mathematical analysis.
This also concerns studying trade-offs with other central utilities, namely, estimation speed, interpretability, and robustness. 
These remain instrumentally the same as for observable inferential goals and can reveal trade-offs with parameter recoverability for the same reason as for predictive performance (see Section~\ref{pp-secondary-utilities}).

\subsubsection{Tertiary Utilities}

Due to the lack of ground-truth latent parameters at real-data inference time, the tertiary, supporting utilities not only aim at speeding up model building but may also function as observable proxies of parameter recoverability. 
These supporting utilities are parsimony, structural faithfulness, and predictive performance. 
The reason for the relevance of the former two is the same as for observable inferential goals (see Section~\ref{pp-tertiary-utilities}), and so only predictive performance requires separate explanation and justification. 

\emph{Predictive Performance}: The relation between predictive performance and parameter recoverability is complicated and using the former as an observable proxy for the latter in a valid way requires great care \citep{shmueli2010explain, scholz_prediction_2022}. 
Most importantly, we should not choose causally inconsistent P models, even if they predict better \citep{mcelreath2020statistical, scholz_prediction_2022}. 
Fortunately, when taking the here-presented hierarchy of utilities seriously, this danger is banished by giving causal consistency priority over almost all other utilities for latent inferential goals. 
Within the class of causally consistent P models, it seems that using predictive performance as a proxy for parameter recoverability in (converged) PAD models represents a valid approach \citep{scholz_prediction_2022}. 
For example, we can utilize predictive performance to determine whether an extra probabilistic structure (e.g., accounting for potential temporal or spatial dependencies; see Section~\ref{probabilistic-structures}) is worth including or not \citep{burkner2021nfloo}. 
This affects the balance between structural faithfulness and parsimony, which in turn serve as proxies for parameter recoverability. 
As another example, the choice of likelihood functions driven by predictive performance can be used to improve parameter recoverability in the context of regression P models \citep{scholz_prediction_2022}.










\section{Conclusion}
\label{conclusion}

We proposed answers to two fundamental questions of Bayesian modeling, namely (1) ``What actually \emph{is} a Bayesian model'' and (2) ``What makes a \emph{good} Bayesian model''? 
Ultimately, we hope that both of these questions and the answers we provided will aid in thinking and talking about Bayesian models, as well as enhance the overarching model-building process, regardless of the specific methods and fields of application. 

As an answer to the first question (Section~\ref{bayesian_model}), we proposed the PAD model taxonomy that defines four different kinds of Bayesian models as subsets of the triple of joint distribution of all involved variables (P), the training data (D), and the posterior approximator (A). 
In this way, we put forward our view that modern Bayesian models are more than just likelihood and prior, but comprise a variety of ``external components'' that influence, and, in turn, are influenced by, the goals and the results of any statistical analysis.

As an answer to the second question (Sections \ref{good_bayesian_model} and \ref{hierarchies_tradeoffs}), we first argued that there are ten utility dimensions along which we can evaluate Bayesian models, namely, (1) causal consistency, (2) parameter recoverability, (3) predictive performance, (4) fairness, (5) structural faithfulness, (6) parsimony, (7) interpretability, (8) convergence, (9) estimation speed, and (10) robustness. 
Then, we proposed two utility trees that embody utility hierarchies and trade-offs depending on the particular inferential goals. 
We hope that our list of utility dimensions and structure of possible inferential goals is exhaustive (up to using synonyms and regrouping sub-utilities differently).
However, it may as well become incomplete in the future, as new ideas are born and rapidly developed, and we will be happy to incorporate these into our taxonomy.

%% file: graphics/simulation_gaps.tex
\begin{tikzpicture}[
every text node part/.style={align=center, font={\Large}},
dot/.style={draw, circle, minimum width=0.5cm},
network-box/.style={draw, rectangle, fill=blue!30, minimum height=5cm, minimum width = 2.5cm, inner sep=0.3cm},
posterior-box/.style={draw, rectangle, rounded corners = .10cm, minimum width = 3cm, inner sep=0.5cm},
arrow/.style = {->, very thick}
]

    \node[draw=none, label={Typical generative set $\mathcal{T}(\text{P})$}] (typical-generative-set) {
    \begin{tikzpicture}[scale=0.6]
    \filldraw[draw=black,fill=darkblue!30]  plot[smooth, tension=.8, fill=darkblue!30] coordinates {(-3.5,0.5) (-3,2.5) (-1,3.5) (1.5,3) (4,3.5) (5,2.5) (5,-2) (2.5,-3) (0.2,-1.5) (-3,-1.5) (-3.5,0.5)};
    \end{tikzpicture}
    };
    \node[dot, fill=darkblue!80, left=1.3, label=above right:{$y \sim p(y)$}] (x-star) at (typical-generative-set) {};
    
    \node[draw, color=darkblue,
    fit={(typical-generative-set)}, 
    label={[anchor=south west]south west:\color{darkblue}$\text{P}$ Model}, 
    rounded corners=.30cm, inner sep=0.8cm
    ] (generative-model) {};

    \node[dot, fill=errorcolor!80, below right = 0.4cm and -2cm of generative-model, label=below right:{$\data \sim p^*(y)$}] (x-obs) {};

    \node[network-box, fill=amortizer!80, right = of generative-model] (summary-network) {$\huge{\mathcal{H}}$\\Summary\\Network};
    
    \node[draw=none, right = of summary-network, label={Latent generative space}] (kde) {
        \begin{tikzpicture}
            \draw[thick,->] (-3.5, -3.5) -- (-3.5, 3);
            \draw[thick,->] (-3.5, -3.5) -- (3, -3.5);
            \foreach \x/\alpha in {3/20, 2/30, 1.2/40, 0.8/60, 0.4/80}
                \fill[darkblue!\alpha] (0, 0) circle (\x);
        \end{tikzpicture}
    };
    
    \node[dot, fill=darkblue!80,
    label=below:{$\mathcal{H}(y)$}
    ] (s-star) at (kde) {};
    \node[dot, 
    fill=errorcolor!80, 
    below left = -1.5cm and -1.4cm of kde,
    label=below:{$\mathcal{H}(\data)$}
    ] (s-obs) {};
    
    \node[network-box, fill=amortizer!80, right = of kde] (inference-network) {$\huge{\mathcal{F}}$\\Inference\\Network};

    \matrix[right = of inference-network, row sep = 0.2cm] {
    \node[posterior-box, fill=correctcolor!20]  (correct-posterior) {Correct\\Posterior}; \\ 
    \node[posterior-box, fill=errorcolor!20] (incorrect-posterior) {Incorrect\\Posterior}; \\
    };

    
    \draw [dashed, thick] (typical-generative-set) -- (x-obs) node [sloped,midway](M){};;
    
    \node[below left = of M] (simulation-gap) {Simulation gap};
    \draw [arrow] (simulation-gap)  -- (M);
    
    \draw [arrow, dashed] (x-star) -- (summary-network);
    \draw [arrow, dashed, color=errorcolor] (x-obs) -- (summary-network);
    
    \draw [arrow, dashed] (summary-network) -- (s-star);
    \draw [arrow, color=errorcolor, dashed] (summary-network) -- (s-obs);
    
    \node[below right = of s-obs] (simulation-gap-detected) {Simulation gap detected};
    \draw [arrow] (simulation-gap-detected) -- (s-obs);
    
    \draw [arrow, dashed] (s-star) -- (inference-network);
    \draw [arrow, dashed, color=errorcolor] (s-obs) -- (inference-network);
    
    \draw [arrow, dashed] (inference-network) -- (correct-posterior.west);
    \draw [arrow, color=errorcolor, dashed] (inference-network) -- (incorrect-posterior.west);
\end{tikzpicture}